\documentclass[12pt]{article}
\usepackage{amsmath,amsthm,amsfonts,amssymb,natbib,graphicx,booktabs,float,caption,subcaption,multirow,dsfont,setspace,chngcntr}
\usepackage{mathrsfs}
\usepackage{enumerate}
\usepackage{url}
\usepackage[utf8]{inputenc}
\usepackage[table]{xcolor}
\usepackage{lscape}
\usepackage{colortbl}
\usepackage{enumitem}
\usepackage{multicol}
\usepackage{longtable}
\usepackage{fancyhdr}
\usepackage{tikz}
\usepackage{cancel}
\usepackage{bibunits}
\usepackage{etoc}
\usepackage[colorlinks=true, linkcolor=blue, citecolor=blue, urlcolor=blue]{hyperref}

\newtheorem{definition}{Definition}
\newtheorem{theorem}{Theorem}

\newtheorem{corollary}[theorem]{Corollary}

\newtheorem{lemma}[theorem]{Lemma}

\newtheorem{assumption}{Assumption}
\newtheorem*{assumption*}{Assumption}
\newtheorem{proposition}[theorem]{Proposition}
\newtheorem{remark}[theorem]{Remark}

\newtheorem*{start*}{Starting Point}
\newtheorem*{proof*}{Proof}
\newtheorem*{note*}{Side Remark}

\def\spacingset#1{\renewcommand{\baselinestretch}{#1}\small\normalsize}
\newcommand{\blind}{0}

\defaultbibliographystyle{chicago}
\defaultbibliography{references/GL1}

\renewcommand{\today}{August 22, 2026}
\begin{document}
\makeatletter
\let\arxivmaketitle\maketitle
\let\arxivinnermaketitle\@maketitle
\let\arxivtitle\title
\let\arxivauthor\author
\let\arxivdate\date
\let\arxivthanks\thanks
\let\arxivand\and
\makeatother

\begin{bibunit}[chicago]
\makeatletter
\def\@extra@b@citeb{.main}
\def\@extra@binfo{.main}
\makeatother
\spacingset{1}
\def\spacingset#1{\renewcommand{\baselinestretch}%
{#1}\small\normalsize} \spacingset{1}

\title{Changes-in-Changes for Ordered Choice Models with Underreporting\thanks{%
We thank the editor, Michal Koles\'{a}r, as well as an Associate Editor and two anonymous referees for insightful comments that led to a much improved version. We are also grateful to  Martin Biewen, D\'{e}sir\'{e} Kedagni, Michael Knaus, David Pacini and seminar participants at T\"{u}bingen University, the North American Summer Meeting of the Econometric Society (Nashville) 2024, the Karlsruhe Institute of Technology (KIT), the University of St. Gallen, the Econometrics Workshop
on ``Causal Inference, Identification, and Measurement Error'' in Frankfurt 2023, the 2023 Asian Meeting of the Econometric Society in Singapore and in Beijing, the International Panel Data Conference (IPDC) 2023 in Amsterdam, the SWETA workshop in Siena 2023, the 9th Annual Conference of the International Association for Applied Econometrics (IAAE) in Oslo, the internal Frankfurt Econometrics workshop 2023  for helpful comments and discussions. Financial support from the DFG (DFG Grant No. 463916029) is gratefully acknowledged.
}}

\author{Daniel Gutknecht\thanks{%
Faculty of Economics and Business, Goethe University Frankfurt, 60629
Frankfurt am Main, Germany. E-mail: \texttt{Gutknecht@wiwi.uni-frankfurt.de.}%
} \and
Cenchen Liu\thanks{%
Faculty of Economics and Business, Goethe University Frankfurt, 60629
Frankfurt am Main, Germany. E-mail: \texttt{Liu@wiwi.uni-frankfurt.de.}%
}}
\date{\today}
\maketitle

\begin{abstract}

\noindent We develop a Difference-in-Differences framework for discrete, ordered outcomes subject to underreporting. Such outcomes commonly arise in self-reported surveys on socially undesirable or stigmatized behaviors, where respondents may conceal their true behavior. For a discrete Changes-in-Changes model that is shown to admit an equivalent threshold-crossing representation, we derive nonparametric bounds for the counterfactual and factual outcome distributions as well as for the associated quantile treatment effects when outcomes are underreported. These bounds are shown to be sharp uniformly across outcome levels under additional support conditions, and we propose suitable estimation and bootstrap inference procedures. In an extension,  we also consider a semiparametric underreporting model that allows to point identify and estimate distributional treatment effects. As an application, we investigate the impact of recreational marijuana legalization on the consumption behavior of 8th-grade students in several U.S. states.

\bigskip



\end{abstract}

\newpage
\spacingset{1.8} 

\etocdepthtag.toc{main}
\section{Introduction}\label{sec:introduction}

The Difference-in-Differences (DiD) method is widely used in applied economics to identify causal effects of policy changes. While the classical DiD framework focuses on mean treatment effects, policy changes may affect different parts of the outcome distribution in distinct ways. This observation has motivated a growing literature on distributional DiD methods that exploit pre- and post-treatment data to identify heterogeneous effects across the outcome distribution \citep[see, e.g.,][for an overview]{rothWhatTrendingDifferenceindifferences2023}. A prominent example of this approach is the Changes-in-Changes (CiC) model developed by \citet{AI2006}.

This paper develops a CiC framework for discrete, ordered outcomes that are potentially underreported, nesting the ``misreporting-to-zero'' or ``false-zero'' scenario as a special case. Underreporting is common in survey data on discrete outcomes of count, ordinal, or binary nature, particularly when socially undesirable or stigmatized behaviors are elicited. In such settings, respondents may conceal their true behavior, posing substantial challenges for identification and inference. Examples include the consumption of ``social bads'' such as marijuana (cannabis) or alcohol, as well as sensitive issues such as abortion or domestic violence.\footnote{For example, \citet{BHSZ2018} found that the actual incidence of marijuana consumption in an Australian survey was nearly double the reported figure (23\% vs. 12.2\%), while \citet{CM2005} identified about 20 percentage points underreporting in cheating behavior among undergraduate students in a U.S. survey. See also \citet{NDT2019} for other examples.}

The paper makes several contributions to the literature. First, we derive, in the presence of underreporting, nonparametric bounds for both the treated group's counterfactual outcome distribution after treatment and the factual outcome distribution. The bounds are shown to be sharp \textit{uniformly} across outcome levels under mild, additional assumptions. Moreover, we demonstrate that, without further restrictions, the bounds on the counterfactual distribution remain uninformative even if an instrumental exclusion restriction for consumption is available. By contrast, leveraging the CiC structure together with an exclusion restriction and an upper bound on the misreporting probability can lead to informative, tight bounds. The latter bound on the misreporting probability may either serve as a sensitivity parameter akin to, e.g., the $c$-dependence parameter in \citet{Masten2018} and the breakdown frontier in \citet{MP2020}, or simply reflect institutional knowledge. We also show that not every candidate value of this bound is admissible: sufficiently small values can be ruled out by the observed relationship between the reported outcome and the instrument. Our bounds extend the partial identification results of \citet{AI2006} for the discrete outcome case to underreporting and give rise to bounds on Quantile Treatment Effects on the Treated (QTTs) as an immediate corollary. { At the cell level, our bounds recover the one-sided underreporting bounds of \citet{MW2024} for binary outcomes without instrumental variation and extend them to general ordered outcomes. Under the standard parallel trends restriction on the mean of the untreated potential outcome, these cellwise bounds also yield bounds on the Average Treatment Effect on the Treated (ATT), as detailed after Corollary~\ref{COR:BOUND}. To the best of our knowledge, corresponding ATT bounds for an ordered outcome subject to one-sided underreporting have not previously been derived. Existing work on measurement error in DiD settings instead focuses on misclassification of treatment status \citep{DK2022,NN2025}. We nevertheless center the analysis on the CiC model because our primary objects are the entire counterfactual distribution and the associated QTTs rather than mean effects alone.} { Second, in the supplement, we show that the discrete CiC model without underreporting admits an equivalent threshold-crossing representation in which the observed outcome is generated by comparing a latent index with time-specific thresholds. Under this representation, the CiC restriction translates into restrictions on the location and scale parameters of the latent index across groups and time, which nest standard parallel-trends restrictions for nonlinear ordered choice models (e.g., Probit or Logit DiD models) as special cases. The representation therefore provides further insights into the discrete CiC model more broadly, and serves as the basis for a point-identified, semiparametric model with underreporting that considers consumption and reporting decisions individually and that allows for the recovery of  Distributional Treatment Effects on the Treated (DTTs)} (cf.\ Section \ref{sec:parametricmodel}).

As a third contribution, we develop estimation and inference theory for all bound functions and resulting QTTs. Under repeated cross-sectional sampling, we establish  weak convergence of the estimated lower and upper bounds for the counterfactual distribution uniformly across outcome levels. Because the bound operators involve nonsmooth transformations such as minimum operators, the relevant maps are only Hadamard directionally differentiable. We therefore follow \citet{FS2019} and \citet{MP2020} and employ bootstrap inference to construct uniformly valid Confidence Sets (CS) for the bounds of the counterfactual and factual distributions. The construction of QTT bounds and corresponding CS then uses the quantile-effect framework of \citet{CFMW2020}. This uniformity is practically useful because it allows the confidence band to be used for a range of functional comparisons without requiring the researcher to commit ex ante to a specific outcome { or quantile} level. With a pre-specified probability, the band covers the entire bound function across all { such} levels, so that any candidate counterfactual distribution function or QTT function that leaves the band at even a single point can be rejected at the corresponding level.

Finally, as an empirical application, we investigate the impact of recreational marijuana legalization for adults in several U.S. states on the short-term marijuana consumption of 8th-grade students. The focus on this group is particularly relevant because early cannabis use is widely viewed as especially consequential for later health and educational outcomes. Unlike existing studies that examined the effects of legalization of medicinal or recreational marijuana use on the mean \citep[cf.][]{AHR2015,WHC2015,CWFKSSH2017,HWB2022}, or on the timing of the first (early age) consumption \citep{WBJ2014}, we consider the effects of the legalization of recreational marijuana use on the entire consumption distribution of this population in the treated states in our sample, thus allowing for the possibility of heterogeneous effects of the reforms across usage levels. Moreover, in contrast to the existing literature, we address concerns that consumption behavior may not always be reported truthfully, and that reporting behavior may have been affected by legalization through shifts in stigma, public perception, or general attitudes towards marijuana consumption. Our analysis reveals that the benchmark specification that ignores underreporting yields little evidence of non-zero effects outside the extreme upper tail of the distribution. Once underreporting is allowed for, however, the picture becomes more asymmetric: at lower upper-tail quantiles the data rule out positive QTTs, whereas at the very top they rule out negative QTTs. A complementary semiparametric specification, reported in Section \ref{sec:empirical_parametric} of the supplement, yields a similar pattern: without covariates, the estimated effects are not significant, whereas including covariates implies a decline in non-consumption of about 2 percentage points and increases of about 1 percentage point in each positive consumption category. Allowing for misreporting further increases these estimated consumption effects by more than 50\%, while yielding little evidence of systematic changes in reporting behavior.

The paper is organized as follows. Section \ref{sec:setup} outlines the model setup and defines the causal parameters of interest. { Section \ref{sec:nounderreporting} studies the benchmark case without underreporting.} Section \ref{sec:Underreporting} extends the framework to account for underreporting and establishes sharp bounds for the counterfactual distribution and the resulting quantile treatment effects. Section \ref{sec:estimation} develops estimation and inference for these bounds. Section \ref{sec:application} empirically investigates the effects of recreational marijuana legalization on the consumption behavior of 8th-grade high-school students. Section \ref{sec:conclusion} concludes.

All proofs and additional material can be found in the Supplementary Material. Section \ref{sec:proofbounds} contains the proofs of the bound results, while Section \ref{sec:proofestinf} contains the proofs for estimation and inference. Section \ref{sec:AddCov} outlines how estimation and inference can be modified when additional covariates are present or some of the variables are continuous. Section \ref{sec:parametricmodel} presents the semiparametric model for underreporting, which delivers point identification and estimation of distributional treatment effects. Finally, Section \ref{sec:MCBounds} reports a Monte Carlo study assessing the finite-sample performance of the bound estimators and the associated bootstrap CS, while Section \ref{sec:ADDEMP} presents additional tables and figures for the empirical application.

\textbf{Notation}. { For random variables $A$ and $B$, we write $A\sim B$ if they are equal in distribution.}
Let $\overline{\mathbb{R}}:=\mathbb{R}\cup\{-\infty,+\infty\}$. For a random variable $Y$, let $F_Y$ denote its cumulative distribution function (CDF), with $F_Y(-\infty)=0$ and $F_Y(+\infty)=1$. The associated generalized inverses are
\begin{align*}
&F_Y^{(-1)} : [0,1] \to \overline{\mathbb{R}}, 
\quad
q \mapsto F_Y^{(-1)}(q)
:= \sup \left\{ y \in \mathrm{supp}(Y) \mid F_Y(y) \le q \right\}, \\
&F_Y^{-1} : [0,1] \to \overline{\mathbb{R}}, 
\quad \ 
q \mapsto F_Y^{-1}(q)
:= \inf \left\{ y \in \mathrm{supp}(Y) \mid F_Y(y) \ge q \right\},
\end{align*}
with the conventions $\sup \emptyset = -\infty$ and $\inf \emptyset = +\infty$. { For $q\in(0,1)$, $F_Y^{-1}(q)$ is the usual quantile function of $Y$ at $q$. The two coincide when $F_Y$ is continuous and strictly increasing, but generally differ for discrete outcomes. For any CDF $F$, let $\operatorname{supp}(F)$ denote the support of a random variable with distribution function $F$} (see Definition \ref{SupportF} in the supplement). The same notation applies to conditional distribution functions.
We adopt the convention that conditional probabilities are defined as zero whenever the conditioning event has probability zero. { This assignment is arbitrary because any two resulting conditional probabilities agree almost surely \citep[pp.~428--429]{Billingsley1995}.}

\section{\texorpdfstring{{ Setup}}{Setup}}\label{sec:setup}

To introduce the CiC framework for discrete, ordered outcomes, let $Y(d)$, $d\in\{0,1\}$, denote the potential outcome with support contained in a finite set $\mathcal{J}$. Motivated by the empirical application on drug consumption frequencies in Section~\ref{sec:application}, we henceforth refer to $Y(d)$ as the “true” potential consumption level. We set $\mathcal{J}:=\{0,1,\ldots,J\}$ with $J\ge 1$.  Let $G\in\{0,1\}$ denote the group membership indicator, where $G=1$ corresponds to the treated group, and let $T\in\{0,1\}$ denote the time period indicator, where $T=1$ corresponds to the post-treatment period. The binary treatment indicator is given by $D=G\cdot T$, so that $D=1$ if and only if $(G,T)=(1,1)$, and $D=0$ otherwise. Throughout the paper, we focus on the case
of two groups and two time periods for notational simplicity. Extensions to
multiple pre-treatment periods are immediate, at the cost of more cumbersome
notation. Finally, to avoid degenerate cases, we assume \(\pi_{gt}:=\mathrm{Pr}\!\left(G=g,T=t\right)>0\) for all \((g,t)\). { For any random variable $V$, write $V_{gt}\sim V\mid G=g,T=t$, as in \citet{AI2006}.}

As emphasized in the introduction, a persistent challenge in analyzing
sensitive behaviors, such as the consumption of ``social bads'', is the
presence of ``false zeros'' or, more broadly, underreporting. Individuals
often conceal their true participation in such behaviors when responding to
retrospective surveys \citep{NDT2019}. This leads to a partially observable
outcome model in which a specific outcome value is observed in the data if
and only if the true outcome corresponds to that value and is reported
truthfully \citep[cf.][]{P1980}. To accommodate this feature, we introduce
potential ``reported'' consumption, denoted by $C(d)$, $d\in\{0,1\}$, which
satisfies $C(d)\le Y(d)$ with probability one. The case $C(d)=Y(d)$
corresponds to truthful reporting, whereas $C(d)<Y(d)$ captures
underreporting. We formalize this in our first assumption within the DiD framework. 

\begin{assumption}[Underreporting]\label{A1}
 For each 
$j\in \mathcal{J}\setminus \{J\}$ and $(d,g,t)\in\{0,1\}\times\{0,1\}\times\{0,1\}$, and for some known constant $\alpha\in[0,1]$, the ``true'' and ``reported'' consumption levels
$Y(d)$ and $C(d)$ satisfy
\begin{align*}
	&\mathrm{Pr}\!\left(C(d)\le Y(d)\mid G=g,T=t\right)=1,\\
&\mathrm{Pr}\!\left(C(d)\le j \,\middle|\, Y(d)\ge j+1,G=g,T=t\right)
\le \alpha.
\end{align*}
\end{assumption}

Assumption~\ref{A1} explicitly characterizes one-sided misreporting, that is, underreporting, for the potential outcome in a given group at a given time. Although it is stated for all \((d,g,t)\), this uniform formulation is adopted for simplicity, and the identification analysis would go through under weaker, cell-specific restrictions on $\alpha$. Such one-sided misreporting is natural in survey settings involving socially undesirable or stigmatized behaviors, including self-reported substance use, where respondents may conceal or downplay true behavior, while overreporting is not really plausible \citep[see, e.g.,][]{BHSZ2018,GHSZ2018}. Alternative scenarios where overreporting is the only likely misreporting form (e.g., a survey question about the frequency of physical exercise) can be accommodated by adapting the assumption accordingly. Thus, the first part of the assumption rules out overreporting, while the second part introduces a uniform upper bound on underreporting probabilities across outcome levels that can reflect institutional knowledge or  serve as a sensitivity parameter. Specifically,
 when \(\alpha=1\), the assumption imposes no restriction on the extent of underreporting, whereas when \(\alpha=0\), reporting is truthful almost surely. Intermediate values of \(\alpha\) therefore impose flexible, nonparametric restrictions on the extent of underreporting. As will be shown in Proposition~\ref{PROP:final2}, in the absence of any restriction on underreporting, the resulting bounds on the counterfactual distribution become uninformative. Hence, \(\alpha\) can serve as an important sensitivity parameter that governs the sharpness and informativeness of the resulting bounds.
 
Finally, the observed outcome is generated according to the standard
observability rule,
\begin{equation}\label{EQOBSERVABILITY}
C = D\cdot C(1) + (1-D)\cdot C(0).
\end{equation}
This identity highlights that, due to underreporting, the observed distribution of $C$ does not generally coincide with the distribution of the true potential outcome $Y(d)$, even within the subpopulation for which $D=d$.

Our primary objects of interest are the QTTs, which characterize how treatment shifts different points of the outcome distribution for treated units, such as the median or the $75$th percentile. In our empirical application, they measure the effect of marijuana legalization on different points of the consumption distribution,  among adolescents residing in treated states. 
To formalize this, we first define the conditional quantile functions of the potential outcomes for the treated population. For each $d\in\{0,1\}$ and quantile level $\tau\in(0,1)$, let
\[
Q_{Y(d)\mid D=1}(\tau)
:=
F^{-1}_{Y(d)\mid D=1}\left(\tau \right)
=
F^{-1}_{Y(d)\mid G=1,T=1}\left(\tau 	\right).
\]
For each $\tau\in(0,1)$, the QTT at quantile level $\tau$ is defined as
\begin{equation}\label{EQQTT}
\Delta_{\mathrm{QTT}}(\tau)
:=
Q_{Y(1)\mid D=1}(\tau)
-
Q_{Y(0)\mid D=1}(\tau).
\end{equation}
In the absence of underreporting, $Q_{Y(1)\mid D=1}(\tau)$ is identified from the observed treated group in the post-treatment period. By contrast, $Q_{Y(0)\mid D=1}(\tau)$ depends on the counterfactual distribution of $Y(0)$ for treated units and is not identified without additional assumptions.

Finally, in Section \ref{sec:parametricmodel} of the supplement, we also consider
Distributional Treatment Effects on the Treated (DTTs), defined for each
$j \in \mathcal{J}$ as
\begin{align}\label{EQQTTDTT}
\Delta_{\mathrm{DTT}}(j)
:=
\Pr\!\left(Y(1)=j \mid D=1\right)
-
\Pr\!\left(Y(0)=j \mid D=1\right).
\end{align}
For discrete outcomes, the DTTs in \eqref{EQQTTDTT} complement the QTTs by tracking how treatment reallocates probability mass across outcome levels. In our empirical application, they capture how marijuana legalization changes the probability that adolescents in treated states fall into each consumption category.\footnote{An analogous extension of the QTT and DTT results to the untreated group is possible. It can be obtained by reversing the roles of treated and untreated groups and imposing the corresponding support conditions; see \citet[Section~3.2]{AI2006}.}

\section{Bounds Without Underreporting}\label{sec:nounderreporting}
 
As a first step, consider the benchmark case without underreporting, $\alpha=0$ in Assumption~\ref{A1}, so that $C_{gt}(d)=Y_{gt}(d)$ a.s. In this setting, the distribution of $Y_{11}(1)$ is identified from the treated group in the post-treatment period, but the counterfactual distribution $F_{Y_{11}(0)}$ is not identified without further restrictions. We therefore require additional structure on the untreated potential outcome $Y(0)$ across groups and time. For discrete, ordered outcomes, we adopt the discrete CiC framework of \citet{AI2006}:

\begin{assumption}[Discrete CiC Model]\label{A2}
There exist a real-valued random variable $U$ and a function $h:\mathbb{R}\times \{0,1\}\rightarrow \mathbb{R}$ such that $Y(0) = h(U,T)$ a.s. The random variable $U$ and the function $h$ satisfy the following conditions: 
\begin{enumerate}[label=(\arabic*)]
	\item $U\mid G=g\text{ is continuously distributed}$ for all $g\in \{0,1\}$,
	\item $U\perp T\mid G$,
	\item $\mathrm{supp}\left(U\mid G=1 \right)\subseteq \mathrm{supp}\left(U \mid G=0\right)$,
	\item $u\mapsto h(u,t)$ is nondecreasing,
\end{enumerate}
\end{assumption}
The function $h$ determines the untreated outcome $Y(0)$ and may depend on the time indicator $T$ and an unobservable scalar $U$ capturing latent heterogeneity. The specification $Y(0)=h(U,T)$ implies that group membership $G$ does not enter the outcome function directly and affects $Y(0)$ only through the distribution of $U$. In particular, conditional on $U=u$, the untreated outcome is identical across groups within a given period $T\in\{0,1\}$.

Condition (1) is standard in nonseparable models and rules out mass points in the latent heterogeneity within a group. Continuity of $U$ is a common assumption in nonlinear discrete choice models, including threshold-crossing specifications.  Condition (3) ensures sufficient support overlap to allow for extrapolation of the counterfactual distribution. The weak monotonicity condition in (4) is natural for discrete, ordered outcomes and aligns with the monotone index structure underlying threshold-crossing models.

Condition (2) is the key identifying restriction of the CiC model. It requires that the distribution of the latent variable $U$, although possibly different across groups, remains stable over time within a group. This conditional independence restriction distinguishes CiC from other distributional DiD approaches. One alternative is the distributional DiD model proposed by \citet{GKM2023}, which is based on time-invariance of the copula linking the untreated potential outcome and treatment assignment. Their identifying assumption imposes stability of the dependence structure between the untreated potential outcome and group membership across periods, while allowing the marginal distributions to evolve freely. The copula-stability restriction and the CiC restriction are not nested in general. They coincide only when the untreated potential outcomes are continuous with strictly increasing marginal distribution functions \citep[cf. Claim D.1 in][]{GKM2023}. { In Section \ref{sec:parametricmodel} of the supplement, we show that the discrete CiC model in Assumption~\ref{A2} admits an equivalent threshold-crossing representation, which clarifies the link with standard nonlinear ordered choice models for DiD (e.g., Probit or Logit DiD models). This representation serves as the basis for a semiparametric model with covariates and underreporting, which is also fully developed in that section.}

Under Assumption \ref{A1} with $\alpha=0$, so that $C_{gt}(d)=Y_{gt}(d)$ with probability one, Assumption \ref{A2} implies that the sharp bounds for the counterfactual distribution $F_{Y_{11}(0)}$ derived in Theorem~4.1 of \citet{AI2006} { can be obtained as}
{\begin{equation}\label{eq:final2-CIC}
F_{C_{10}}\!\left(F_{C_{00}}^{(-1)}\!\left(F_{C_{01}}\!\left(y\right)\right)\right)
\le
F_{Y_{11}(0)}\!\left(y\right)
\le
F_{C_{10}}\!\left(F_{C_{00}}^{-1}\!\left(F_{C_{01}}\!\left(y\right)\right)\right),
\end{equation}
for all $y\in\left[\underline{y}_{01},\overline{y}_{01}\right]$, where
$\underline{y}_{01}:=\inf \mathrm{supp}\!\left(C_{01}\right)$ and
$\overline{y}_{01}:=\sup \mathrm{supp}\!\left(C_{01}\right)$.
Moreover, $F_{Y_{11}(0)}(y)=0$ for all $y<\underline{y}_{01}$ and 
$F_{Y_{11}(0)}(y)=1$ for all $y>\overline{y}_{01}$.
}

\section{Bounds for Underreported Outcomes}\label{sec:Underreporting}

We now allow for underreporting by considering $\alpha\ge 0$, relaxing the benchmark case $\alpha=0$ in which $Y(d)=C(d)$ a.s. { In that benchmark, each observed distribution $F_{C_{gt}}$ identifies $F_{Y_{gt}(d)}$ for the realized treatment state $d=g\cdot t$, and the bounds in \eqref{eq:final2-CIC} apply. With underreporting, these distributions are only partially identified. Below, Proposition~\ref{PROP:misr_func_main} therefore first derives bounds on $F_{Y_{gt}(d)}$ in each cell from the observed reported outcomes, and Proposition~\ref{PROP:final2} then combines these bounds with the CiC structure to bound $F_{Y_{11}(0)}$. As Proposition~\ref{PROP:final2} shows, absent any restriction on underreporting ($\alpha=1$), the resulting bounds become uninformative, even with the instrument introduced below.} To possibly tighten these bounds, we introduce an exclusion restriction on an observable instrument that does not directly affect consumption behavior. In what follows, we exploit this instrument together with the underreporting parameter $\alpha$, as the two will play distinct roles in tightening the bounds.

\begin{assumption}[Exclusion Restriction]\label{A3}
Let $\mathcal{Z}_{gt}:=\mathrm{supp}\left(Z_{gt}\right)$. 
For every $y\in\overline{\mathbb{R}}$, $z\in\mathcal{Z}_{gt}$, $(g,t)\in\{0,1\}\times\{0,1\}$, and $d=g\cdot t$,\[
\mathrm{Pr}\!\left(Y_{gt}(d)\le y \mid Z_{gt}=z\right)
=
\mathrm{Pr}\!\left(Y_{gt}(d)\le y\right).
\]
\end{assumption}

Under the observability rule in \eqref{EQOBSERVABILITY}, the variables $(C,Z,G,T,D)$ are observed. Since treatment status satisfies $D=G\cdot T$, in each $(g,t)$ cell only the potential outcome corresponding to the realized treatment status, $Y_{gt}(d)$ with $d=g\cdot t$, is relevant for identification. The exclusion restriction is therefore imposed on $Y_{11}\left(1\right)$ in the treated post-treatment cell and on $Y_{gt}\left(0\right)$ in the remaining cells. This feature reflects the combination of underreporting and the DiD treatment assignment structure.

Assumption~\ref{A3} is used to tighten the nonparametric bounds and requires that $Z$ be independent of the true potential outcome $Y(d)$, so that any dependence between $Z$ and the observed outcome $C$ arises from reporting behavior rather than from an effect of $Z$ on true consumption. In our empirical setting, a natural example is a survey cooperation indicator measured after consumption has occurred. As such, this indicator is plausibly related to reporting behavior but is assumed to have no direct effect on the true consumption of marijuana. Finally, note that, if one were concerned that a single instrument does not satisfy the exclusion restriction for both potential outcomes, it would in principle be possible to use different instrumental variables for different potential outcomes.

Under Assumptions~\ref{A1} and \ref{A3} alone, that is, without imposing any CiC structure across groups and time, the exclusion restriction yields partial identification of the distribution of the true outcome within each \((g,t)\) cell. In particular, for the true potential outcome distribution \(F_{Y_{gt}(d)}\) with \(d=g\cdot t\), the sharp bounds are given by { the following result.}

{\begin{proposition}[Bounds for $F_{Y_{gt}(d)}$ without CiC Structure]\label{PROP:misr_func_main}
Suppose Assumptions~\ref{A1} and \ref{A3} hold for some $\alpha\in[0,1]$. Let $(g,t)\in \{0,1\}\times \{0,1\}$ and $d=g\cdot t$. If
$ \alpha_{gt}^*\leq \alpha\leq 1$, then
\begin{align}\label{EQBOUNDS}
L_{gt,\alpha}(y)\le F_{Y_{gt}(d)}(y)\le U_{gt}(y)
,\quad y\in\overline{\mathbb{R}}
,
\end{align}
where
\begin{align}
\begin{aligned}
&\alpha_{g t}^*:=\max \left\{0, \sup _{j \in \mathcal{J} \backslash\{J\}, z \in \mathcal{Z}_{g t}, F_{C_{g t} \mid Z_{g t}}(j \mid z)<1} \frac{F_{C_{g t}}(j)-F_{C_{g t} \mid Z_{g t}}(j \mid z)}{1-F_{C_{g t} \mid Z_{g t}}(j \mid z)}\right\} ,\\
&L_{gt,\alpha}(y):=
\mathbb{I}\{\alpha_{gt}^*\le \alpha<1\}
\frac{F_{C_{gt}}(y)-\min\{F_{C_{gt}}(y),\alpha\}}{1-\min\{F_{C_{gt}}(y),\alpha\}}
+\mathbb{I}\{\alpha=1\}\,\mathbb{I}\{y\ge J\},\\
&U_{gt}(y):=
\inf_{z\in \mathcal{Z}_{gt}}
F_{C_{gt}\mid Z_{gt}}(y\mid z).
\end{aligned}\label{EQLUBOUND}
\end{align}
These bounds are sharp. Moreover,
if $0\leq \alpha< \alpha_{gt}^*$,
then there exists no data-generating process that is compatible with the maintained assumptions and yields the observed distributions.
\end{proposition}
}

{ The above bounds are sharp in the sense that each of $L_{gt,\alpha}$ and $U_{gt}$ is attained as the distribution function of $Y_{gt}(d)$ under a data-generating process that satisfies the maintained assumptions and yields the observed distribution of $(C_{gt},Z_{gt})$.} The upper bound \(U_{gt}\) exploits variation in the instrument by taking the infimum of the conditional CDF of \(C_{gt}\) across values of \(Z_{gt}\). When \(Z_{gt}\) has no variation, this bound coincides with the unconditional CDF \(F_{C_{gt}}\). The lower bound $L_{gt,\alpha}$ captures worst-case one-sided underreporting. When $\alpha=1$, it collapses to a degenerate distribution concentrated at $J$, rendering the lower bound uninformative. { To clarify the relationship with \citet{MW2024}, who study binary choice models with misreported outcomes, note that in the special case of a binary outcome ($J=1$) for a given $(g,t)$ cell, Proposition~\ref{PROP:misr_func_main} yields, for any admissible $\alpha\in[\alpha_{gt}^*,1)$,
\begin{equation}\label{eq:binary-cellwise}
\sup_{z\in\mathcal{Z}_{gt}}
\Pr\!\left(C_{gt}=1\mid Z_{gt}=z\right)
\leq
\Pr\!\left(Y_{gt}(d)=1\right)
\leq
\min\left\{
1,
\frac{\Pr\!\left(C_{gt}=1\right)}{1-\alpha}
\right\}.
\end{equation}
If $Z_{gt}$ is degenerate, the bounds in \eqref{eq:binary-cellwise} coincide with the one-sided-underreporting specialization of Proposition~4 of \citet{MW2024} since the lower bound reduces to $\Pr\!\left(C_{gt}=1\right)$. With nondegenerate instrumental variation, however, the bounds generally differ because the reporting-instrument specification in \citet{MW2024} bounds underreporting conditional on the instrument and imposes monotonicity across its values.
}

Finally, to understand why underreporting must exceed a threshold $\alpha_{gt}^*$, consider $\alpha=0$, which corresponds to truthful reporting. Under Assumption~\ref{A1} and the law of total probability,
$F_{C_{gt}\mid Z_{gt}}(j\mid z)=\Pr(Y_{gt}(d)\le j\mid Z_{gt}=z)$ and hence $F_{C_{gt}}(j)=\Pr(Y_{gt}(d)\le j)$. By Assumption~\ref{A3}, the distribution of $Y_{gt}(d)$ does not depend on $Z_{gt}$, implying $F_{C_{gt}\mid Z_{gt}}(j\mid z)=F_{C_{gt}}(j)$. Hence, in the presence of a valid instrument, truthful reporting has a strong and testable implication: the reported outcome must be independent of the instrument. Any observed dependence rules out $\alpha=0$ and implies some degree of underreporting. More generally, values of $\alpha$ below $\alpha_{gt}^*$ are incompatible with the joint distribution of $(C_{gt},Z_{gt})$.

Combining Assumptions \ref{A1}--\ref{A3}, we obtain the following result:
\begin{proposition}[Bounds for $F_{Y_{11}(0)}$]\label{PROP:final2}
Suppose Assumptions \ref{A1}--\ref{A3} hold for some $\alpha\in[0,1]$. If
$ \alpha^*\leq \alpha\leq 1$, then
\begin{align}\label{final2BOUND1}
	L_{11,\alpha}^{(0)}\left(y\right)\leq F_{Y_{11}(0)}(y)\leq U_{11,\alpha}^{(0)}\left(y\right),\quad y\in\overline{\mathbb{R}},
\end{align}
where 
\begin{align}
\begin{aligned}
&\alpha^*=\sup_{\left(g,t\right)\in  \{0,1\}\times\{0,1\}}\alpha^*_{gt},\\
&	L_{11,\alpha}^{(0)}\left(y\right)
:=
\mathbb{I}\!\left\{\underline{y}_{L,01}\le y\leq \overline{y}_{L,01}\right\}\,
L_{10,\alpha}\!\left(
U_{00}^{(-1)}\!\left(L_{01,\alpha}(y)\right)
\right)
+\mathbb{I}\{y> \overline{y}_{L,01}\},\\
&	U_{11,\alpha}^{(0)}\left(y\right)
:=
\mathbb{I}\!\left\{\underline{y}_{U,01}\le y\leq \overline{y}_{U,01}\right\}\,
U_{10}\!\left(
L_{00,\alpha}^{-1}\!\left(U_{01}(y)\right)
\right)
+\mathbb{I}\{y> \overline{y}_{U,01}\},
\end{aligned}\label{final2BOUND2}
\end{align}
with $\underline{y}_{L,01}:=\inf \mathrm{supp}(L_{01,\alpha})$, $\overline{y}_{L,01}:=\sup \mathrm{supp}(L_{01,\alpha})$, $\underline{y}_{U,01}:=\inf \mathrm{supp}(U_{01})$, $\overline{y}_{U,01}:=\sup \mathrm{supp}(U_{01})$, and $\alpha^*_{gt}$, $L_{gt,\alpha}$ and $U_{gt}$ defined in \eqref{EQLUBOUND}.
 The bounds are { sharp} if\\
$\inf _{z \in \mathcal{Z}_{00}} \mathrm{Pr}\left(C_{00}=j \mid Z_{00}=z\right)>0$ for all $j\in \mathcal{J}$ and $\alpha<F_{C_{00}}(0)$.

Moreover, 
if $0\leq \alpha< \alpha^*$,
then there exists no data-generating process that is compatible with the maintained assumptions and yields the observed distributions.
\end{proposition}

Proposition~\ref{PROP:final2} characterizes sharp bounds on the counterfactual distribution \(F_{Y_{11}(0)}\) under the discrete CiC structure in the presence of underreporting. { The bounds combine the cellwise bounds in Proposition~\ref{PROP:misr_func_main} with the CiC structure on the true potential outcome \(Y(0)\).} As in the no-misreporting case discussed in Section~\ref{sec:nounderreporting}, the counterfactual distribution \(F_{Y_{11}(0)}\) is truncated outside the support of $Y_{01}(0)$, as implied by the support overlap condition $\mathrm{supp}\!\left(Y_{11}(0)\right)\subseteq \mathrm{supp}\!\left(Y_{01}(0)\right)$ under Assumption~\ref{A2}. With underreporting, however, this truncation is governed by the bounding distributions $L_{01,\alpha}$ and $U_{01}$.

The exclusion restriction enters through the operators $U_{gt}$ and tightens the bounds by exploiting variation in $Z_{gt}$. At the same time, the bounds depend on the sensitivity parameter $\alpha$: larger values widen the bounds, and when $\alpha=1$ they become uninformative even with the instrument, since $L_{gt,\alpha}$ collapses to a degenerate distribution concentrated at $J$, so that $L_{gt,\alpha}^{-1}(q)=J$ for all $q\in[0,1]$. By contrast, smaller values of $\alpha$ preserve informative variation in $L_{gt,\alpha}$ and yield tighter restrictions on $F_{Y_{11}(0)}$.

The proposition also establishes { sharpness} under mild conditions, ensuring that the lower and upper bound functions are attainable as counterfactual distributions under the maintained assumptions. Here, these conditions ensure $\operatorname{supp}\!\left(L_{10,\alpha}\right)\subseteq \operatorname{supp}\!\left(U_{00}\right)$ and $\operatorname{supp}\!\left(U_{10}\right)\subseteq \operatorname{supp}\!\left(L_{00,\alpha}\right)$, which parallel the support overlap condition in the no-misreporting case, where $\mathrm{supp}\!\left(Y_{10}(0)\right)\subseteq \mathrm{supp}\!\left(Y_{00}(0)\right)$ under Assumption~\ref{A2}. In our application, these conditions are mild: all outcome levels occur with positive probability for each instrument value across all $(g,t)$ cells, including the control group prior to treatment. Moreover, the mass at zero exceeds $90\%$, so that the restriction $\alpha<F_{C_{00}}(0)$ is not binding in economically relevant ranges and still permits substantial underreporting.

Finally, the restriction $\alpha \ge \alpha^*$ ensures joint feasibility of the cellwise restrictions under the DiD structure. When $\alpha<\alpha^*$, the observed distributions violate the exclusion restriction and no compatible data-generating process exists. Importantly, this restriction is entirely driven by variation in the instrument: if $Z$ is degenerate and exhibits no variation, then $\alpha^*=0$ and the restriction disappears. {In this case, $U_{gt}=F_{C_{gt}}$ in every cell. For
$y\in[\underline{y}_{L,01},\overline{y}_{L,01}]$ and
$y\in[\underline{y}_{U,01},\overline{y}_{U,01}]$, respectively, the lower and upper bounds in Proposition~\ref{PROP:final2} simplify to
\begin{equation}
L_{11,\alpha}^{(0)}(y)
 =L_{10,\alpha}\!\left(F_{C_{00}}^{(-1)}\!\left(L_{01,\alpha}(y)\right)\right),
\qquad
U_{11,\alpha}^{(0)}(y)
 =F_{C_{10}}\!\left(L_{00,\alpha}^{-1}\!\left(F_{C_{01}}(y)\right)\right).
\label{eq:noinstrument}
\end{equation}
At $\alpha=0$, $L_{gt,0}=F_{C_{gt}}$, and equation~\eqref{eq:noinstrument} reduces to the no-misreporting bounds in \eqref{eq:final2-CIC}.
}

Given the bounds for the factual CDF 
$F_{Y_{11}(1)}$ in \eqref{EQBOUNDS} and the counterfactual CDF 
$F_{Y_{11}(0)}$ in \eqref{final2BOUND1}, we construct bounds (and later confidence bands) for the 
QTTs defined in \eqref{EQQTT}. 
To this end, we invert the bounds for the distribution functions 
of $Y_{11}(0)$ and $Y_{11}(1)$ using the generalized inverses defined 
at the end of Section \ref{sec:introduction}. Because generalized inverses reverse the ordering of distribution functions, lower and upper bounds for the CDFs induce upper and lower bounds, respectively, for the corresponding quantile functions. These quantile bounds in turn imply bounds for $\Delta_{\mathrm{QTT}}(\tau)$ obtained via the pointwise Minkowski difference of the corresponding quantile bounds.\footnote{\label{FOOTMINKOWSKI}
If $V=[v_1,v_2]$ and $U=[u_1,u_2]$ are intervals, their pointwise
Minkowski difference is
\(
V\ominus U
:=
[v_1-u_2,\, v_2-u_1].
\)
} This construction parallels the general quantile-effect framework of \citet{CFMW2020}.

The resulting bounds are summarized in the following corollary.

\begin{corollary}[Bounds for QTT]\label{COR:BOUND}
Suppose Assumptions \ref{A1}--\ref{A3} hold for some $\alpha\in\left[\alpha^*,1\right]$.
Then
$$
L_{\mathrm{QTT},\alpha}(\tau)
\le
\Delta_{\mathrm{QTT}}(\tau)
\le
U_{\mathrm{QTT},\alpha}(\tau),\quad \tau\in(0,1),
$$
where 
$$
L_{\mathrm{QTT},\alpha}\left(\tau\right)
:=
U_{11}^{(1),-1}\left(\tau\right)
-
L_{11,\alpha}^{(0),-1}\left(\tau\right),
\qquad
U_{\mathrm{QTT},\alpha}\left(\tau\right)
:=
L_{11,\alpha}^{(1),-1}\left(\tau\right)
-
U_{11,\alpha}^{(0),-1}\left(\tau\right),
$$
with $L_{11,\alpha}^{(1)} := L_{11,\alpha}$ and $U_{11}^{(1)} := U_{11}$,
and $L_{11,\alpha}$, $U_{11}$, $L_{11,\alpha}^{(0)}$, and $U_{11,\alpha}^{(0)}$
defined in \eqref{EQLUBOUND} and \eqref{final2BOUND2}.
\end{corollary}

Corollary \ref{COR:BOUND} establishes bounds for the QTTs defined in 
\eqref{EQQTT}. The lower bound subtracts the upper quantile bound of 
$Y_{11}(0)$ from the lower quantile bound of $Y_{11}(1)$, while the 
upper bound subtracts the lower quantile bound of $Y_{11}(0)$ from the 
upper quantile bound of $Y_{11}(1)$. 

{ Although our main objects are the distributional effects summarized in Corollary~\ref{COR:BOUND}, the cellwise bounds in \eqref{EQBOUNDS} also yield ATT bounds under the standard parallel trends restriction on the mean of the untreated potential outcome. Because the mean of a discrete outcome equals the sum of its upper-tail probabilities, these bounds imply bounds on $\mathrm{E}[Y_{gt}(d)]$ in each group-time cell. This restriction then maps the bounds for the three untreated cells into bounds on the treated group's counterfactual mean after treatment. Combining this interval with the bounds on the treated post-treatment mean yields bounds on the ATT without imposing the discrete CiC structure in Assumption~\ref{A2}. Remark~\ref{REM:MEAN_DID} in the supplement provides explicit expressions.}

\section{Estimation and Inference}\label{sec:estimation}

In this section, we describe how to estimate the bounds from Section \ref{sec:Underreporting} and conduct inference on them. In Section \ref{sec:AddCov} of the supplement, we outline how the bounds and corresponding estimators can be extended to allow for additional exogenous covariates.

For estimation and inference, we rely on repeated cross-sectional sampling. Let the sample $\{(C_i, Z_i, G_i, T_i)\}_{i=1}^N$ be i.i.d.\ draws from the population, where $G_i, T_i \in \{0,1\}$ denote group and time indicators. Within each cell $(g,t) \in \{0,1\}\times\{0,1\}$, the observations with $G_i = g$ and $T_i = t$ are random draws from the corresponding subpopulation. We re-index these observations as $\{(C_{i,gt}, Z_{i,gt})\}_{i=1}^{N_{gt}}$, where $N_{gt} := \sum_{i=1}^N \mathbb{I}\{G_i = g, T_i = t\}$ denotes the number of observations in cell $(g,t)$.

To construct estimators of the bounds, consider $\alpha\in [\alpha^*,1)$, with $\alpha^*$ defined in \eqref{final2BOUND2}. We exclude the endpoint $\alpha=1$ and focus on the interior case for ease of exposition.  For each $y\in\mathcal{J}$, define
\begin{align}
\overline{k}_{\alpha}\left(y\right)
:=
L_{01,\alpha}^{-1}\!\left(U_{00}\left(y\right)\right),
\qquad
\underline{k}_{\alpha}\left(y\right)
:=
U_{01}^{-1}\!\left(L_{00,\alpha}\left(y-1\right)\right),
\label{klowup} 
\end{align}
where $L_{00,\alpha}\left(y\right)=0$ for all $y<0$ by definition.  In the spirit of Section~5.2 of \citet{AI2006}, Lemma \ref{BOUNDEQV} in the supplement shows that, under mild regularity conditions, the bounds in \eqref{final2BOUND2} admit an equivalent representation: for each $y\in\mathcal{J}$,
\begin{align}
\begin{aligned}
L_{11,\alpha}^{(0)}(y)
&= \frac{\Pr\left(\overline{k}_{\alpha}\left(C_{10}\right)\leq  y\right)-\min\left\{\Pr\left(\overline{k}_{\alpha}\left(C_{10}\right)\leq  y\right),\alpha\right\}}{1- \min\left\{\Pr\left(\overline{k}_{\alpha}\left(C_{10}\right)\leq  y\right),\alpha\right\}} ,\\
U_{11,\alpha}^{(0)}(y)
&=
\min_{z\in \mathcal{Z}_{10}}
\Pr\!\left(\underline{k}_{\alpha}\left(C_{10}\right)\le y \mid Z_{10}=z\right).
\end{aligned}\label{EQLU11FUNC}
\end{align}
Thus, the bounds on $F_{Y_{11}(0)}$ can be obtained by first mapping $C_{10}$ through $\overline{k}_{\alpha}$ or $\underline{k}_{\alpha}$ and then evaluating the bound functionals in \eqref{EQLUBOUND} at the resulting distribution functions. Replacing the probabilities in \eqref{EQLU11FUNC} by their empirical counterparts yields the following estimators:
\begin{align}
\begin{aligned}
\widehat{L}_{11,\alpha}^{(0)}(y)
&:=
\frac{
\frac{1}{N_{10}}\sum_{i=1}^{N_{10}}
\mathbb{I}\left\{\widehat{\overline{k}}_{\alpha}(C_{i,10})\le y\right\}
-
\min\!\left\{
\frac{1}{N_{10}}\sum_{i=1}^{N_{10}}
\mathbb{I}\left\{\widehat{\overline{k}}_{\alpha}(C_{i,10})\le y\right\},
\alpha
\right\}
}{
1-
\min\!\left\{
\frac{1}{N_{10}}\sum_{i=1}^{N_{10}}
\mathbb{I}\left\{\widehat{\overline{k}}_{\alpha}(C_{i,10})\le y\right\},
\alpha
\right\}
},
\\
\widehat{U}_{11,\alpha}^{(0)}(y)
&:=
\min_{z\in \mathcal{Z}_{10}}
\frac{
\sum_{i=1}^{N_{10}}
\mathbb{I}\left\{\widehat{\underline{k}}_{\alpha}(C_{i,10})\le y\right\}
\mathbb{I}\left\{Z_{i,10}=z\right\}
}{
\sum_{i=1}^{N_{10}}
\mathbb{I}\left\{Z_{i,10}=z\right\}
},
\end{aligned}\label{ESTLUBOUND}
\end{align}
 where 
\[
\widehat{\overline{k}}_{\alpha}(y)
:=
\widehat{L}_{01,\alpha}^{-1}\!\left(\widehat{U}_{00}(y)\right),
\qquad \widehat{\underline{k}}_{\alpha}(y)
:=
\widehat{U}_{01}^{-1}\!\left(\widehat{{L}}_{00,\alpha}(y-1)\right),
\]
and for each $(g,t)\in \{(0,1),(0,0)\}$,
\begin{align*}
\widehat{L}_{gt,\alpha}(y)
&:=
\frac{
\frac{1}{N_{gt}}\sum_{i=1}^{N_{gt}}\mathbb{I}\{C_{i,gt}\le y\}
-
\min\!\left\{
\frac{1}{N_{gt}}\sum_{i=1}^{N_{gt}}\mathbb{I}\{C_{i,gt}\le y\},
\alpha
\right\}
}{
1-
\min\!\left\{
\frac{1}{N_{gt}}\sum_{i=1}^{N_{gt}}\mathbb{I}\{C_{i,gt}\le y\},
\alpha
\right\}
}, 
\\
\widehat{U}_{gt}(y)
&:=
\min_{z\in \mathcal{Z}_{gt}}
\frac{
\sum_{i=1}^{N_{gt}}
\mathbb{I}\{C_{i,gt}\le y\}
\mathbb{I}\{Z_{i,gt}=z\}
}{
\sum_{i=1}^{N_{gt}}
\mathbb{I}\{Z_{i,gt}=z\}
}.
\end{align*}
We set $\widehat{U}_{11,\alpha}^{(0)}(y)$ and $\widehat{U}_{gt}(y)$ equal to one whenever their denominators are zero. 

We impose the following sampling and support conditions:

\begin{assumption}[Sampling]\label{AE0} $\{(C_i, Z_i, G_i, T_i)\}_{i=1}^N$ are i.i.d.\ copies of $(C, Z, G, T)$.
\end{assumption}

\begin{assumption}[Discrete Instrument]\label{AE0A}
 $\mathcal{Z}_{gt}$ is finite for all $(g,t)\in\{0,1\}\times\{0,1\}$.
\end{assumption}

\begin{assumption}[Regularity]\label{AE0_TIES}
There exists a nonempty set $\mathcal{A}\subseteq [\alpha^*,1)$ such that for all $\alpha\in\mathcal{A}$:
\begin{enumerate}[label=(\roman*)]
\item $U_{01}(y)\neq L_{00,\alpha}(y')$ and $L_{01,\alpha}(y)\neq U_{00}(y')$ for all  $y,y'\in\mathcal{J}\setminus \{J\}$.
\item $\inf _{y\in\mathcal{J}, z \in \mathcal{Z}_{00}} \mathrm{Pr}\left(C_{00}=y \mid Z_{00}=z\right)>0$  and $\alpha<F_{C_{00}}(0)$.
\end{enumerate}
\end{assumption}	

Assumption \ref{AE0} is a standard sampling condition for repeated cross-sectional data. Combined with the maintained condition $\pi_{gt}=\Pr(G=g,T=t)>0$ for all $(g,t)$, this implies $N_{gt}>0$ with probability approaching one.

Assumption \ref{AE0A} requires that the instrumental variable has finite support across all cells.
Allowing for a continuous instrument would introduce additional technical complications for inference \citep[cf.][]{MP2020,MPZ2024}. In applications where $Z_{gt}$ is continuous or takes on many values, it can be discretized on a finite grid without altering the structure of the estimators (see Section \ref{sec:application}).  

Assumption \ref{AE0_TIES}(i) imposes a no–tie condition and adapts Assumption 5.2 in \citet{AI2006} to our setting. 
The set $\mathcal{A}\subseteq [\alpha^*,1)$ indexes the sensitivity parameters over which the bounds are evaluated, typically a finite grid chosen by the researcher.
For every $\alpha\in\mathcal{A}$, the assumption requires that the lower and upper bound functions do not coincide across subpopulations.
This rules out boundary cases in which the bounds coincide and the estimator becomes nonregular.\footnote{\label{Median} As discussed in \citet[p.~466]{AI2006}, a similar issue arises for the sample median of i.i.d.\ binary variables $X_i \in \{0,1\}$, which converges to the population median only if $\Pr(X_i = 1) \neq 0.5$. At the boundary case $\Pr(X_i = 1) = 0.5$, the estimator fails to converge.}

Assumption~\ref{AE0_TIES}(ii) coincides with the condition for sharpness of the bounds in Proposition~\ref{PROP:final2}. Here, it ensures $\operatorname{supp}(C_{10})\subseteq \operatorname{supp}(L_{00,\alpha})$ and $\operatorname{supp}(C_{10})\subseteq \operatorname{supp}(U_{00})$, which are analogous to Assumption~5.1(iv) in \citet{AI2006}. 
With underreporting, the CiC transformation  is constructed from the bounding distributions $L_{00,\alpha}$ and $U_{00}$ rather than from a single observed control distribution. Since these distributions may differ and, in the case of $L_{00,\alpha}$, depend on $\alpha$, the support inclusion must hold simultaneously for both distributions entering the bounds. 

We establish weak convergence of the bound estimators for the counterfactual distribution $F_{Y_{11}(0)}$, uniformly over $y \in \mathcal{J}$.

\begin{proposition}[Convergence of Bound Estimators for $F_{Y_{11}(0)}$]\label{PROPLIMIT}
Suppose Assumptions \ref{AE0}--\ref{AE0_TIES} hold. Then, for each $\alpha\in\mathcal{A}$,
\[
\sqrt{N}
\begin{pmatrix}
\widehat{L}_{11,\alpha}^{(0)}
-
L_{11,\alpha}^{(0)}
\\
\widehat{U}_{11,\alpha}^{(0)}
-
U_{11,\alpha}^{(0)}
\end{pmatrix}
\rightsquigarrow
\begin{pmatrix}
G_{L,11,\alpha}^{(0)}
\\
G_{U,11,\alpha}^{(0)}
\end{pmatrix}\quad
\text{in }
\ell^{\infty}\!\left(\mathcal{J},\mathbb{R}^2\right),
\]
with $L_{11,\alpha}^{(0)}$, $U_{11,\alpha}^{(0)}$, $\widehat{L}_{11,\alpha}^{(0)}$, and $\widehat{U}_{11,\alpha}^{(0)}$ defined in \eqref{final2BOUND2} and \eqref{ESTLUBOUND}, and $G_{L,11,\alpha}^{(0)}$ and $G_{U,11,\alpha}^{(0)}$ are limit processes defined in Lemma \ref{BOUNDDIST} of the supplement.\footnote{Let $\mathscr{A}$ be an arbitrary set and $\mathscr{B}$ a Banach space. Then $\ell^{\infty}(\mathscr{A},\mathscr{B})$ denotes the space of all bounded functions $f:\mathscr{A}\to\mathscr{B}$ such that $\sup_{a\in\mathscr{A}}\|f(a)\|_{\mathscr{B}}<\infty$, equipped with the sup-norm $\|f\|_{\infty} := \sup_{a\in\mathscr{A}}\|f(a)\|_{\mathscr{B}}$ (see, e.g., \citet{VanderVaart1996}).}
\end{proposition}

Proposition \ref{PROPLIMIT} can be reformulated as a directional delta-method result for the bound estimators. To see this, let \begin{align}\label{eq:phi_def}
\left(\theta_{L,\alpha}^{(0)},\theta_{U,\alpha}^{(0)}\right):=\left(F_{\overline{k}_{\alpha}(C_{10})},F_{\underline{k}_{\alpha}(C_{10})\mid Z_{10}}\right)\in \ell^{\infty}\!\left(\mathcal{J},\mathbb{R}\right)
\times
\ell^{\infty}\!\left(\mathcal{J}\times\mathcal{Z}_{10},\mathbb{R}\right),
\end{align}
and define the functional
$
\boldsymbol{\phi}_{\alpha}^{(0)}
:
\ell^{\infty}\!\left(\mathcal{J},\mathbb{R}\right)
\times
\ell^{\infty}\!\left(\mathcal{J}\times\mathcal{Z}_{10},\mathbb{R}\right)
\to
\ell^{\infty}\!\left(\mathcal{J},\mathbb{R}^{2}\right)
$ 
by
\begin{align}\label{eq:phi_defs}
\left[\boldsymbol{\phi}_{\alpha}^{(0)}\left(\theta_{1},\theta_{2}\right)\right](y)
:=
\begin{pmatrix}
\dfrac{\theta_{1}(y)-\min\{\theta_{1}(y),\alpha\}}
{1-\min\{\theta_{1}(y),\alpha\}}
\\
\inf_{z\in\mathcal{Z}_{10}}\theta_{2}(y,z)
\end{pmatrix}.
\end{align}
Then, by the representation in \eqref{EQLU11FUNC},
\[
\begin{pmatrix}
L_{11,\alpha}^{(0)}
\\
U_{11,\alpha}^{(0)}
\end{pmatrix}
=
\boldsymbol{\phi}_{\alpha}^{(0)}\!\left(\theta_{L,\alpha}^{(0)},\theta_{U,\alpha}^{(0)}\right)\in  \ell^{\infty}(\mathcal{J},\mathbb{R}^{2}).
\]
As shown in the supplement, the weak limit in Proposition 
\ref{PROPLIMIT} is obtained by applying the Hadamard directional
derivative of $\boldsymbol{\phi}_{\alpha}^{(0)}$ at 
$(\theta_{L,\alpha}^{(0)},\theta_{U,\alpha}^{(0)})$ to the weak limit of the
underlying empirical distribution functions.
Although the latter limit is Gaussian, the resulting limit
process is generally non-Gaussian and non-pivotal because
$\boldsymbol{\phi}_{\alpha}^{(0)}$ is only Hadamard directionally
differentiable: its first component is kinked at $\theta_{1}(y)=\alpha$, and its second component involves a pointwise infimum.

We therefore follow \citet{FS2019} and \citet{MP2020} and use a bootstrap
procedure for directionally Hadamard differentiable functionals, implemented
via the numerical directional derivative of \citet{HL2018}. For
$\left(g,t\right)=\left(1,0\right)$, let
$\left\{\left(C_{i,10}^{*},Z_{i,10}^{*}\right)\right\}_{i=1}^{N_{10}}$
denote a generic nonparametric i.i.d.\ bootstrap sample drawn with replacement
from $\left\{\left(C_{i,10},Z_{i,10}\right)\right\}_{i=1}^{N_{10}}$. 
For each bootstrap sample, the resampled observations are transformed using the
mappings $\widehat{\underline{k}}_{\alpha}$ and $\widehat{\overline{k}}_{\alpha}$
obtained from the original sample. As shown in Lemma \ref{LEMPEE} in the
supplement, with probability approaching one,
\[
\widehat{\underline{k}}_\alpha(y)=\underline{k}_\alpha(y), \qquad
\widehat{\overline{k}}_\alpha(y)=\overline{k}_\alpha(y), \qquad
y\in\mathcal{J}.
\]
Consequently, recomputing these transformations within the bootstrap sample
would not affect first-order asymptotics, so we keep them fixed in the
bootstrap procedure.

To construct empirical and bootstrap counterparts of \eqref{eq:phi_def}, denote
\small
\begin{align*}
\widehat{F}_{\widehat{\overline{k}}_{\alpha}(C_{10})}(y)
&:=
\frac{1}{N_{10}}
\sum_{i=1}^{N_{10}}
\mathbb{I}\!\left\{
\widehat{\overline{k}}_{\alpha}(C_{i,10}) \le y
\right\},\quad
\widehat{F}_{\widehat{\underline{k}}_{\alpha}(C_{10})\mid Z_{10}}(y\mid z)
:=
\frac{
\frac{1}{N_{10}}
\sum_{i=1}^{N_{10}}
\mathbb{I}\!\left\{
\widehat{\underline{k}}_{\alpha}(C_{i,10}) \le y
\right\}
\mathbb{I}\!\left\{
Z_{i,10}=z
\right\}
}{
\frac{1}{N_{10}}
\sum_{i=1}^{N_{10}}
\mathbb{I}\!\left\{
Z_{i,10}=z
\right\}
},\\
\widehat{F}^{\,*}_{\widehat{\overline{k}}_{\alpha}(C_{10})}(y)
&:=
\frac{1}{N_{10}}
\sum_{i=1}^{N_{10}}
\mathbb{I}\!\left\{
\widehat{\overline{k}}_{\alpha}(C_{i,10}^{*}) \le y
\right\},
\quad
\widehat{F}^{\,*}_{\widehat{\underline{k}}_{\alpha}(C_{10})\mid Z_{10}}(y\mid z)
:=
\frac{
\frac{1}{N_{10}}
\sum_{i=1}^{N_{10}}
\mathbb{I}\!\left\{
\widehat{\underline{k}}_{\alpha}(C_{i,10}^{*}) \le y
\right\}
\mathbb{I}\!\left\{
Z_{i,10}^{*}=z
\right\}
}{
\frac{1}{N_{10}}
\sum_{i=1}^{N_{10}}
\mathbb{I}\!\left\{
Z_{i,10}^{*}=z
\right\}
}.
\end{align*}
\normalsize
for all $y\in\mathcal{J}$ and $z\in\mathcal{Z}_{10}$.  With this notation, define
\[
\left(\widehat{\theta}_{L,\alpha}^{(0)},\widehat{\theta}_{U,\alpha}^{(0)}\right)
:=
\left(
\widehat{F}_{\widehat{\overline{k}}_{\alpha}(C_{10})},
\widehat{F}_{\widehat{\underline{k}}_{\alpha}(C_{10})\mid Z_{10}}
\right),
\qquad
\left(\widehat{\theta}_{L,\alpha}^{(0),*},\widehat{\theta}_{U,\alpha}^{(0),*}\right)
:=
\left(
\widehat{F}^{\,*}_{\widehat{\overline{k}}_{\alpha}(C_{10})},
\widehat{F}^{\,*}_{\widehat{\underline{k}}_{\alpha}(C_{10})\mid Z_{10}}
\right).
\]
 To implement the numerical directional derivative,
define for $h_L \in \ell^{\infty}\!\left(\mathcal{J},\mathbb{R}\right)$ and
$h_U \in \ell^{\infty}\!\left(\mathcal{J}\times\mathcal{Z}_{10},\mathbb{R}\right)$
\[
\widehat{\boldsymbol{\phi}}_{\alpha}^{(0)'}\left(h_L,h_U\right)
:=
\frac{
\boldsymbol{\phi}_{\alpha}^{(0)}\!\left(
\widehat{\theta}_{L,\alpha}^{(0)}+\epsilon_N h_L,
\widehat{\theta}_{U,\alpha}^{(0)}+\epsilon_N h_U
\right)
-
\boldsymbol{\phi}_{\alpha}^{(0)}\!\left(
\widehat{\theta}_{L,\alpha}^{(0)},
\widehat{\theta}_{U,\alpha}^{(0)}
\right)
}{\epsilon_N},
\]
where $\epsilon_N \to 0$ and $\sqrt{N}\epsilon_N \to \infty$.
 We then  approximate the limiting distribution of
$
\sqrt{N}\left(
\boldsymbol{\phi}_{\alpha}^{(0)}\left(\widehat{\theta}_{L,\alpha}^{(0)},\widehat{\theta}_{U,\alpha}^{(0)}\right)
-
\boldsymbol{\phi}_{\alpha}^{(0)}\left(\theta_{L,\alpha}^{(0)},\theta_{U,\alpha}^{(0)}\right)
\right)
$
in Proposition~\ref{PROPLIMIT} by the distribution of the bootstrap process
\[
\begin{pmatrix}
T_{1,\alpha}^{(0),*}
\\
T_{2,\alpha}^{(0),*}
\end{pmatrix}
:=
\widehat{\boldsymbol{\phi}}_{\alpha}^{(0)'}\!\left(
\sqrt{N}\left(
\widehat{\theta}_{L,\alpha}^{(0),*}-\widehat{\theta}_{L,\alpha}^{(0)}
\right),
\sqrt{N}\left(
\widehat{\theta}_{U,\alpha}^{(0),*}-\widehat{\theta}_{U,\alpha}^{(0)}
\right)
\right)
\in \ell^{\infty}\!\left(\mathcal{J},\mathbb{R}^{2}\right).
\]
This construction corresponds to the numerical directional derivative
proposed by \citet{HL2018}. Under the maintained conditions, Theorem~3.1 of \citet{HL2018} implies
that the bootstrap consistently estimates the limiting distribution in Proposition~\ref{PROPLIMIT}.

To construct a symmetric $100\cdot\left(1-\gamma\right)\%$ confidence band that is
uniform over $j\in\mathcal{J}\setminus\{J\}$, define the functional
$m: \ell^{\infty}\!\left(\mathcal{J},\mathbb{R}^{2}\right) \to \mathbb{R}$ by
\begin{equation}\label{EQmfunction}
 m\left(\begin{pmatrix}
h_1\\h_2	
\end{pmatrix}
\right)
:=
\max\left\{
\max_{y\in\mathcal{J}\setminus\{J\}}
\frac{h_{1}(y)}{\sigma(y)},
\ 
\max_{y\in\mathcal{J}\setminus\{J\}}
\left(
-\frac{h_{2}(y)}{\sigma(y)}
\right)
\right\},\end{equation}
where $\sigma(j)>0$ is a known, bounded weighting function. Define the bootstrap statistic and its conditional $\left(1-\gamma\right)$-quantile:
\begin{align*}
M^{(0),*}_{\alpha}
&:=
m\!\left(\begin{pmatrix}
T_{1,\alpha}^{(0),*}
\\
T_{2,\alpha}^{(0),*}
\end{pmatrix}
\right)
=
\max\left\{
\max_{j\in\mathcal{J}\setminus\{J\}}
\frac{T_{1,\alpha}^{(0),*}(j)}{\sigma(j)},
\ 
\max_{j\in\mathcal{J}\setminus\{J\}}
\left(
-\frac{T_{2,\alpha}^{(0),*}(j)}{\sigma(j)}
\right)
\right\},\\
\widehat z_{\alpha}^{(0)}\left(1-\gamma\right)
&:=
\inf\left\{
z\in\mathbb{R}
\mid
\Pr\!\left(
M^{(0),*}_{\alpha}\le z
\mid 
\left\{\left(C_i, Z_i, G_i, T_i\right)\right\}_{i=1}^N
\right)
\ge 1-\gamma
\right\}.
\end{align*}
 As outlined in \cite{FR2018}, many choices of $\sigma$ yield valid
uniform confidence bands. In our application, we use the
simple choice $\sigma(j)=1$, which produces an equal-width band. 
This yields an asymptotically valid $100\cdot\left(1-\gamma\right)\%$ confidence band
for the counterfactual bounds, uniform over $y\in\mathcal{J}\setminus\{J\}$,
defined by
\begin{align}\label{EQCI}
\mathrm{CS}_{\alpha}^{(0)}(y;1-\gamma)
:=
\left[
\widehat{L}_{11,\alpha}^{(0)}(y)-\frac{\sigma(y)\widehat z^{(0)}_{\alpha}\left(1-\gamma\right)}{\sqrt{N}},
\,
\widehat{U}_{11,\alpha}^{(0)}(y)+\frac{\sigma(y)\widehat z^{(0)}_{\alpha}\left(1-\gamma\right)}{\sqrt{N}}
\right]\cap[0,1],
\end{align}
for $y\in\mathcal{J}\setminus\{J\}$, while we set
$\mathrm{CS}_{\alpha}^{(0)}(J;1-\gamma):=\{1\}$. 
Since
$L_{11,\alpha}^{(0)}(y),U_{11,\alpha}^{(0)}(y)\in[0,1]$ for all
$y\in\mathcal{J}$ by definition, intersection with $[0,1]$ only enforces
the natural range of a CDF and does not affect asymptotic coverage.

Proposition \ref{PROPZGAMMA} establishes the asymptotic validity of this
$100\cdot\left(1-\gamma\right)\%$ confidence band uniformly over $y\in\mathcal{J}$.

\begin{proposition}[{ Counterfactual Confidence Bands via Numerical Delta Method}]\label{PROPZGAMMA}
Suppose Assumptions \ref{AE0}--\ref{AE0_TIES} hold, and let
$\epsilon_{N}\to 0$ satisfy $\sqrt{N}\epsilon_{N}\to \infty$.
Let  $\alpha\in\mathcal{A}$ and $\gamma\in\left(0,1\right)$. Define
\[
M_{\alpha}^{(0)}
:=m\!\left(
\begin{pmatrix}
G_{L,11,\alpha}^{(0)}
\\
G_{U,11,\alpha}^{(0)}
\end{pmatrix}
\right), \quad z_{\alpha}^{(0)}\left(1-\gamma\right)
:=
\inf\left\{
z\in\mathbb{R}
\mid
\mathrm{Pr}\left(M_{\alpha}^{(0)}\le z\right)\ge 1-\gamma
\right\},
\]
where $\left(G_{L,11,\alpha}^{(0)},G_{U,11,\alpha}^{(0)}\right)$ is the weak limit in
Proposition \ref{PROPLIMIT}.  Assume that the distribution function of
$M_{\alpha}^{(0)}$ is continuous and strictly increasing in a neighborhood of
$z_{\alpha}^{(0)}\left(1-\gamma\right)$. Then
\[
\lim_{N\to\infty}
\mathrm{Pr}\left(
\left[
L_{11,\alpha}^{(0)}(y),
U_{11,\alpha}^{(0)}(y)
\right]
\subseteq
\mathrm{CS}_{\alpha}^{(0)}(y;1-\gamma)
\ \text{for all } y\in\mathcal{J}
\right)
= 1-\gamma.
\]
\end{proposition}

The continuity assumption on the distribution of $M_{\alpha}^{(0)}$ ensures that
the population quantile $z_{\alpha}^{(0)}\left(1-\gamma\right)$ of $M_{\alpha}^{(0)}$ is well defined
and that the bootstrap critical value $\widehat z_{\alpha}^{(0)}\left(1-\gamma\right)$
consistently estimates it, see Corollary~3.2 of \citet{FS2015}. Moreover, the uniformity of the confidence band over $y \in \mathcal{J}$ yields a straightforward interpretation: with a pre-specified probability, for example $90\%$, it covers the counterfactual bound functions at all values of $y \in \mathcal{J}$ simultaneously. This is practically useful because it does not require the researcher to decide in advance which particular outcome level or functional feature is of interest. Hence, any candidate function that lies outside $\mathrm{CS}_{\alpha}^{(0)}(y;0.9)$ at even a single value of $y$ can be rejected at the corresponding level, namely $10\%$.

As with inference on the distribution bounds, we can also construct uniform confidence
bands for the QTT bounds. Specifically, we invert the confidence bands for
$\left[L_{11,\alpha}^{(1)},U_{11}^{(1)}\right]$ and
$\left[L_{11,\alpha}^{(0)},U_{11,\alpha}^{(0)}\right]$, each at level $1-\gamma/2$, 
and then take the pointwise Minkowski difference of the resulting quantile bands,
following \citet{CFMW2020}. Let
$\mathrm{CS}_{\mathrm{QTT},\alpha}\left(\tau;1-\gamma\right)$
denote the resulting confidence band, whose construction is given  in the supplement. Section \ref{sec:parametricmodel} { also presents a parametric estimation and inference approach that does not require the tuning parameter $\epsilon_N$, but imposes parametric restrictions instead.}

We obtain the following corollary:

\begin{corollary}[{ QTT Confidence Bands via Numerical Delta Method}]\label{COR:CIINFERENCE}
Suppose Assumptions \ref{AE0}--\ref{AE0_TIES} hold, and let
$\epsilon_{N}\to 0$ satisfy $\sqrt{N}\epsilon_{N}\to \infty$.
Let $\alpha\in\mathcal{A}$ and $\gamma\in\left(0,1\right)$.  Assume that the distribution function of
$M_{\alpha}^{(d)}$ is continuous and strictly increasing in a neighborhood of
$z_{\alpha}^{(d)}\left(1-\gamma/2\right)$ for all $d\in \{0,1\}$, with $M_{\alpha}^{(0)}$ and $z_{\alpha}^{(0)}$ defined in Proposition \ref{PROPZGAMMA}, and $M_{\alpha}^{(1)}$ and $z_{\alpha}^{(1)}$ in the supplement. Then
\[
\liminf_{N\to\infty}
\Pr\!\left(
\left[
L_{\mathrm{QTT},\alpha}\left(\tau\right),
U_{\mathrm{QTT},\alpha}\left(\tau\right)
\right]
\subseteq
\mathrm{CS}_{\mathrm{QTT},\alpha}\left(\tau;1-\gamma\right)
\text{ for all }\tau\in(0,1)
\right)
\ge 1-\gamma,
\]
with $L_{\mathrm{QTT},\alpha}$ and $U_{\mathrm{QTT},\alpha}$ defined in Corollary \ref{COR:BOUND}, and $\mathrm{CS}_{\mathrm{QTT},\alpha}$ in the supplement.
\end{corollary}

\section{Empirical Application}\label{sec:application}

The intense debate surrounding the legalization of marijuana (cannabis) and its impact on individual consumption behavior, and more generally, on society, has been a pivotal policy issue in Western countries for decades. In many of these nations, the possession and consumption of marijuana are illegal, primarily due to concerns over its contribution to crime escalation and the associated societal and economic costs \citep{BML2019}. Additionally, there are health-related concerns, with evidence suggesting that early-age cannabis consumption can lead to lower educational attainment \citep{VOW2009, MZLBHB2007, MOCCEHOSS2004} and cognitive deficits in verbal learning and memory tasks \citep{SJRDCHL2011}. The potential of cannabis serving as a gateway to more harmful drug use has also been documented \citep[see][and references therein]{P2010}.

In this section, we investigate the effects of recreational marijuana legalization for adults in several U.S. states on the short-term consumption behavior of 8th-grade high-school students in those states during our sample period. The focus on these minors, typically aged 13 to 14, is critical due to their heightened vulnerability to the health risks associated with marijuana consumption. Unlike existing studies, which predominantly examined legalization effects on average usage or consumption \citep[see][]{AHR2015, WHC2015, CWFKSSH2017, HWB2022}, or on the timing of the first (early age) consumption \citep{WBJ2014}, we investigate the distributional effects of recreational marijuana legalization on the entire consumption distribution within the aforementioned high-school cohort in various states.\footnote{Unlike \citet{HWB2022}, who consider mean effects for the legalization of recreational marijuana use on underage adolescents among others, the survey data used in this analysis exclusively cover enrolled high-school students (see next paragraph). Thus, they do not provide insights into effects on high-school drop-outs.} Additionally, we account for the possibility that consumption behavior may not always be reported truthfully, and more importantly, that reporting behavior itself may be affected by the legalization, possibly due to shifts in stigma, public perception, or attitudes towards marijuana consumption.

The data are composed of several cross-sectional waves of the ``Monitoring the Future'' survey, an annual survey conducted in the United States gathering data on attitudes, behaviors, and values of American adolescents currently enrolled in high school.\footnote{The ``Monitoring the Future'' survey is funded by the National Institute on Drug Abuse (NIDA). For more information, see Monitoring the Future, Inter-university Consortium for Political and Social Research (ICPSR), University of Michigan (https://monitoringthefuture.org/).} Specifically, the survey for 8th graders, which is conducted annually, covers a wide range of topics including substance (ab)use such as the consumption of marijuana and variants thereof. The primary sample of our empirical analysis consists of 46,472 high-school students, sampled as repeated cross-sections across all contiguous U.S. states. With this sample, our focus is on recreational marijuana legalization for adults aged 21 or older, examining its impact on 8th-grade students' (short-term) consumption behavior. We use observations from states that legalized marijuana for adults during our sampling period as treated group ($g=1$), and label observations as coming from the control group otherwise ($g=0$).\footnote{Due to data privacy restrictions of the ICPSR, we are not allowed to publish information that allows to identify single states either directly or indirectly. Information about the specific legalization year used in the analysis as well as about the states that entered the treatment group can be obtained from the authors upon request.} Similarly, we record observations from the two years before the legalization event as coming from the pre-treatment period ($t=0$), while observations from the two years thereafter are labeled as post-treatment observations ($t=1$). We exclude observations from states that had already passed recreational marijuana laws prior to the sampling window. Moreover, all states from the treated group had previously legalized medicinal marijuana use before the observation period. It is also important to note that we focus on a specific legalization year, and no other major marijuana-related legalization, either for medicinal or recreational purposes, took place in the U.S. during that period. Finally, given that surveys were conducted in the spring of each year, while recreational marijuana legalization was enacted in the last quarter for all treated states considered in the analysis, we deem the risk of major anticipation effects to be small. The final subsamples consist of $N_{00}=19940$, $N_{01}=17769$, $N_{10}=4710$, and $N_{11}=4053$ observations.

The outcome variable of interest (\texttt{cons\_30days}), based on the survey responses to the question \textit{``On how many occasions have you used marijuana (weed, pot) or hashish (hash, hash oil) ... in the last 30 days?''}, is categorized into three levels: ``0'' (never used), ``1'' (used 1--2 times), and ``2'' (used more than 2 times). Because this measure is based on students' self-reports of recent marijuana use, truthful reporting is not automatic in this setting. Legalization for adults may affect not only actual consumption but also the perceived cost of admitting use in a survey, so allowing treatment to affect reporting is empirically important for interpretation. We also use a discretized ``survey cooperation'' measure (\texttt{trust}) as an instrumental variable for reporting, following \citet{GHSZ2018} and \citet{BHSZ2018}. Here, ``survey cooperation'', defined as the proportion of nonmissing survey responses and based solely on questions common to all questionnaire forms and not directly related to drug use, is assumed to provide an indication of the respondents' willingness to cooperate in the survey, unrelated to marijuana consumption itself, as the reporting decision occurs post-consumption only. We discretize this variable into four categories based on the 25th, 50th, and 75th percentiles of its unconditional distribution. Descriptive statistics for \texttt{cons\_30days}, \texttt{trust}, and other control variables used in the analysis of Section \ref{sec:parametricmodel} can be found in Section \ref{sec:ADDEMP} of the supplementary material.

\begin{figure}[htb]
  \begin{center}
  \caption{Estimated Bounds and CDFs W/ and W/O Underreporting.}\label{FIG1} 

    \includegraphics[width=0.49\linewidth]{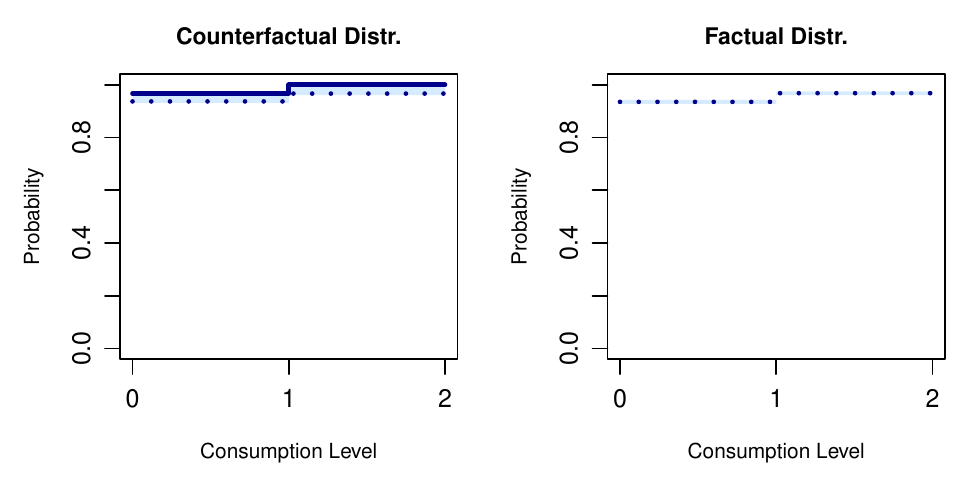}\hfill
    \includegraphics[width=0.49\linewidth]{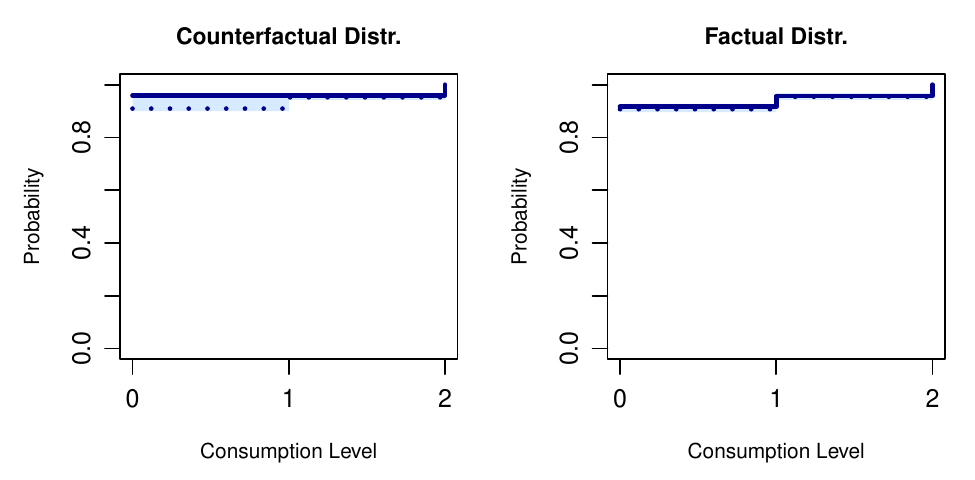}
   \\
    \includegraphics[width=0.49\linewidth]{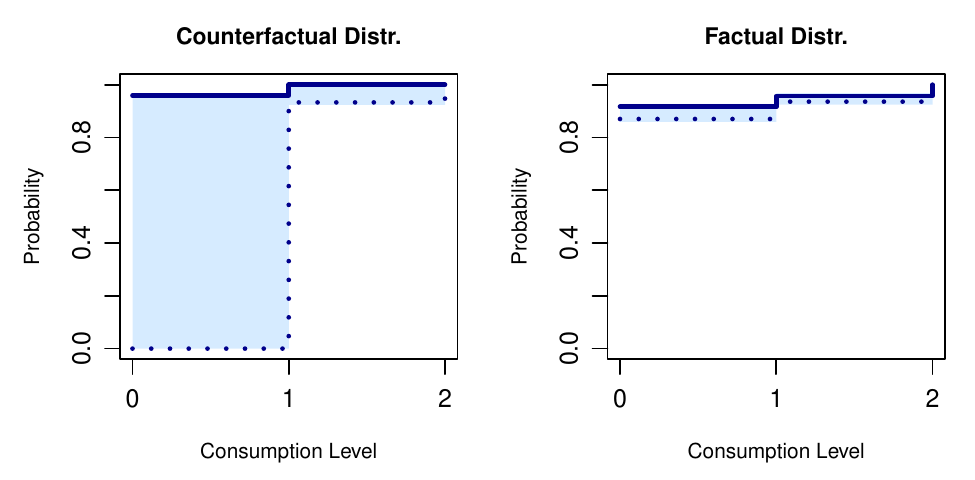}\hfill
    \includegraphics[width=0.49\linewidth]{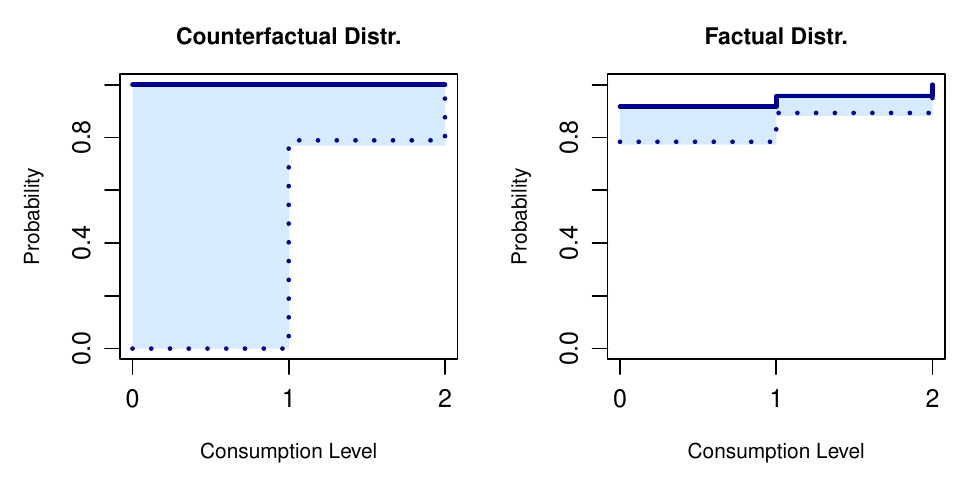}
  \end{center}

{\footnotesize Note: estimated bounds and CDFs with and without underreporting. The upper-left pair of panels displays the estimated bounds from \citet{AI2006} for the counterfactual distribution and the empirical CDF for the factual distribution in the case without underreporting. The upper-right, lower-left, and lower-right pairs of panels exhibit \(\widehat{L}_{11,\alpha}^{(0)}\) and \(\widehat{U}_{11,\alpha}^{(0)}\) for the counterfactual distribution, and \(\widehat{L}_{11,\alpha}^{(1)}\) and \(\widehat{U}_{11}^{(1)}\) for the factual distribution, for \(\alpha=0.3\), \(\alpha=0.5\), and \(\alpha=0.7\), respectively. Lower bound estimates appear as dotted lines, upper bound estimates as solid lines. The shaded areas illustrate uniform Confidence Sets \(\mathrm{CS}_{\alpha}^{(0)}(\cdot;0.95)\) as defined in (\ref{EQCI}) and \(\mathrm{CS}_{\alpha}^{(1)}(\cdot;0.95)\) as defined in Subsection \ref{ssec:CSQTT} of the supplement.}
\end{figure}

We first present estimated bounds for the CDF of the counterfactual and factual distribution for the case without and with underreporting. Specifically, the upper-left pair of panels in Figure \ref{FIG1} displays the estimated bounds from \citet{AI2006} for the counterfactual and the empirical CDF for the factual distribution for the case without misreporting. By contrast, the upper-right, lower-left, and lower-right pairs of panels contain the estimated bounds $\widehat{L}_{11,\alpha}^{(0)}$, $\widehat{U}_{11,\alpha}^{(0)}$ for the counterfactual, and $\widehat{L}_{11,\alpha}^{(1)}$, $\widehat{U}_{11}^{(1)}$ for the factual distribution at values of $\alpha$ equal to $0.3$, $0.5$, and $0.7$, respectively. Note that values with $\alpha<0.3$ lead to a crossing of the bounds, suggesting that such values are not compatible with the observed distribution under the maintained exclusion restriction. In this sense, the data imply a minimal feasible degree of underreporting in the application. The shaded areas illustrate uniform 95\% Confidence Sets, $\mathrm{CS}_{\alpha}^{(0)}(\cdot;0.95)$, as defined in (\ref{EQCI}) for the counterfactual distributions, and in Subsection \ref{ssec:CSQTT} of the supplement for the factual distribution. We compute these Confidence Sets using the nonparametric bootstrap together with the numerical derivative method outlined in Section \ref{sec:estimation} using $B=1000$ bootstrap replications.\footnote{We construct the Confidence Set for the case without underreporting using the same bootstrap method.} For simplicity, we display all results for the ordinary nonparametric bootstrap { using $\epsilon_N=N_{10}^{-1/2}$, the benchmark tuning choice that performed reasonably well overall in the Monte Carlo simulations reported in} Section \ref{sec:MCBounds} { of the supplement. The simulations also show that finite-sample coverage is sensitive to $\epsilon_N$, as values that are too small may lead to undercoverage, whereas values that are too large may lead to overcoverage.}

Turning to the results in Figure \ref{FIG1}, we first observe that $\alpha$ plays an important role in tightening the bounds for the counterfactual distribution, and that different values can lead to a large heterogeneity in the informativeness of the bounds. As expected, this holds true for the bounds of the factual distribution to a much lesser extent, as the latter are much tighter in general. Second, note that choosing a sufficiently small $\alpha$ level leads to bound estimates that are very similar to the case without misreporting. { Third, we note that the bounds are estimated very precisely since the Confidence Sets are very tight throughout.}

\begin{figure}[htb]
  \begin{center}
  \caption{Estimated QTT Bounds W/ and W/O Underreporting.}\label{FIG2} 

    \includegraphics[width=0.48\linewidth]{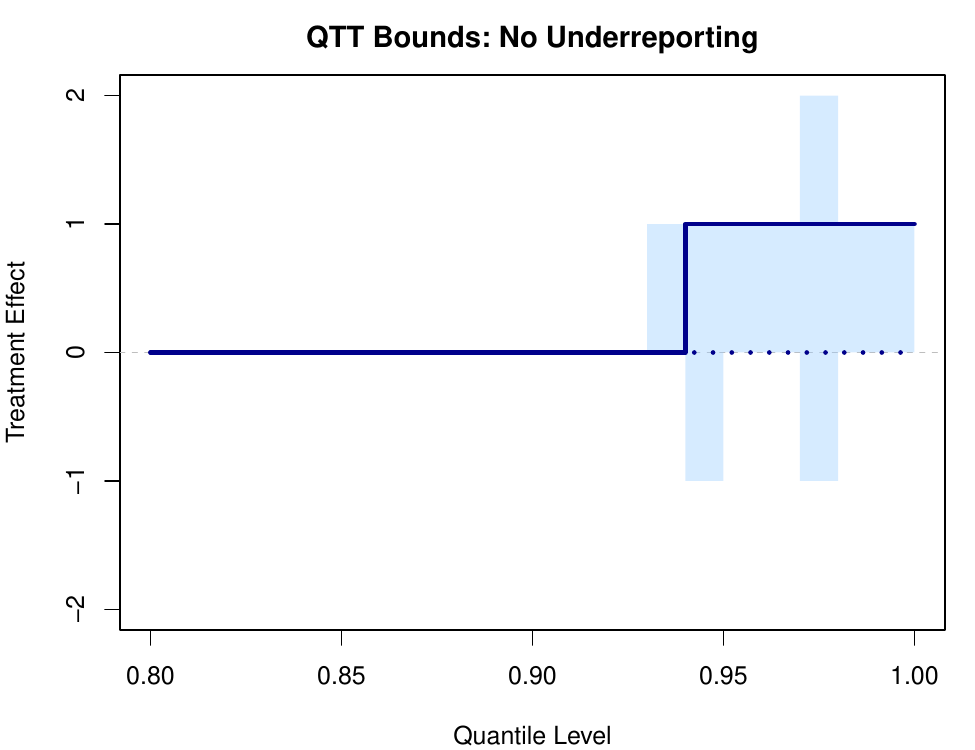}\hfill
    \includegraphics[width=0.48\linewidth]{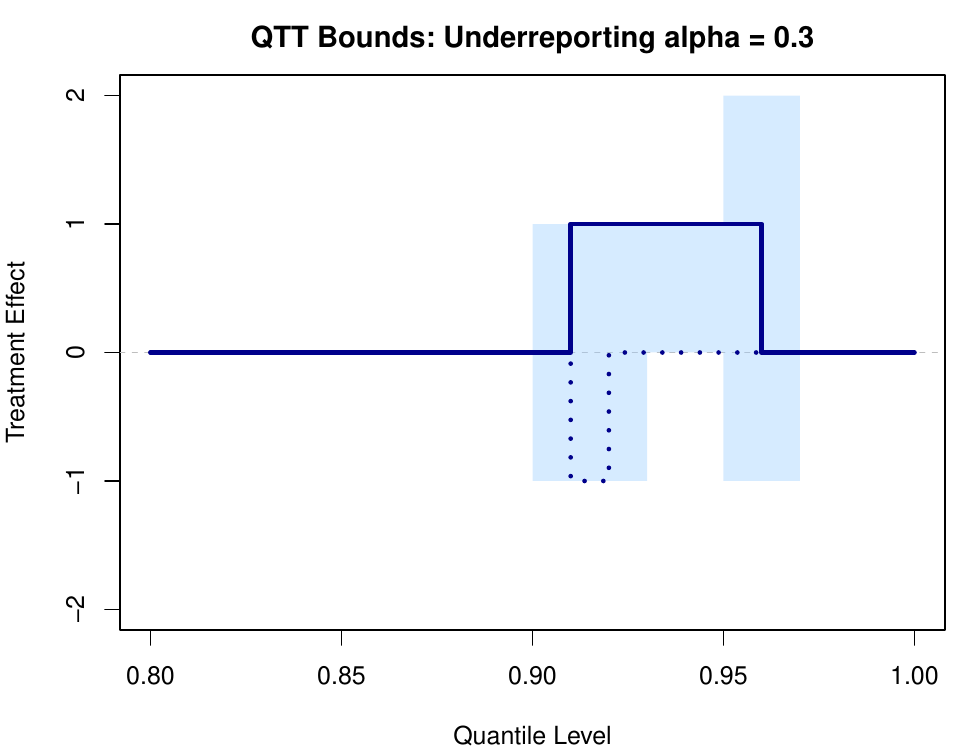}
   \\
    \includegraphics[width=0.48\linewidth]{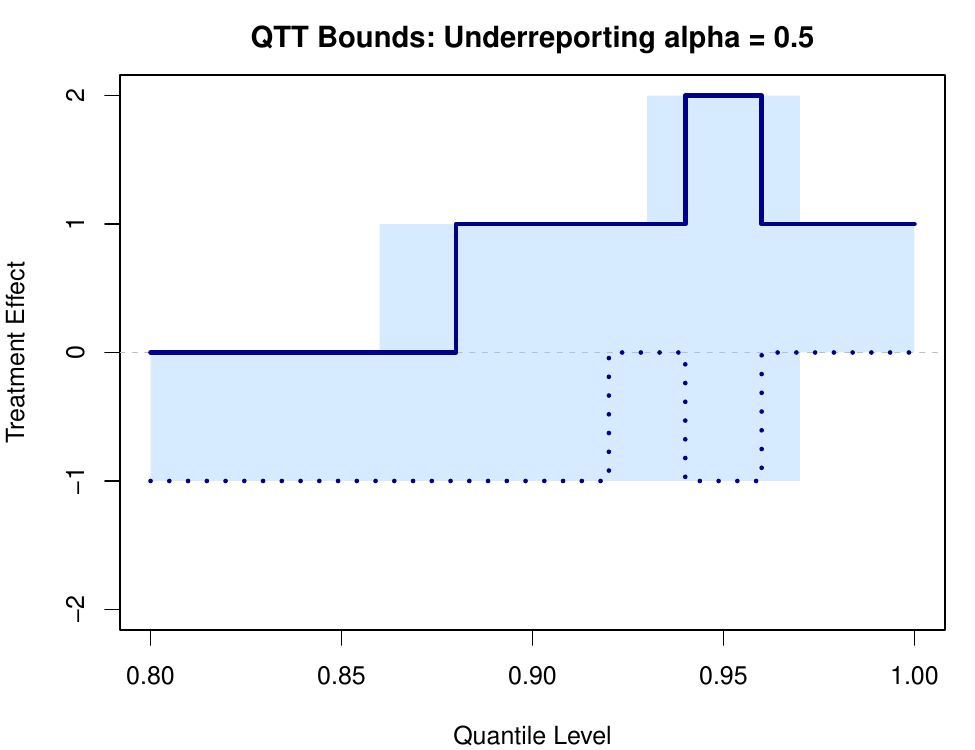}\hfill
    \includegraphics[width=0.48\linewidth]{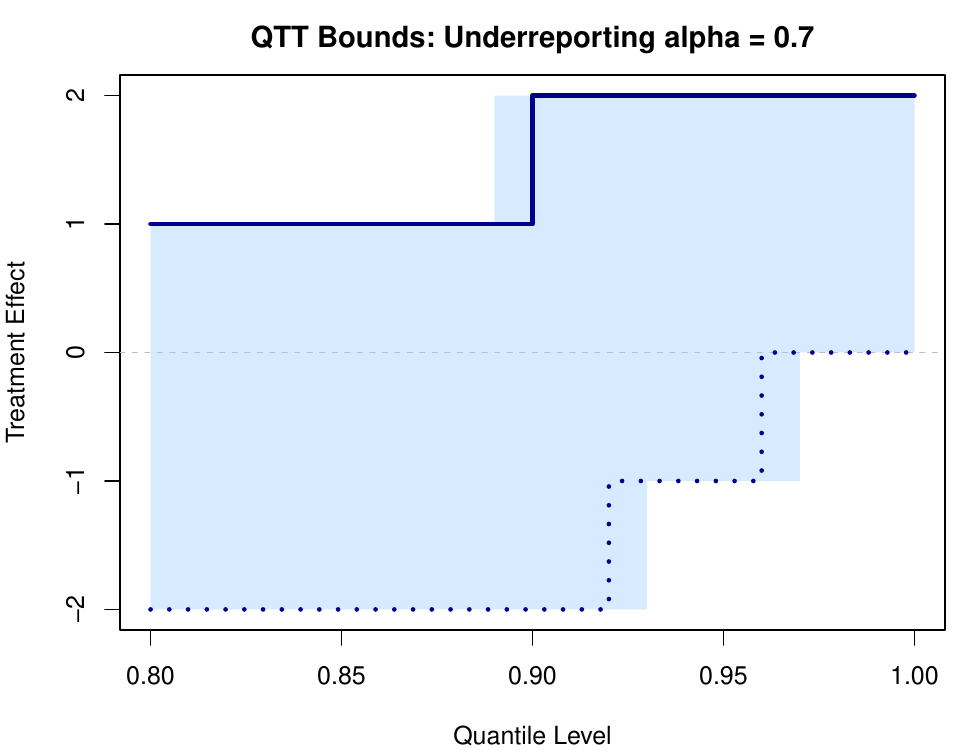}
  \end{center}

{\footnotesize Note: The figure contains estimated bounds for the QTTs $\Delta_{\mathrm{QTT}}(\tau)$, $\tau\in(0.8,1)$, with and without underreporting. The upper left panel displays the estimated QTT bounds w/o underreporting, the upper right panel exhibits the estimated QTT bounds w/ underreporting ($\alpha=0.3$), the lower left panel w/ underreporting ($\alpha=0.5$), and the lower right panel w/ underreporting ($\alpha=0.7$). Lower bound estimates appear in all figures as dotted lines, upper bound estimates as solid lines. Shaded areas illustrate uniform Confidence Sets $\mathrm{CS}_{\mathrm{QTT},\alpha}\left(\cdot;0.9\right)$. }
\end{figure}

Next, we turn to the QTT estimates in Figure \ref{FIG2}. We present these estimates alongside uniform 90\% Confidence Sets, $\mathrm{CS}_{\mathrm{QTT},\alpha}(\cdot;0.9)$, as outlined in Corollary \ref{COR:CIINFERENCE}.  Moreover, as more than 90\% of the 8th graders in the sample do not report consumption of marijuana in the past 30 days (cf. also Figure \ref{FIG1}), we concentrate our analysis on quantile levels of the upper part of the distribution, namely $\tau\in(0.8,1)$. Examining Figure \ref{FIG2},  we observe that without misreporting there is only evidence for possibly non-zero QTTs at quantile levels larger than approximately $0.95$. By contrast, we can reject the null hypothesis of positive (or negative) treatment effects for smaller quantile levels at the 10\% significance level. This suggests that, without accounting for misreporting, there is no evidence at the 10\% significance level that legalization of marijuana had any impact on the short-term consumption frequency of 8th-graders in treated states. On the other hand, moving to the case with underreporting, we first note that, as expected, the evidence for potentially non-zero effects shifts downwards to lower quantile levels. For instance, at $\alpha=0.5$, we cannot reject either positive or negative QTTs for quantile levels between $0.86$ and $0.97$. Interestingly, both at $\alpha=0.5$ and $\alpha=0.7$, we can reject negative QTTs for the highest quantile levels, while positive (but not negative) QTTs can be rejected at $\alpha=0.5$ for quantile levels below $\tau=0.86$. This evidence is in line with our estimates from the parametric model in Section \ref{sec:parametricmodel} of the supplement, { where allowing for misreporting leads to a significant negative DTT at the zero consumption level and significant increases at both positive consumption levels. By contrast, the estimates without covariates are insignificant throughout.}

Summarizing the empirical analysis above, there is little overall evidence of either positive or negative QTTs from marijuana legalization on the short-term consumption frequency of 8th-grade high-school students in treated states. Once underreporting is taken into account, however, the picture becomes more nuanced: at lower quantile levels, only positive QTTs are rejected, whereas at upper quantile levels, only negative QTTs can be rejected. This pattern is in line with the estimation results from the semiparametric model in Section \ref{sec:parametricmodel} of the supplement. In Section \ref{sec:ADDEMP} of the supplement, we also examine QTT bounds for subsamples of the primary dataset. In particular, we consider four subsamples defined by self-reported ethnicity and gender (\texttt{white}=1 if the student reported being white, \texttt{white}=0 otherwise; \texttt{male}=1 if the student reported being male, \texttt{male}=0 otherwise). The results remain qualitatively similar, especially for subsamples with \texttt{white}=1, but the bound estimates and Confidence Sets are generally much less informative due to substantially smaller sample sizes. Finally, as noted above, choosing larger values of $\epsilon_{N}$ yields wider and thus less informative Confidence Sets for the bounds under underreporting, while leaving the qualitative conclusions unchanged.

\section{Conclusion}\label{sec:conclusion}

This paper develops a Difference-in-Differences framework for discrete, ordered outcomes that may be underreported. The analysis builds on the discrete CiC model of \citet{AI2006} and extends it to settings with partially observable outcomes. A key feature of our setup is that reporting behavior may be correlated with the underlying outcome and may also vary with treatment status. Under an exclusion restriction and an upper bound on underreporting, we derive sharp nonparametric bounds, and we provide corresponding estimators and bootstrap-based uniform Confidence Sets. In the supplement, we complement this nonparametric analysis with a point-identified semiparametric model based on a real analytic extrapolation assumption, and we propose a flexible estimator for a parametric version of that model.

We apply our methodology to investigate QTTs of the legalization of recreational marijuana use for adults on the short-term consumption behavior of 8th-grade high-school students in the affected states. Our findings suggest little evidence of non-zero effects outside the extreme upper tail of the distribution when underreporting is ignored. Once underreporting is allowed for, however, only positive QTTs are ruled out at moderate upper-tail quantiles, whereas at the very top the reverse is true.

\medskip
\spacingset{1}

\putbib[references/GL1]
\end{bibunit}

\clearpage
\setcounter{page}{1}
\setcounter{section}{0}
\setcounter{equation}{0}
\setcounter{assumption}{0}
\setcounter{theorem}{0}
\setcounter{table}{0}
\setcounter{figure}{0}
\setcounter{footnote}{0}
\makeatletter
\let\c@definition\c@theorem
\makeatother
\renewcommand{\thedefinition}{\thetheorem}
\renewcommand{\labelenumi}{(\arabic{enumi})}
\renewcommand{\theequation}{S.\arabic{equation}}
\renewcommand{\thesection}{S.\arabic{section}}
\renewcommand{\thepage}{S.\arabic{page}}
\renewcommand{\thetheorem}{S.\arabic{theorem}}
\renewcommand{\thelemma}{S.\arabic{lemma}}
\renewcommand{\theassumption}{S.\arabic{assumption}}
\renewcommand{\theHsection}{supplement.\arabic{section}}
\renewcommand{\theHequation}{supplement.\arabic{equation}}
\renewcommand{\theHtheorem}{supplement.\arabic{theorem}}
\renewcommand{\theHdefinition}{supplement.\arabic{theorem}}
\renewcommand{\theHassumption}{supplement.\arabic{assumption}}
\renewcommand{\theHtable}{supplement.\arabic{table}}
\renewcommand{\theHfigure}{supplement.\arabic{figure}}
\setlength{\parindent}{0in}
\setlength{\parskip}{1em}
\setlength{\bibsep}{0pt plus 0.3ex}

\begin{bibunit}[chicago]
\makeatletter
\def\@extra@b@citeb{.supplement}
\def\@extra@binfo{.supplement}
\makeatother
\spacingset{1}
\etocdepthtag.toc{supplement}
\makeatletter
\let\maketitle\arxivmaketitle
\let\@maketitle\arxivinnermaketitle
\let\title\arxivtitle
\let\author\arxivauthor
\let\date\arxivdate
\let\thanks\arxivthanks
\let\and\arxivand
\makeatother
\def\spacingset#1{\renewcommand{\baselinestretch}%
{#1}\small\normalsize} \spacingset{1}


\if0\blind
{
  \title{\bf Supplementary Material to \textit{Changes-in-Changes for Ordered Choice Models with Underreporting}}
  \author{Daniel Gutknecht\\
    Faculty of Economics and Business, Goethe University Frankfurt\\
   Cenchen Liu \\
   Faculty of Economics and Business, Goethe University Frankfurt }
  \date{\today}
  \maketitle
} \fi

\if1\blind
{
  \bigskip
  \bigskip
  \bigskip
  \begin{center}
    {\LARGE\bf Supplementary Material to \textit{Changes-in-Changes for Ordered Choice Models with Underreporting}}
\end{center}
  \medskip
} \fi

\vfill

\newpage

\setcounter{assumption}{0}
\setcounter{theorem}{0}

\renewcommand{\theassumption}{S.\arabic{assumption}}
\renewcommand{\thetheorem}{S.\arabic{theorem}}

\etocsettagdepth{main}{none}
\etocsettagdepth{supplement}{subsection}
\tableofcontents

\section{Proofs of Bounds}\label{sec:proofbounds}

Before presenting the auxiliary results, we introduce notation for supports and generalized inverses defined directly at the level of CDFs. 

\medskip

\begin{definition}[Support of a CDF]\label{SupportF}
Let $F:\overline{\mathbb{R}}\to[0,1]$ be a CDF. Its support is
\[
\operatorname{supp}(F):=\{y\in\overline{\mathbb{R}}\mid \forall \varepsilon>0,\ F(y+\varepsilon)-F(y-\varepsilon)>0\}.
\]
\end{definition}
Thus, $\operatorname{supp}(F)$ consists of all real points such that every open neighborhood of the point carries positive probability mass. It coincides with the usual notion of $\operatorname{supp}(Y)$ when $F$ is the distribution of a random variable $Y$.

For $q\in[0,1]$, define the generalized inverses of $F$ by
\begin{align*}
&F^{(-1)}:[0,1]\to\overline{\mathbb{R}}, 
\quad q\mapsto F^{(-1)}(q):=\sup\{y\in \operatorname{supp}(F)\mid F(y)\le q\},\\
&F^{-1}:[0,1]\to\overline{\mathbb{R}}, 
\quad q\mapsto F^{-1}(q):=\inf\{y\in \operatorname{supp}(F)\mid F(y)\ge q\}.
\end{align*}
These definitions agree with the generalized inverses in the main text when $F=F_Y$ for some random variable $Y$, since $\operatorname{supp}(F_Y)=\operatorname{supp}(Y)$ under Definition~\ref{SupportF}. Formulating them directly in terms of $\operatorname{supp}(F)$ is useful because, in Proposition \ref{PROP:final2}, the bounding objects are themselves distribution functions with potentially different supports.

%

\subsection{Auxiliary Results}\label{ssec:1.1}

\begin{remark}[A Mean-DiD Approach]\label{REM:MEAN_DID}
Suppose Assumptions \ref{A1} and \ref{A3} { hold for some $\alpha\in[\alpha^*,1]$, where $\alpha^*:=\max_{(g,t)\in\{0,1\}\times\{0,1\}}\alpha_{gt}^*$.} Under these conditions, the cellwise distribution
bounds also imply corresponding bounds for cellwise means. Hence, if one imposes
a standard parallel trends restriction only on the mean of the true untreated outcome,
these cellwise mean bounds immediately imply bounds on the treated group's counterfactual
mean after treatment and, in turn, on the corresponding average treatment effect on the treated.

To see this, note that since $Y\in\mathcal{J}=\{0,\ldots,J\}$, we have
$
\mathrm{E}\left[Y\right]
=
\sum_{j=0}^{J-1}\left(1-F_Y\left(j\right)\right).
$
Hence, for each observed cell $\left(g,t\right)$ with realized treatment state $d=g\cdot t$,
Proposition~\ref{PROP:misr_func_main} yields the mean bounds
\[
\underline{\mu}_{gt}
\le
\mathrm{E}\!\left[Y_{gt}\left(d\right)\right]
\le
\overline{\mu}_{gt,\alpha},
\]
with
\[
\underline{\mu}_{gt}
:=
\sum_{j=0}^{J-1}\left(1-U_{gt}\left(j\right)\right),
\qquad
\overline{\mu}_{gt,\alpha}
:=
\sum_{j=0}^{J-1}\left(1-L_{gt,\alpha}\left(j\right)\right).
\]
If, in addition, one imposes the standard parallel trends restriction on the mean of
the true untreated outcome,
\[
\mathrm{E}\!\left[Y_{11}\left(0\right)\right]-\mathrm{E}\!\left[Y_{10}\left(0\right)\right]
=
\mathrm{E}\!\left[Y_{01}\left(0\right)\right]-\mathrm{E}\!\left[Y_{00}\left(0\right)\right],
\]
then the untreated counterfactual mean for treated units after treatment satisfies
\[
\mathrm{E}\!\left[Y_{11}\left(0\right)\right]
\in
\left[\underline{\mu}^{(0)}_{11,\alpha},\overline{\mu}^{(0)}_{11,\alpha}\right],
\]
where
{\[
\underline{\mu}^{(0)}_{11,\alpha}
:=
\max\!\left\{0,\,
\underline{\mu}_{10}
+
\underline{\mu}_{01}
-
\overline{\mu}_{00,\alpha}\right\},
\qquad
\overline{\mu}^{(0)}_{11,\alpha}
:=
\min\!\left\{J,\,
\overline{\mu}_{10,\alpha}
+
\overline{\mu}_{01,\alpha}
-
\underline{\mu}_{00}\right\}.
\]
}
Combining this with the cellwise bounds for $\mathrm{E}\!\left[Y_{11}\left(1\right)\right]$ gives
\[
\mathrm{ATT}
:=
\mathrm{E}\!\left[Y_{11}\left(1\right)\right]
-
\mathrm{E}\!\left[Y_{11}\left(0\right)\right]
\in
\left[
\underline{\mu}_{11}
-
\overline{\mu}^{(0)}_{11,\alpha},
\;
\overline{\mu}_{11,\alpha}
-
\underline{\mu}^{(0)}_{11,\alpha}
\right].
\]
In the no-underreporting case $\alpha=0$, the cellwise mean bounds collapse, and the argument reduces to the usual mean-DiD expression.

The additional DiD restriction above is imposed only on the first moment of the
untreated counterfactual outcome $Y_{11}\left(0\right)$. It therefore yields information
about $\mathrm{E}\!\left[Y_{11}\left(0\right)\right]$ and the corresponding ATT, but not about
the full counterfactual distribution. This is a clear limitation in the present ordered
discrete setting because different distributions on $\mathcal{J}$ can share the same mean. By contrast, Proposition \ref{PROP:final2} and Corollary \ref{COR:BOUND} use the discrete CiC structure to bound the entire
counterfactual distribution and the associated QTTs.
\end{remark}

Next, we present a lemma that collects useful properties of generalized inverses. Parts (i)--(iii) establish monotonicity results used in the proof of Proposition \ref{PROP:final2}, while parts (iv)--(v) provide equivalences used in the proof of Lemma~\ref{BOUNDEQV}.

\medskip

\begin{lemma}[Properties of Generalized Inverses]
\label{lem:inv_mon}
Let $\mathcal{F}$ denote the set of all CDFs concentrated on $\mathcal{J}$, that is,
$$
	\mathcal{F}:= \left\{F: \overline{\mathbb{R}} \rightarrow[0,1] \mid \exists\left(p_0, \ldots, p_J\right) \in[0,1]^{J+1}, \sum_{j=0}^J p_j=1, F(y)=\sum_{j=0}^J p_j \mathbb{I}\left(j \leq y\right) \right\}.
$$
Then:
\begin{enumerate}[label=(\roman*)]
\item If $F\in \mathcal{F}$,  then $F^{\left(-1\right)}$ and $F^{-1}$ are nondecreasing on $\left[0,1\right]$.

\item If $F_1,F_2,F_3\in \mathcal{F}$ with 
$F_1\left(y\right)\le F_2\left(y\right)$ for all $y\in\overline{\mathbb{R}}$ and $\mathrm{supp}\left(F_3\right)\subseteq \mathrm{supp}\left(F_1\right)$, then for each $q\in\left[0,1\right]$,
\[
F_3\left(F_1^{\left(-1\right)}\left(q\right)\right)\ge F_3\left(F_2^{\left(-1\right)}\left(q\right)\right).
\]
\item If $F_1,F_2\in \mathcal{F}$ with
$F_1\left(y\right)\le F_2\left(y\right)$ for all $y\in\overline{\mathbb{R}}$, then for each $q\in\left[0,1\right]$,
\[
F_1^{-1}\left(q\right)\ge F_2^{-1}\left(q\right).
\]
\item If $F\in \mathcal{F}$,  then for each $q\in\left(0,1\right]$ and $x\in\overline{\mathbb{R}}$,
$$
F^{-1}\left(q\right)\le x
\iff
q\le F\left(x\right).
$$

\item If $F\in \mathcal{F}$,  then for each $q\in\left[0,1\right]$ and $x\in\operatorname{supp}\left(F\right)$,
$$
F\left(x\right)\le q
\iff
x\le F^{(-1)}\left(q\right).
$$
\end{enumerate}
\end{lemma}

\subsection{Proofs of Main Results}\label{ssec:MainResults}

\begin{proof}[Proof of Proposition~\ref{PROP:misr_func_main}]
Fix $(g,t)\in\{0,1\}\times\{0,1\}$ and let $d=g\cdot t$. By equation~(\ref{EQOBSERVABILITY}), we have $C_{gt}(d)\sim C\mid G=g,T=t$, so within the $(g,t)$ cell the reported outcome $C_{gt}(d)$ is observed, whereas the true outcome $Y_{gt}(d)$ remains latent. To simplify notation, we suppress $(g,t,d)$ and write $Y:=Y_{gt}(d)$, $C:=C_{gt}(d)$, $Z:=Z_{gt}$, $\mathcal{Z}:=\mathcal{Z}_{gt}$, $\alpha^*:=\alpha_{gt}^*$, $L_{\alpha}:=L_{gt,\alpha}$, and $U:=U_{gt}$. All probabilities are understood as conditional on $(G,T)=(g,t)$ unless conditioning on $Z$ is stated explicitly.

By the law of total probability, the exclusion restriction in Assumption~\ref{A3}, and $C\leq Y$ a.s.\ from Assumption~\ref{A1}, we have for each $z\in \mathcal{Z}$ and $j\in\mathcal{J}\setminus \{J\}$,
\begin{align}\label{eq:misr_func1}
\begin{aligned}
&\quad \ \Pr\!\left(C\leq j\,\middle|\,Z=z\right)
		\\
	&=\mathrm{Pr}\!\left(C\le j,\,Y\le j\,\middle|\,Z=z\right)
	+\mathrm{Pr}\!\left(C\le j,\,Y\ge j+1\,\middle|\,Z=z\right)
	\\
	&=\mathrm{Pr}\!\left(Y\le j\,\middle|\,Z=z\right)
	+\mathrm{Pr}\!\left(C\le j\,\middle|\,Y\ge j+1,\,Z=z\right)
	\,\mathrm{Pr}\!\left(Y\ge j+1\,\middle|\,Z=z\right)\\
	&=\mathrm{Pr}\!\left(Y\le j\right)
	+\mathrm{Pr}\!\left(C\le j\,\middle|\,Y\ge j+1,\,Z=z\right)
	\,\left(1-\mathrm{Pr}\!\left(Y\leq j\right)\right)
\end{aligned}
\end{align}
Applying the same decomposition to $\Pr(C\le j)$ yields the unconditional identity
\begin{align}\label{eq:misr_func2}
\Pr\!\left(C\leq j\right)=\mathrm{Pr}\!\left(Y\le j\right)
	+\mathrm{Pr}\!\left(C\le j\,\middle|\,Y\ge j+1\right)
	\,\left(1-\mathrm{Pr}\!\left(Y\leq j\right)\right).
\end{align}
We define
\[
j(y)
:=
\max\left\{j\in\left\{-\infty\right\}\cup \mathcal{J}\mid j\le y\right\},
\quad y\in \overline{\mathbb{R}},
\]
that is, \(j(y)=-\infty\) for \(y<0\), \(j(y)=k\) for \(y\in\left[k,k+1\right)\) with \(k=0,1,\ldots,J-1\), and \(j(y)=J\) for \(y\ge J\).
Since \(C,Y\in \mathcal{J}\) a.s., their distribution functions only depend on \(y\) through \(j(y)\). That is, for any $y\in \overline{\mathbb{R}}$ and $z\in \mathcal{Z}$,
\begin{align}
\begin{aligned}\label{eq:misr_func3}
	&F_C\left(y\right)=\Pr\!\left(C\leq y\right)=\Pr\!\left(C\leq j(y)\right),\\
&F_Y\left(y\right)=\Pr\!\left(Y\leq y\right)=\Pr\!\left(Y\leq j(y)\right),\\
&F_{C\mid Z}\left(y\mid z\right)=\Pr\!\left(C\leq y\mid Z=z\right)=\Pr\!\left(C\leq j(y)\mid Z=z\right).
\end{aligned}
\end{align}

\noindent\textbf{\underline{Case 1}: $\alpha\in \left[ \alpha^*,1\right)$.}

\noindent\underline{\textit{Validity.}}  
From \eqref{eq:misr_func2}, for each $j\in\mathcal{J}\setminus\{J\}$,
\[
0\le \Pr\!\left(C\le j\,\middle|\,Y\ge j+1\right)\le \alpha<1.
\]
Since the function $x\mapsto \frac{\Pr\!\left(C\le j\right)-x}{1-x}$ is nonincreasing on $[0,1)$, we have
$$
\Pr\!\left(Y\leq j\right)=\frac{\Pr\!\left(C\leq j\right)- \mathrm{Pr}\!\left(C\le j\,\middle|\,Y\ge j+1\right)}{1- \mathrm{Pr}\!\left(C\le j\,\middle|\,Y\ge j+1\right)} \geq \frac{\Pr\!\left(C\leq j\right)-\alpha}{1-\alpha}, \quad j \in\mathcal{J}\setminus \{J\} .
$$
From \eqref{eq:misr_func1} and $\mathrm{Pr}\!\left(C\le j\,\middle|\,Y\ge j+1,Z=z\right)\ge 0$, it follows that $\mathrm{Pr}\!\left(Y\le j\right)\le \mathrm{Pr}\!\left(C\le j\,\middle|\,Z=z\right)$ for all $z\in\mathcal{Z}$ and $j\in\mathcal{J}\setminus\{J\}$. Taking the infimum yields
$$
\Pr\!\left(Y\leq j\right) \leq \inf _{z \in \mathcal{Z}} \Pr\!\left(C\leq j\,\middle|\,Z=z\right), \quad j \in\mathcal{J}\setminus \{J\}.
$$
Combining this with $\Pr(Y\le j)\ge 0$ and \eqref{eq:misr_func3}, we obtain 
\begin{align*}
&\max\!\left\{0,\frac{F_C\left(y\right)-\alpha}{1-\alpha}\right\}=\max\!\left\{0,\frac{\Pr\!\left(C\leq j(y)\right)-\alpha}{1-\alpha}\right\}\\
&\le \Pr\!\left(Y\leq j(y)\right)=F_Y\left(y\right)\\
&\leq\inf_{z\in\mathcal{Z}}\Pr\!\left(C\leq j(y)\,\middle|\,Z=z\right)=  \inf_{z\in \mathcal{Z}}F_{C\mid Z}(y\mid z), \quad y\in [0,J).
\end{align*}
Since $Y\in\mathcal{J}$ a.s., we have $F_Y(y)=0$ for $y<0$ and $F_Y(y)=1$ for $y\ge J$, and the same holds for the bounds. Hence, the inequalities extend to all $y\in\overline{\mathbb{R}}$. These expressions coincide with $L_\alpha(y)$ and $U(y)$, noting that, for each $p\in[0,1]$,
$$
\frac{p-\min\left\{p,\alpha\right\}}{1-\min\left\{p,\alpha\right\}}
=
\mathbb{I}\left\{p\le \alpha\right\}\frac{p-p}{1-p}
+
\mathbb{I}\left\{p>\alpha\right\}\frac{p-\alpha}{1-\alpha}
=\max\!\left\{0,\frac{p-\alpha}{1-\alpha}\right\}.
$$

\noindent\underline{\textit{Sharpness.}}
We only show that the lower bound $L_{\alpha}$ is attainable.
The same choice of construction applies to the upper bound $U$.

Let $F_Y:=L_{\alpha}$. It suffices to construct a joint distribution of $(C,Y,Z)$ that 
reproduces the observed law of $(C,Z)$ and satisfies all model restrictions:
\begin{enumerate}[label=(\alph*)]
	\item Target parameters: $\Pr(Y\le y)=F_Y(y)$ for all $y\in \overline{\mathbb{R}}$.
	\item Known distributions: $\Pr(C\le y)=F_{C}(y)$ and $\Pr(C\le y\mid Z=z)=F_{C\mid Z}(y\mid z)$ for all $y\in \overline{\mathbb{R}}$ and  $z\in\mathcal{Z}$.
	\item No overreporting: $C,Y\in\mathcal{J}$ and $C\le Y$ a.s.
	\item Exclusion restriction: $\Pr(Y\le y\mid Z=z)=\Pr(Y\le y)$  for all \(y\in \overline{\mathbb{R}}\) and \(z\in\mathcal{Z}\),
	\item Underreporting bound: $\Pr(C\le j\mid Y\ge j+1)\le\alpha$ for all $j\in\mathcal{J}\setminus \{J\}$.
\end{enumerate}
For each $z\in\mathcal{Z}$, define 
$$
\Pr\left(C\leq c,Y\leq y\mid Z=z\right)
:=
\min\!\left\{F_{C\mid Z}(c\mid z),\,F_Y(y)\right\},
\qquad c,y\in\overline{\mathbb{R}}.
$$
This corresponds to the Fr\'{e}chet--Hoeffding upper bound for the conditional joint distribution of $(C,Y)$ given $Z$ with conditional marginals $F_{C\mid Z}(\cdot\mid z)$ and $F_Y$. As is well known, the Fr\'{e}chet--Hoeffding upper bound provides a valid conditional bivariate distribution with
\small\begin{align*}
	\Pr\left(C\leq c\mid Z=z\right)&=\Pr\left(C\leq c,Y\leq +\infty \mid Z=z\right)=  \min\!\left\{F_{C\mid Z}(c\mid z),\,1\right\}=F_{C\mid Z}(c\mid z),\\ 
\Pr\left(Y\leq y\mid Z=z\right)&= \Pr\left(C\leq +\infty,Y\leq y\mid Z=z\right)=  \min\!\left\{1,\,F_Y(y)\right\}=F_Y(y),\quad  c,y\in\overline{\mathbb{R}},\ z\in \mathcal{Z},
\end{align*}\normalsize
and integrating over $Z$ yields
\[
\Pr\left(C\leq c\right)=F_{C}(c),\qquad 
\Pr\left(Y\leq y\right)=F_Y(y),\qquad c,y\in\overline{\mathbb{R}}.
\]
This shows that (a), (b) and (d) are satisfied. Moreover, \begin{align*}
&\quad \ \Pr(C>y,\,Y\le y\mid Z=z)
=
\Pr(Y\le y\mid Z=z)
-
\Pr(C\le y,\,Y\le y\mid Z=z)
\\
&=
F_Y(y)
-
\min\!\left\{F_{C\mid Z}(y\mid z),\,F_Y(y)\right\}
=
F_Y(y)-F_Y(y)
=
0,\qquad y\in \overline{\mathbb{R}},\ z\in\mathcal{Z},
\end{align*}
where we used  $\{Y\le y\}=\{C\le y,\,Y\le y\}\uplus \{C>y,\,Y\le y\}$  and $F_Y(y)\le 
U(y)=\inf_{z\in \mathcal{Z}}F_{C\mid Z}(y\mid z)$. Since $C,Y\in\mathcal{J}$ a.s., it holds that
 $
\{C>Y\}
=\bigcup_{j=0}^{J-1}\{C>j,\,Y\le j\}
$ a.s.
Hence,
$$
\Pr(C>Y)
\le
\sum_{j=0}^{J-1}\Pr(C>j,\,Y\le j)
=
0,
$$
so that $C\le Y$ a.s. Besides, $C,Y\in\mathcal{J}$ a.s. is ensured by construction, so (c) is satisfied. Finally, the underreporting probabilities satisfy
\begin{align*}
	\Pr(C\le j\mid Y\ge j+1)
	&=\frac{\Pr(C\le j, Y\ge j+1)}{\Pr(Y\ge j+1)}=\frac{\Pr(C \leq j)-\Pr(C \leq j, Y \leq j)}{1-\Pr(Y \leq j)}\\
	&=\frac{F_C(j)-F_Y(j)}{1-F_Y(j)}, \quad j\in \mathcal{J}\setminus \{J\},\ F_Y(j)<1.
\end{align*}
where we again used $F_Y(j)\le
U(j)=\inf_{z\in \mathcal{Z}}F_{C\mid Z}(j\mid z)$. If $F_Y(j)=1$, then by convention
$
\Pr\!\left(C\le j \,\middle|\, Y\ge j+1\right)=0\le \alpha.
$ Because $F_Y(j)\geq L_{\alpha}(j)= \max\!\left\{0,\frac{F_{C}(j)-\alpha}{1-\alpha}\right\}$, we have
\begin{align*}
	F_Y(j)\geq \frac{F_{C}(j)-\alpha}{1-\alpha}  \iff \frac{F_C(j)-F_Y(j)}{1-F_Y(j)}\leq \alpha, \quad j\in \mathcal{J}\setminus \{J\},\ F_Y(j)<1.
\end{align*}
Hence,
$$\mathrm{Pr}\!\left(C\le j \mid Y\ge j+1\right)\le \alpha,\quad j\in\mathcal{J}\setminus \{J\},$$
so (e) is satisfied.

\medskip
\noindent\textbf{\underline{Case 2}: $\alpha=1$.} 

\noindent\underline{\textit{Validity.}} The underreporting restriction is vacuous.
Since $C\le Y$ a.s., we have 
$\{Y\le j\}\subseteq\{C\le j\}$ (up to a null set) and hence, from  (\ref{eq:misr_func3}),
$$F_Y(y)=\Pr(Y\le j(y))\le \Pr(C\le j(y)\mid Z=z)=F_{C\mid Z}(y\mid z),\quad y\in \overline{\mathbb{R}},\ z\in\mathcal{Z}.$$  
Since $C,Y\in \mathcal{J}$ a.s., $F_Y(y)\geq 0$ for $y< J$ and $F_Y(y)=1$ for $y\geq J$, so we obtain
$$\mathbb{I}\{y\ge J\}\leq F_Y(y)\leq \inf_{z\in \mathcal{Z}}F_{C\mid Z}(y\mid z),\quad y\in \overline{\mathbb{R}}.$$
These bounds for $F_{Y}(y)$ coincide with the definitions of $ L_{\alpha}(y)$ and $U(y)$ for $\alpha=1$.

\noindent\underline{\textit{Sharpness.}} The construction in Case~1 also applies when $\alpha=1$.

\noindent\textbf{\underline{Case 3}: $\alpha\in \left[0, \alpha^*\right)$.}

Suppose 
$\alpha<\alpha^*.$ 
We show that no $F_Y$ can be generated by any conditional laws of $(C,Y)$ given $Z=z$ that satisfy all imposed restrictions.  
Assume, towards a contradiction, that such $F_Y$ and conditional laws exist. 
From
 (\ref{eq:misr_func2}),
\[
\mathrm{Pr}\!\left(C\le j\,\middle|\,Y\ge j+1\right)=\frac{\Pr\!\left(C\leq j\right)-\Pr\!\left(Y\leq j\right)}{1-\Pr\!\left(Y\leq j\right)},
\quad  j\in\mathcal{J}\setminus \{J\},\ \Pr\!\left(Y\leq j\right)<1
\]\normalsize
(noting $\Pr\left(Y\geq j+1\right)=1-\Pr\left(Y\leq  j\right)>0$, so the fraction is well defined).
 Since $C\le Y$ a.s., we have $\{Y\le j\}\subseteq\{C\le j\}$ (up to a null set), hence, from exclusion restriction,
 $$ \Pr\!\left(Y\leq j\right)=\Pr\!\left(Y\leq j\mid Z=z\right)\le \Pr\!\left(C\leq j\,\middle|\,Z=z\right),\quad j\in\mathcal{J}\setminus \{J\},\ z\in\mathcal{Z}.$$
From this inequality, we note $\Pr\!\left(C\leq j\mid Z=z\right)<1$ implies $\Pr\!\left(Y\leq j\right)<1$. Moreover, since  the function $x\mapsto\frac{\Pr\!\left(C\leq j\right)-x}{1-x}$ is nonincreasing on $[0,1)$, it follows that
\footnotesize$$
 \mathrm{Pr}\!\left(C\le j\,\middle|\,Y\ge j+1\right)
 \ge
\frac{\Pr\!\left(C\leq j\right) - \Pr\!\left(C\leq j\mid Z=z\right)}{1 - \Pr\!\left(C\leq j\mid Z=z\right)},
\quad j\in\mathcal{J}\setminus \{J\},\ z\in\mathcal{Z},\ \Pr\!\left(C\leq j\mid Z=z\right)<1.
$$\normalsize
Taking $\sup_{z\in\mathcal{Z},\,\Pr\!\left(C\leq j\,\middle|\,Z=z\right)<1}$ and then $\max_{j\in\mathcal{J}\setminus \{J\}}$ yields
$$
 \max_{j\in\mathcal{J}\setminus \{J\}} \mathrm{Pr}\!\left(C\le j\,\middle|\,Y\ge j+1\right)
\ge\
\sup_{j\in \mathcal{J}\setminus \{J\},z \in \mathcal{Z},F_{C\mid Z}(j\mid z)<1} \frac{F_{C}(j) - F_{C\mid Z}(j\mid z)}{1 - F_{C\mid Z}(j\mid z)}=\alpha^*.
$$
However, from the underreporting bound, $\max_{j\in\mathcal{J}\setminus \{J\}}\mathrm{Pr}\!\left(C\le j\,\middle|\,Y\ge j+1\right)\le\alpha$. Combining this with the previous inequality yields
$
	\alpha\geq \alpha^*
$,
which contradicts the assumption 
$\alpha<
\alpha^*.$ 
\end{proof}

\begin{proof}[Proof of Proposition \ref{PROP:final2}]

\medskip
\noindent\textbf{\underline{Case 1}: $\alpha\in\left[\alpha^*,1\right]$.}

\noindent\underline{\textit{Validity.}}
Fix $\left(g,t\right)\neq\left(1,1\right)$. In these three cells we have $d=g\cdot t=0$. Since $\alpha\ge \alpha^*\ge \alpha^*_{gt}$, Proposition~\ref{PROP:misr_func_main} (applied conditional on $\left\{G=g,T=t\right\}$) implies that
\begin{equation}\label{eq:final2-UR}
L_{gt,\alpha}\left(y\right)\le F_{Y_{gt}(0)}\left(y\right)\le U_{gt}\left(y\right),
\qquad y\in\overline{\mathbb{R}},\ \left(g,t\right)\neq\left(1,1\right).
\end{equation}

On the other hand, by Theorem~4.1 of \citet{AI2006}, we have
\begin{equation*}
F_{Y_{10}(0)}\!\left(F_{Y_{00}(0)}^{(-1)}\!\left(F_{Y_{01}(0)}\!\left(y\right)\right)\right)
\le
F_{Y_{11}(0)}\!\left(y\right)
\le
F_{Y_{10}(0)}\!\left(F_{Y_{00}(0)}^{-1}\!\left(F_{Y_{01}(0)}\!\left(y\right)\right)\right)
\quad \text{for } y\in\left[\underline{y}_{01},\overline{y}_{01}\right],
\end{equation*}
where 
$\underline{y}_{01}:=\inf \mathrm{supp}\left(Y_{01}(0)\right)$ and 
$\overline{y}_{01}:=\sup \mathrm{supp}\left(Y_{01}(0)\right)$, and
\[
F_{Y_{11}(0)}\!\left(y\right)=0 \quad \text{for } y<\underline{y}_{01},
\qquad
F_{Y_{11}(0)}\!\left(y\right)=1 \quad \text{for } y>\overline{y}_{01}.
\]
We claim the inequality extends to all $y\in\overline{\mathbb{R}}$:
\begin{equation}\label{eq:final2-CIC2}
F_{Y_{10}(0)}\!\left(F_{Y_{00}(0)}^{(-1)}\!\left(F_{Y_{01}(0)}\!\left(y\right)\right)\right)
\le
F_{Y_{11}(0)}\!\left(y\right)
\le
F_{Y_{10}(0)}\!\left(F_{Y_{00}(0)}^{-1}\!\left(F_{Y_{01}(0)}\!\left(y\right)\right)\right),
\quad y\in\overline{\mathbb{R}}.
\end{equation}
To verify that \eqref{eq:final2-CIC2} also holds outside $\left[\underline{y}_{01},\overline{y}_{01}\right]$, note that condition~(3) of Assumption \ref{A2}, together with the monotonicity of $u\mapsto h(u,t)$, implies
\begin{equation}\label{eq:final2-support}
\mathrm{supp}\!\left(Y_{1t}(0)\right)\subseteq \mathrm{supp}\!\left(Y_{0t}(0)\right),
\qquad t\in\{0,1\}.
\end{equation}
Hence, if $y<\underline{y}_{01}$, we have $F_{Y_{11}(0)}(y)=0$, while $F_{Y_{01}(0)}(y)=0$, so
\[
F_{Y_{10}(0)}\!\left(F_{Y_{00}(0)}^{(-1)}(0)\right)=0,
\qquad
F_{Y_{10}(0)}\!\left(F_{Y_{00}(0)}^{-1}(0)\right)
=
F_{Y_{10}(0)}\!\left(\underline{y}_{00}\right),
\]
where $\underline{y}_{00}:=\inf \mathrm{supp}\!\left(Y_{00}(0)\right)$.  If $y>\overline{y}_{01}$, we have $F_{Y_{11}(0)}(y)=1$, while $F_{Y_{01}(0)}(y)=1$, so
\[
F_{Y_{10}(0)}\!\left(F_{Y_{00}(0)}^{(-1)}(1)\right)
=
F_{Y_{10}(0)}\!\left(F_{Y_{00}(0)}^{-1}(1)\right)
=
F_{Y_{10}(0)}\!\left(\overline{y}_{00}\right)=1,
\]
where $\overline{y}_{00}:=\sup \mathrm{supp}\!\left(Y_{00}(0)\right)$. Thus, the inequality holds.

\noindent\underline{Lower bound.}
Let $y\in\overline{\mathbb{R}}$. If $y<\underline{y}_{L,01}$, then $L_{11,\alpha}^{(0)}\left(y\right)=0$ by definition, and therefore
$$
L_{11,\alpha}^{(0)}\left(y\right)\le F_{Y_{11}(0)}\left(y\right).
$$ If $y>\overline{y}_{L,01}$, then $L_{01,\alpha}\left(y\right)=1$ by definition, and \eqref{eq:final2-UR} in cell $\left(g,t\right)=\left(0,1\right)$ yields
$F_{Y_{01}(0)}\left(y\right)=1$. By \eqref{eq:final2-support} with $t=1$, this implies $F_{Y_{11}(0)}\left(y\right)=1$. Since
$L_{11,\alpha}^{(0)}\left(y\right)=1$ for $y>\overline{y}_{L,01}$, we obtain
\[
L_{11,\alpha}^{(0)}\left(y\right)\le F_{Y_{11}(0)}\left(y\right).
\]
Now, suppose $y\in\left[\underline{y}_{L,01},\overline{y}_{L,01}\right]$.
From \eqref{eq:final2-CIC2},
\[
F_{Y_{11}(0)}\left(y\right)
\ge
F_{Y_{10}(0)}\!\left(F_{Y_{00}(0)}^{(-1)}\!\left(F_{Y_{01}(0)}\!\left(y\right)\right)\right).
\]
By \eqref{eq:final2-UR} in cell $\left(g,t\right)=\left(0,0\right)$, we have
$F_{Y_{00}(0)}\left(\cdot\right)\le U_{00}\left(\cdot\right)$.
Together with $\mathrm{supp}\!\left(Y_{10}(0)\right)\subseteq \mathrm{supp}\!\left(Y_{00}(0)\right)$ from \eqref{eq:final2-support} with $t=0$,
Lemma~\ref{lem:inv_mon}(ii) applied with
$F_1:=F_{Y_{00}(0)}$, $F_2:=U_{00}$, and $F_3:=F_{Y_{10}(0)}$
implies that, for each $q\in\left[0,1\right]$,
$
F_{Y_{10}(0)}\!\left(F_{Y_{00}(0)}^{(-1)}\!\left(q\right)\right)
\ge
F_{Y_{10}(0)}\!\left(U_{00}^{(-1)}\!\left(q\right)\right).
$
Hence,
\[
F_{Y_{10}(0)}\!\left(F_{Y_{00}(0)}^{(-1)}\!\left(F_{Y_{01}(0)}\!\left(y\right)\right)\right)
\ge
F_{Y_{10}(0)}\!\left(U_{00}^{(-1)}\!\left(F_{Y_{01}(0)}\left(y\right)\right)\right).
\]
Next, \eqref{eq:final2-UR} in cell $\left(g,t\right)=\left(1,0\right)$ gives
$L_{10,\alpha}\left(\cdot\right)\le F_{Y_{10}(0)}\left(\cdot\right)$, so
\[
F_{Y_{10}(0)}\!\left(U_{00}^{(-1)}\!\left(F_{Y_{01}(0)}\left(y\right)\right)\right)
\ge
L_{10,\alpha}\!\left(U_{00}^{(-1)}\!\left(F_{Y_{01}(0)}\left(y\right)\right)\right).
\]
Finally, by \eqref{eq:final2-UR} in cell $\left(g,t\right)=\left(0,1\right)$, we have
$F_{Y_{01}(0)}\left(y\right)\ge L_{01,\alpha}\left(y\right)$. Lemma~\ref{lem:inv_mon}(i) implies that $U_{00}^{(-1)}$ is nondecreasing, and $L_{10,\alpha}$ is also nondecreasing. Hence
\[
L_{10,\alpha}\!\left(U_{00}^{(-1)}\!\left(F_{Y_{01}(0)}\left(y\right)\right)\right)
\ge
L_{10,\alpha}\!\left(U_{00}^{(-1)}\!\left(L_{01,\alpha}\left(y\right)\right)\right).
\]
Combining the above inequalities yields
\[
F_{Y_{11}(0)}\left(y\right)
\ge
L_{10,\alpha}\!\left(U_{00}^{(-1)}\!\left(L_{01,\alpha}\left(y\right)\right)\right)
=
L_{11,\alpha}^{(0)}\left(y\right),
\qquad y\in\left[\underline{y}_{L,01},\overline{y}_{L,01}\right].
\]

\noindent\underline{Upper bound.}
Let $y\in\overline{\mathbb{R}}$. If $y<\underline{y}_{U,01}$, then $U_{01}\left(y\right)=0$ by definition, and \eqref{eq:final2-UR} in cell $\left(g,t\right)=\left(0,1\right)$ yields
$F_{Y_{01}(0)}\left(y\right)=0$. By \eqref{eq:final2-support} with $t=1$, this implies
$F_{Y_{11}(0)}\left(y\right)=0$. Since
$U_{11,\alpha}^{(0)}\left(y\right)=0$ for $y<\underline{y}_{U,01}$, we obtain
\[
F_{Y_{11}(0)}\left(y\right)\le U_{11,\alpha}^{(0)}\left(y\right).
\]
If $y>\overline{y}_{U,01}$, then $U_{11,\alpha}^{(0)}\left(y\right)=1$ by definition, and therefore
\[
F_{Y_{11}(0)}\left(y\right)\le U_{11,\alpha}^{(0)}\left(y\right).
\]
Now, suppose $y\in\left[\underline{y}_{U,01},\overline{y}_{U,01}\right]$.
From \eqref{eq:final2-CIC2},
\[
F_{Y_{11}(0)}\left(y\right)
\le
F_{Y_{10}(0)}\!\left(F_{Y_{00}(0)}^{-1}\!\left(F_{Y_{01}(0)}\!\left(y\right)\right)\right).
\]
By \eqref{eq:final2-UR} in cell $\left(g,t\right)=\left(0,1\right)$, we have
$F_{Y_{01}(0)}\left(y\right)\le U_{01}\left(y\right)$.
Since both $F_{Y_{00}(0)}^{-1}$ and $F_{Y_{10}(0)}$ are nondecreasing,
\[
F_{Y_{10}(0)}\!\left(F_{Y_{00}(0)}^{-1}\!\left(F_{Y_{01}(0)}\!\left(y\right)\right)\right)
\le
F_{Y_{10}(0)}\!\left(F_{Y_{00}(0)}^{-1}\!\left(U_{01}\left(y\right)\right)\right).
\]
Moreover, \eqref{eq:final2-UR} in cell $\left(g,t\right)=\left(0,0\right)$ gives
$L_{00,\alpha}\left(\cdot\right)\le F_{Y_{00}(0)}\left(\cdot\right)$. Lemma~\ref{lem:inv_mon}(iii) therefore implies that for each $q\in\left[0,1\right]$,
$
L_{00,\alpha}^{-1}\left(q\right)\ge F_{Y_{00}(0)}^{-1}\left(q\right).
$
Since $F_{Y_{10}(0)}$ is nondecreasing, 
\[
F_{Y_{10}(0)}\!\left(F_{Y_{00}(0)}^{-1}\!\left(U_{01}\left(y\right)\right)\right)
\le
F_{Y_{10}(0)}\!\left(L_{00,\alpha}^{-1}\!\left(U_{01}\left(y\right)\right)\right).
\]
Finally, \eqref{eq:final2-UR} in cell $\left(g,t\right)=\left(1,0\right)$ yields
$F_{Y_{10}(0)}\left(\cdot\right)\le U_{10}\left(\cdot\right)$, so
\[
F_{Y_{10}(0)}\!\left(L_{00,\alpha}^{-1}\!\left(U_{01}\left(y\right)\right)\right)
\le
U_{10}\!\left(L_{00,\alpha}^{-1}\!\left(U_{01}\left(y\right)\right)\right)
=
U_{11,\alpha}^{(0)}\left(y\right).
\]
Combining the above inequalities yields
\[
F_{Y_{11}(0)}\left(y\right)
\le
U_{10}\!\left(L_{00,\alpha}^{-1}\!\left(U_{01}\left(y\right)\right)\right)
=
U_{11,\alpha}^{(0)}\left(y\right),
\qquad
y\in\left[\underline{y}_{U,01},\overline{y}_{U,01}\right].
\]

\noindent\underline{\textit{Sharpness.}}
We show that, under the stated sufficient conditions, both the lower and upper bound functions are attainable. First, we note that, under
\(
\inf_{z\in \mathcal{Z}_{00}}
\mathrm{Pr}\!\left(C_{00}=j \,\middle|\, Z_{00}=z\right)>0
\)
for all \(j\in\mathcal{J}\) and \(\alpha<F_{C_{00}}(0)\), we have $\mathrm{supp}\!\left(L_{00,\alpha}\right)=\mathrm{supp}\!\left(U_{00}\right)
=
\mathcal{J}$. This implies
\begin{align}\label{eq:final2-support2}
	\mathrm{supp}\!\left(U_{10}\right)
\subseteq
\mathrm{supp}\!\left(L_{00,\alpha}\right),
\quad \mathrm{supp}\!\left(L_{10,\alpha}\right)
\subseteq
\mathrm{supp}\!\left(U_{00}\right).
\end{align}

\noindent\underline{Lower bound.}
Define the target untreated potential outcome distributions by
\[
F_{Y_{10}(0)}:=L_{10,\alpha},\qquad 
F_{Y_{00}(0)}:=U_{00},\qquad 
F_{Y_{01}(0)}:=L_{01,\alpha}.
\]
By \eqref{eq:final2-support2}, the support-overlap condition in \eqref{eq:final2-support} holds. Hence, these marginal distributions are admissible under Assumption \ref{A2}.

By the sharpness statement in Theorem~4.1 of \citet{AI2006}, applied to the triple 
\(\left(U_{00},L_{01,\alpha},L_{10,\alpha}\right)\), there exists a discrete CiC model satisfying Assumption \ref{A2}, whose induced marginals for $Y_{00}(0)$, $Y_{01}(0)$, and $Y_{10}(0)$ equal the chosen targets and for which
\[
F_{Y_{11}(0)}(y)
=
L_{10,\alpha}\!\left(
U_{00}^{(-1)}\!\left(L_{01,\alpha}(y)\right)
\right),
\]
with the corresponding boundary values outside 
\([\underline{y}_{L,01},\overline{y}_{L,01}]\).

Moreover, for each untreated cell $(g,t)\neq(1,1)$, the sharpness part of Proposition~\ref{PROP:misr_func_main} yields a joint distribution of $(C_{gt}(0),Y_{gt}(0),Z_{gt})$ that reproduces the observed law of $(C_{gt}(0),Z_{gt})$, satisfies Assumptions \ref{A1} and \ref{A3}, and preserves the constructed marginal $F_{Y_{gt}(0)}$.
Since the CiC model specifies a joint distribution of $(Y(0),G,T)$ and the cellwise construction specifies the conditional distribution of $(C(0),Z)$ given $(Y(0),G,T)$, there exists a joint distribution of $(C(0),Y(0),Z,G,T)$ consistent with both constructions. 

This joint distribution satisfies Assumptions \ref{A1}--\ref{A3}, preserves the CiC structure under Assumption \ref{A2}, reproduces the observed distributions in every cell, and attains $F_{Y_{11}(0)}(y)=L_{11,\alpha}^{(0)}(y)$ for all $y\in\overline{\mathbb{R}}$. Hence, the lower bound is sharp.

\noindent\underline{Upper bound.} Define the target untreated potential outcome distributions by
\[
F_{Y_{10}(0)}:=U_{10},\qquad 
F_{Y_{00}(0)}:=L_{00,\alpha},\qquad 
F_{Y_{01}(0)}:=U_{01}.
\]
By \eqref{eq:final2-support2}, the support-overlap condition in \eqref{eq:final2-support} holds. Hence, these marginal distributions are admissible under Assumption \ref{A2}. The remainder of the argument is identical to the lower-bound case.

\medskip
\noindent\textbf{\underline{Case 2}: $\alpha\in\left[0,\alpha^*\right)$.}
Since $\alpha<\alpha^*=\sup_{(g,t)\in\{0,1\}\times\{0,1\}}\alpha^*_{gt}$, there exists $(g,t)\in\{0,1\}\times\{0,1\}$ such that $\alpha<\alpha^*_{gt}$. 
By Proposition~\ref{PROP:misr_func_main}, there exists no data-generating process for $(C_{gt},Y_{gt}(d),Z_{gt})$ with $d=g\cdot t$ that is compatible with Assumptions \ref{A1} and \ref{A3} and reproduces the observed distribution of $(C_{gt},Z_{gt})$ in that cell. Hence, no data-generating process satisfying Assumptions \ref{A1}--\ref{A3} exists when $\alpha<\alpha^*$.\end{proof}

Before proving Corollary~\ref{COR:BOUND}, we relate the left-inverse in \citet{CFMW2020} to our generalized inverse. In \citet{CFMW2020}, distribution and band functions are defined on a closed subinterval of $\overline{\mathbb{R}}$. In our setting, $Y_{11}(d)\in\mathcal{J}$ a.s.\ for $d\in\{0,1\}$, so $F_{Y_{11}(d)}(y)=0$ for $y<0$ and $F_{Y_{11}(d)}(y)=1$ for $y\ge J$, and the same holds for the bounding functions in equations~(\ref{EQBOUNDS})--(\ref{EQLUBOUND}) and (\ref{final2BOUND1})--(\ref{final2BOUND2}). Hence, we may take $[0,J]$ as the relevant closed subinterval.

For any nondecreasing $F:[0,J]\to[0,1]$, \citet[Definition~2]{CFMW2020} define
\[
F^{\leftarrow}(\tau):=\inf\{y\in[0,J]\mid F(y)\ge \tau\},\quad \tau\in(0,1),
\]
whenever $\sup_{y\in[0,J]}F(y)\ge\tau$, and $F^{\leftarrow}(\tau)=J$ otherwise. In our application, $\sup_{y\in[0,J]}F(y)=1$, so the latter case does not arise. Since all relevant functions are proper distribution functions supported on $\mathcal{J}$, the infimum is attained at a support point, implying
\[
F^{\leftarrow}(\tau)
=
\inf\{y\in\operatorname{supp}(F)\mid F(y)\ge\tau\}
=
F^{-1}(\tau),\quad \tau\in(0,1).
\]
Thus, in the proof below we use the generalized inverse $F^{-1}$ in place of $F^{\leftarrow}$.
\begin{proof}[Proof of Corollary~\ref{COR:BOUND}]
Fix $\tau\in(0,1)$. Under Assumptions~\ref{A1}--\ref{A3} with $\alpha\in[\alpha^*,1]$, the bounds in equations~(\ref{EQBOUNDS}) and~(\ref{final2BOUND1}) yield
\[
L_{11,\alpha}^{(1)}(y)\le F_{Y_{11}(1)}(y)\le U_{11}^{(1)}(y),
\qquad 
L_{11,\alpha}^{(0)}(y)\le F_{Y_{11}(0)}(y)\le U_{11,\alpha}^{(0)}(y),
\qquad y\in[0,J].
\]
Applying Theorem~1 of \citet{CFMW2020} (with $p=1$) to each band and using $F^{\leftarrow}=F^{-1}$ on $(0,1)$ gives
\[
U_{11}^{(1),-1}(\tau)\le F_{Y_{11}(1)}^{-1}(\tau)\le L_{11,\alpha}^{(1),-1}(\tau),
\qquad 
U_{11,\alpha}^{(0),-1}(\tau)\le F_{Y_{11}(0)}^{-1}(\tau)\le L_{11,\alpha}^{(0),-1}(\tau).
\]
Since $\Delta_{\mathrm{QTT}}(\tau)=F_{Y_{11}(1)}^{-1}(\tau)-F_{Y_{11}(0)}^{-1}(\tau)$, Theorem~2 of \citet{CFMW2020} implies that $\Delta_{\mathrm{QTT}}(\tau)$ is contained in the pointwise Minkowski difference of the two quantile function bands. That is,
\[
U_{11}^{(1),-1}(\tau)-L_{11,\alpha}^{(0),-1}(\tau)
\le
\Delta_{\mathrm{QTT}}(\tau)
\le
L_{11,\alpha}^{(1),-1}(\tau)-U_{11,\alpha}^{(0),-1}(\tau),\qquad \tau\in(0,1).
\]
\end{proof}

\subsection{Proofs of Auxiliary Results}


\begin{proof}[Proof of Lemma \ref{lem:inv_mon}]
\noindent\textit{(i)} Let $q_1,q_2\in\left[0,1\right]$ with $q_1\le q_2$. 
Since $F(y)\le q_1$ implies $F(y)\le q_2$, we have
$
\left\{y\in\operatorname{supp}\left(F\right)\mid F\left(y\right)\le q_1\right\}
\subseteq
\left\{y\in\operatorname{supp}\left(F\right)\mid F\left(y\right)\le q_2\right\}.
$
Taking suprema on both sides yields
\[
F^{\left(-1\right)}\left(q_1\right)
\leq
F^{\left(-1\right)}\left(q_2\right).
\]
Likewise,
$
\left\{y\in\operatorname{supp}\left(F\right)\mid F\left(y\right)\ge q_2\right\}
\subseteq
\left\{y\in\operatorname{supp}\left(F\right)\mid F\left(y\right)\ge q_1\right\},
$
so taking infima yields
\[
F^{-1}\left(q_2\right)
\geq
F^{-1}\left(q_1\right).
\]
Hence, $F^{\left(-1\right)}$ and $F^{-1}$ are nondecreasing on $\left[0,1\right]$.

\noindent\textit{(ii)} Let $q\in\left[0,1\right]$.  If $F_1^{\left(-1\right)}\left(q\right)\ge F_2^{\left(-1\right)}\left(q\right)$, it immediately follows $F_3\left(F_1^{\left(-1\right)}\left(q\right)\right)\ge F_3\left(F_2^{\left(-1\right)}\left(q\right)\right)$ by monotonicity of $F_3$. Suppose $F_1^{\left(-1\right)}\left(q\right)<F_2^{\left(-1\right)}\left(q\right)$.
 We claim that
$$
\operatorname{supp}\left(F_{3}\right)\cap\left(F_1^{\left(-1\right)}\left(q\right),F_2^{\left(-1\right)}\left(q\right)\right]=\emptyset.
$$
Since $F_3$ can increase only at points in $\operatorname{supp}\left(F_3\right)$, the claim implies
$
F_3\!\left(F_1^{\left(-1\right)}\left(q\right)\right)=F_3\!\left(F_2^{\left(-1\right)}\left(q\right)\right).
$ To prove the claim, suppose toward a contradiction that there exists
$
y\in \operatorname{supp}\left(F_{3}\right)\cap\left(F_1^{\left(-1\right)}\left(q\right),F_2^{\left(-1\right)}\left(q\right)\right].
$
Because $\operatorname{supp}\left(F_3\right)\subseteq \operatorname{supp}\left(F_1\right)$, it follows that
\[
y\in \operatorname{supp}\left(F_{1}\right)\cap\left(F_1^{\left(-1\right)}\left(q\right),F_2^{\left(-1\right)}\left(q\right)\right].
\]
Using monotonicity of $F_1$, the inequality $F_1(y)\le F_2(y)$ for all $y\in\overline{\mathbb{R}}$, and the property $F_2\!\left(F_2^{(-1)}(q)\right)\le q$ for all $q\in[0,1]$ (see, e.g., \cite{AI2006}, p.~453), we obtain
\[
F_1(y)\le F_1\!\left(F_2^{\left(-1\right)}(q)\right)\le F_2\!\left(F_2^{\left(-1\right)}(q)\right)\le q.
\]
Hence, $y\in \left\{y\in\operatorname{supp}\left(F_1\right)\mid F_1(y)\le q\right\}$, which implies that
\[
y\le \sup\left\{y\in\operatorname{supp}\left(F_1\right)\mid F_1(y)\le q\right\}
=F_1^{\left(-1\right)}(q),
\]
contradicting $y>F_1^{\left(-1\right)}(q)$.

\noindent\textit{(iii)}  Let $q\in\left[0,1\right]$. If $q=0$, then $F_j^{-1}(0)=\inf \operatorname{supp}\left(F_j\right)$, and 
$F_1(y)\le F_2(y)$ for all $y\in\overline{\mathbb{R}}$ implies
\[
F_1^{-1}(0)=\inf \operatorname{supp}\left(F_1\right) 
\ge 
\inf \operatorname{supp}\left(F_2\right)
=F_2^{-1}(0).
\]
Suppose $q \in(0,1]$. Since $F_1(y)\le F_2(y)$ for all $y\in\overline{\mathbb{R}}$, we have 
$
\left\{j \in \mathcal{J}\mid  F_1(j) \geq q\right\} 
\subseteq 
\left\{j \in \mathcal{J}\mid  F_2(j) \geq q\right\}.
$
Taking infima yields
\[
\inf\left\{j \in \mathcal{J}\mid  F_1(j) \geq q\right\} 
\ge 
\inf\left\{j \in \mathcal{J}\mid  F_2(j) \geq q\right\}.
\]
To conclude the proof, we claim that for each $j\in\{1,2\}$,
\[
F_j^{-1}\left(q\right)
=
\inf\left\{y\in\mathcal{J}\mid F_j(y)\ge q\right\}.
\]
To prove the claim, first note that since $\operatorname{supp}(F_j)\subseteq \mathcal{J}$,
\[
F_j^{-1}(q)
=
\inf\left\{y\in \operatorname{supp}(F_j)\mid F_j(y)\ge q\right\}
\ge
\inf\left\{y\in \mathcal{J}\mid F_j(y)\ge q\right\}.
\]
For the reverse inequality, let
$
y^*:=\inf\left\{y\in \mathcal{J}\mid F_j(y)\ge q\right\}.
$
Since $\mathcal{J}$ is finite, $y^*\in\mathcal{J}$ and $F_j(y^*)\ge q$. Hence,
$
y^*\in \left\{y\in \operatorname{supp}(F_j)\mid F_j(y)\ge q\right\},
$
so
\[
F_j^{-1}(q)
=
\inf\left\{y\in \operatorname{supp}(F_j)\mid F_j(y)\ge q\right\}
\le y^*.
\]

\noindent\textit{(iv)} Let $q\in\left(0,1\right]$ and $x\in\overline{\mathbb{R}}$. Using $F\left(F^{-1}\left(q\right)\right)\geq q$ for all $q\in[0,1]$ (see, e.g., \cite{AI2006}, p.~453), we have
\begin{align*}
F^{-1}\left(q\right)\le x
\;\implies\;
q\le F\left(F^{-1}\left(q\right)\right)\le F(x).
\end{align*}
For the reverse direction, assume $q \leq F(x)$. Since $q>0$, we have $F(x)>0$. 
Because $F \in \mathcal{F}$ is concentrated on $\mathcal{J}$, this implies
$
x \ge \inf \operatorname{supp}(F).
$
Hence, by definition of the generalized inverse,
\begin{align*}
F^{-1}\left(F(x)\right)
=
\inf \left\{ y \in \operatorname{supp}(F) \mid F(y) \ge F(x) \right\}
\le x.
\end{align*}
Since $F^{-1}$ is nondecreasing on $\left[0,1\right]$ by Lemma \ref{lem:inv_mon}(i),
\begin{align*}
F^{-1}\left(q\right) \le F^{-1}\left(F(x)\right) \le x.
\end{align*}

\noindent\textit{(v)} Let $q\in\left[0,1\right]$ and $x\in\operatorname{supp}\left(F\right)$. If $F(x)\le q$, then by definition of $F^{(-1)}$,
\[
x\in\left\{y\in\operatorname{supp}\left(F\right)\mid F(y)\le q\right\}
\quad\implies\quad
x\le F^{(-1)}(q).
\]
Conversely, if $x\le F^{(-1)}(q)$, then by monotonicity of $F$ and using $F\left(F^{\left(-1\right)}\left(q\right)\right)\le q$ for all $q\in[0,1]$ (see, e.g., \cite{AI2006}, p.~453),
\[
F(x)\le F\!\left(F^{(-1)}(q)\right)\le q.
\]
\end{proof}

\section{Proofs of Estimation and Inference}\label{sec:proofestinf}

\subsection{Construction of Confidence Bands for QTT Bounds}\label{ssec:CSQTT}

To construct confidence bands for the QTT bounds, we first introduce the treated
post-treatment analogue of the objects used in the main text. Since $F_{Y_{11}(1)}$ is bounded directly from the observed distribution of
$\left(C_{11},Z_{11}\right)$ under Assumptions~\ref{A1} and~\ref{A3}, the corresponding
construction is simpler and does not involve the CiC transformations
$\underline{k}_{\alpha}$ and $\overline{k}_{\alpha}$.

Let $\alpha\in\mathcal{A}$. As in equations~(\ref{eq:phi_def})--(\ref{eq:phi_defs}), let
$$
\left(\theta_{L}^{(1)},\theta_{U}^{(1)}\right)
:=
\left(
F_{C_{11}},
F_{C_{11}\mid Z_{11}}
\right)
\in
\ell^{\infty}\!\left(\mathcal{J},\mathbb{R}\right)
\times
\ell^{\infty}\!\left(\mathcal{J}\times\mathcal{Z}_{11},\mathbb{R}\right),
$$
and define the functional
$
\boldsymbol{\phi}_{\alpha}^{(1)}
:
\ell^{\infty}\!\left(\mathcal{J},\mathbb{R}\right)
\times
\ell^{\infty}\!\left(\mathcal{J}\times\mathcal{Z}_{11},\mathbb{R}\right)
\to
\ell^{\infty}\!\left(\mathcal{J},\mathbb{R}^{2}\right)
$
by
\[
\left[
\boldsymbol{\phi}_{\alpha}^{(1)}\left(\theta_{1},\theta_{2}\right)
\right](y)
=
\begin{pmatrix}
\dfrac{
\theta_{1}(y)-\min\{\theta_{1}(y),\alpha\}
}{
1-\min\{\theta_{1}(y),\alpha\}
}
\\
\inf_{z\in\mathcal{Z}_{11}} \theta_{2}(y,z)
\end{pmatrix}.
\]

To construct empirical and bootstrap counterparts, denote
\small
\begin{align*}
\widehat{F}_{C_{11}}(y)
&:=
\frac{1}{N_{11}}
\sum_{i=1}^{N_{11}}
\mathbb{I}\!\left\{
C_{i,11}\le y
\right\},
\qquad
\widehat{F}_{C_{11}\mid Z_{11}}(y\mid z)
:=
\frac{
\frac{1}{N_{11}}
\sum_{i=1}^{N_{11}}
\mathbb{I}\!\left\{
C_{i,11}\le y
\right\}
\mathbb{I}\!\left\{
Z_{i,11}=z
\right\}
}{
\frac{1}{N_{11}}
\sum_{i=1}^{N_{11}}
\mathbb{I}\!\left\{
Z_{i,11}=z
\right\}
},\\
\widehat{F}^{\,*}_{C_{11}}(y)
&:=
\frac{1}{N_{11}}
\sum_{i=1}^{N_{11}}
\mathbb{I}\!\left\{
C_{i,11}^{*}\le y
\right\},
\qquad
\widehat{F}^{\,*}_{C_{11}\mid Z_{11}}(y\mid z)
:=
\frac{
\frac{1}{N_{11}}
\sum_{i=1}^{N_{11}}
\mathbb{I}\!\left\{
C_{i,11}^{*}\le y
\right\}
\mathbb{I}\!\left\{
Z_{i,11}^{*}=z
\right\}
}{
\frac{1}{N_{11}}
\sum_{i=1}^{N_{11}}
\mathbb{I}\!\left\{
Z_{i,11}^{*}=z
\right\}
}.
\end{align*}
\normalsize
for all $y\in\mathcal{J}$ and $z\in\mathcal{Z}_{11}$, where $\left\{\left(C_{i, 11}^*, Z_{i, 11}^*\right)\right\}_{i=1}^{N_{11}}$ is a nonparametric bootstrap sample drawn with replacement from $\left\{\left(C_{i, 11}, Z_{i, 11}\right)\right\}_{i=1}^{N_{11}}$. Set
$$
\left(\widehat{\theta}_{L}^{(1)},\widehat{\theta}_{U}^{(1)}\right)
:=
\left(
\widehat{F}_{C_{11}},
\widehat{F}_{C_{11}\mid Z_{11}}
\right),
\qquad
\left(\widehat{\theta}_{L}^{(1),*},\widehat{\theta}_{U}^{(1),*}\right)
:=
\left(
\widehat{F}^{\,*}_{C_{11}},
\widehat{F}^{\,*}_{C_{11}\mid Z_{11}}
\right).
$$
Then
$$
\begin{pmatrix}
L_{11,\alpha}^{(1)}
\\
U_{11}^{(1)}
\end{pmatrix}
=
\boldsymbol{\phi}_{\alpha}^{(1)}
\left(
\theta_{L}^{(1)},
\theta_{U}^{(1)}
\right),
\qquad
\begin{pmatrix}
\widehat{L}_{11,\alpha}^{(1)}
\\
\widehat{U}_{11}^{(1)}
\end{pmatrix}
=
\boldsymbol{\phi}_{\alpha}^{(1)}
\left(
\widehat{\theta}_{L}^{(1)},
\widehat{\theta}_{U}^{(1)}
\right).
$$

To implement the numerical directional derivative bootstrap, define
$$
\widehat{\boldsymbol{\phi}}_{\alpha}^{(1)\prime}
\left(
h_{L},h_{U}
\right)
:=
\frac{
\boldsymbol{\phi}_{\alpha}^{(1)}
\left(
\widehat{\theta}_{L}^{(1)}+\epsilon_{N}h_{L},
\widehat{\theta}_{U}^{(1)}+\epsilon_{N}h_{U}
\right)
-
\boldsymbol{\phi}_{\alpha}^{(1)}
\left(
\widehat{\theta}_{L}^{(1)},
\widehat{\theta}_{U}^{(1)}
\right)
}{
\epsilon_{N}
},
$$
where $\epsilon_{N}\to0$ and $\sqrt{N}\epsilon_{N}\to\infty$. As in the main text, we
then approximate the limiting distribution of the bound estimator by applying this
numerical directional derivative to the bootstrap empirical process. Specifically,
define
$$
\begin{pmatrix}
T_{1,\alpha}^{(1),*}
\\
T_{2,\alpha}^{(1),*}
\end{pmatrix}
:=
\widehat{\boldsymbol{\phi}}_{\alpha}^{(1)\prime}
\left(
\sqrt{N}
\left(
\widehat{\theta}_{L}^{(1),*}-\widehat{\theta}_{L}^{(1)}
\right),
\sqrt{N}
\left(
\widehat{\theta}_{U}^{(1),*}-\widehat{\theta}_{U}^{(1)}
\right)
\right)
\in
\ell^{\infty}\!\left(\mathcal{J},\mathbb{R}^{2}\right).
$$

To describe the corresponding population limit, let
$$
\begin{pmatrix}
G_{L,11,\alpha}^{(1)}
\\
G_{U,11}^{(1)}
\end{pmatrix}
:=
\boldsymbol{\phi}_{\alpha,\left(\theta_{L}^{(1)},\theta_{U}^{(1)}\right)}^{(1)\prime}
\!\left(
G_{C_{11}},
G_{C_{11}\mid Z_{11}}
\right),
$$
where $\boldsymbol{\phi}_{\alpha,\left(\theta_{L}^{(1)},\theta_{U}^{(1)}\right)}^{(1)\prime}$
is the Hadamard directional derivative of $\boldsymbol{\phi}_{\alpha}^{(1)}$ at
$\left(\theta_{L}^{(1)},\theta_{U}^{(1)}\right)$, and
$
\left(
G_{C_{11}},
G_{C_{11}\mid Z_{11}}
\right)
$
is the weak limit in Lemma~\ref{LEMDIST}, applied with
$\left(g,t\right)=\left(1,1\right)$ and $\psi_{1}=\psi_{2}=\mathrm{id}$.
Moreover,
$$
\left[
\boldsymbol{\phi}_{\alpha,\left(\theta_{L}^{(1)},\theta_{U}^{(1)}\right)}^{(1)\prime}
\left(
h_{1},h_{2}
\right)
\right]\left(y\right)
=
\begin{pmatrix}
\dfrac{1}{1-\alpha}
h_{1}\left(y\right)
\left(
\mathbb{I}\!\left\{
\theta_{L}^{(1)}\left(y\right)>\alpha
\right\}
+
\mathbb{I}\!\left\{
\theta_{L}^{(1)}\left(y\right)=\alpha
\right\}
\mathbb{I}\!\left\{
h_{1}\left(y\right)>0
\right\}
\right)
\\
\min_{z\in\Psi^{(1)}\left(y\right)}
h_{2}\left(y,z\right)
\end{pmatrix},
$$
for all
$h_{1}\in \ell^{\infty}\!\left(\mathcal{J},\mathbb{R}\right)$ and
$h_{2}\in \ell^{\infty}\!\left(\mathcal{J}\times\mathcal{Z}_{11},\mathbb{R}\right)$,
with
$
\Psi^{(1)}\left(y\right)
:=
\operatorname*{arg\,inf}_{z\in\mathcal{Z}_{11}}
\theta_{U}^{(1)}\left(y,z\right).
$

Using the same max functional as in Proposition~\ref{PROPZGAMMA}, define
\begin{align*}
M_{\alpha}^{(1),*}
&:=
m\!\left(
\begin{pmatrix}
T_{1,\alpha}^{(1),*}
\\
T_{2,\alpha}^{(1),*}
\end{pmatrix}
\right)
=
\max\left\{
\max_{j\in\mathcal{J}\setminus\{J\}}
\frac{T_{1,\alpha}^{(1),*}(j)}{\sigma(j)},
\ 
\max_{j\in\mathcal{J}\setminus\{J\}}
\left(
-\frac{T_{2,\alpha}^{(1),*}(j)}{\sigma(j)}
\right)
\right\},\\
\widehat z_{\alpha}^{(1)}\left(1-\gamma\right)
&:=
\inf\left\{
z\in\mathbb{R}
\mid
\Pr\!\left(
M_{\alpha}^{(1),*}\le z
\mid
\left\{\left(C_i,Z_i,G_i,T_i\right)\right\}_{i=1}^{N}
\right)
\ge 1-\gamma
\right\},
\end{align*}
and
$$
M_{\alpha}^{(1)}
:=
m\!\left(
\begin{pmatrix}
G_{L,11,\alpha}^{(1)}
\\
G_{U,11}^{(1)}
\end{pmatrix}
\right),
\qquad
z_{\alpha}^{(1)}\left(1-\gamma\right)
:=
\inf\left\{
z\in\mathbb{R}
\mid
\mathrm{Pr}\!\left(
M_{\alpha}^{(1)}\le z
\right)
\ge 1-\gamma
\right\}.
$$
This yields a confidence band for the bounds on $F_{Y_{11}(1)}$. For $y\in\mathcal{J}\setminus\{J\}$, define
$$
\mathrm{CS}_{\alpha}^{(1)}\left(y;1-\gamma\right)
:=
\left[
\widehat{L}_{11,\alpha}^{(1)}\left(y\right)
-
\frac{
\sigma\left(y\right)\widehat z_{\alpha}^{(1)}\left(1-\gamma\right)
}{
\sqrt{N}
},
\;
\widehat{U}_{11}^{(1)}\left(y\right)
+
\frac{
\sigma\left(y\right)\widehat z_{\alpha}^{(1)}\left(1-\gamma\right)
}{
\sqrt{N}
}
\right]
\cap
\left[0,1\right],
$$
and set
$
\mathrm{CS}_{\alpha}^{(1)}\left(J;1-\gamma\right):=\left\{1\right\}.
$
Finally, for $\tau\in\left(0,1\right)$, define the QTT confidence band by
$$
	\mathrm{CS}_{\mathrm{QTT},\alpha}\left(\tau;1-\gamma\right)
:=
\mathrm{CS}_{\alpha}^{(1),-1}\left(\tau;1-\gamma/2\right)
\ominus
\mathrm{CS}_{\alpha}^{(0),-1}\left(\tau;1-\gamma/2\right),
$$
that is, the pointwise Minkowski difference of the two quantile confidence bands, as defined in Footnote~\ref{FOOTMINKOWSKI} in the main text.

\subsection{Auxiliary Results}\label{AuxLemmas}

The first lemma establishes joint weak convergence of the empirical distribution functions defined below, both unconditionally and conditional on $Z_{gt}$.

Let
$
\psi:\mathcal{J}\to\mathcal{J}
$ be a function.
In our application, $\psi$ is either the identity map or $\underline{k}_{\alpha}$ or $\overline{k}_{\alpha}$,
where $\underline{k}_{\alpha}$ and $\overline{k}_{\alpha}$ are defined in
equation~(\ref{klowup}). For each $(g,t)\in \{0,1\}\times\{0,1\}$, $y\in\mathcal{J}$ and $z\in\mathcal{Z}_{gt}$, define
\begin{align*}
F_{\psi(C_{gt})}(y)
&:=
\mathrm{Pr}\!\left(\psi(C_{gt})\le y\right),\qquad F_{\psi(C_{gt})\mid Z_{gt}}(y\mid z)
:=
\mathrm{Pr}\!\left(\psi(C_{gt})\le y \mid Z_{gt}=z\right).
\end{align*}
Their empirical counterparts are
\begin{align*}
\widehat{F}_{\psi(C_{gt})}(y)
&:=
\frac{1}{N_{gt}}
\sum_{i=1}^{N_{gt}}
\mathbb{I}\!\left\{\psi(C_{i,gt})\le y\right\},\qquad 
\widehat{F}_{\psi(C_{gt})\mid Z_{gt}}(y\mid z)
:=
\frac{
\frac{1}{N_{gt}}\sum_{i=1}^{N_{gt}}
\mathbb{I}\!\left\{\psi(C_{i,gt})\le y\right\}
\mathbb{I}\!\left\{Z_{i,gt}=z\right\}
}{
\frac{1}{N_{gt}}\sum_{i=1}^{N_{gt}}
\mathbb{I}\!\left\{Z_{i,gt}=z\right\}
}.
\end{align*}
We may set $\widehat{F}_{\psi(C_{gt})\mid Z_{gt}}(y\mid z)$ equal to an arbitrary value in $[0,1]$ whenever the denominator is zero, an event that occurs with probability approaching zero under Assumptions~\ref{AE0}--\ref{AE0A}. The same convention applies to $\widetilde{U}_{11,\alpha}^{(0)}(y)$ defined below.


\medskip

\begin{lemma}[Convergence of Empirical CDFs of $\psi(C_{gt})$]\label{LEMDIST}
Suppose Assumptions~\ref{AE0}--\ref{AE0A} hold. Let $(g,t)\in \{0,1\}\times \{0,1\}$,
$\psi_1,\psi_2:\mathcal{J}\rightarrow \mathcal{J}$ be functions. Then
\small$$
\sqrt{N}
\left(
\widehat{F}_{\psi_1(C_{gt})}-F_{\psi_1(C_{gt})},
\widehat{F}_{\psi_2(C_{gt})\mid Z_{gt}}-F_{\psi_2(C_{gt})\mid Z_{gt}}
\right)
\rightsquigarrow
\left(
G_{\psi_1(C_{gt})},
G_{\psi_2(C_{gt})\mid Z_{gt}}
\right)$$
in $\ell^{\infty}\!\left(\mathcal{J},\mathbb{R}\right)
\times
\ell^{\infty}\!\left(\mathcal{J}\times\mathcal{Z}_{gt},\mathbb{R}\right)$,
where $\bigl(G_{\psi_1(C_{gt})},G_{\psi_2(C_{gt})\mid Z_{gt}}\bigr)$ is a mean-zero Gaussian
process with covariance kernel characterized by, for $y,y'\in\mathcal{J}$ and
$z,z'\in\mathcal{Z}_{gt}$,
\fontsize{9}{10}\selectfont
\begin{align*}
\mathrm{E}\!\left[G_{\psi_1(C_{gt})}(y)G_{\psi_1(C_{gt})}(y')\right]
&=
\frac{1}{\pi_{gt}}
\left(
F_{\psi_1(C_{gt})}\!\left(\min\left\{y,y'\right\}\right)
-
F_{\psi_1(C_{gt})}(y)F_{\psi_1(C_{gt})}(y')
\right),\\
\mathrm{E}\!\left[
G_{\psi_2(C_{gt})\mid Z_{gt}}(y,z)
G_{\psi_2(C_{gt})\mid Z_{gt}}(y',z')
\right]
&=
\frac{1}{\pi_{gt}}
\frac{\mathbb{I}\left\{z=z'\right\}}{\Pr\left(Z_{gt}=z\right)}
\left(
F_{\psi_2(C_{gt})\mid Z_{gt}}\!\left(\min\left\{y,y'\right\}\mid z\right)
-
F_{\psi_2(C_{gt})\mid Z_{gt}}(y\mid z)\,
F_{\psi_2(C_{gt})\mid Z_{gt}}(y'\mid z)
\right),\\
\mathrm{E}\!\left[
G_{\psi_1(C_{gt})}(y)
G_{\psi_2(C_{gt})\mid Z_{gt}}(y',z')
\right]
&=
\frac{1}{\pi_{gt}}
\left(
F_{\psi_1(C_{gt}),\psi_2(C_{gt})\mid Z_{gt}}\!\left(y,y'\mid z'\right)
-
F_{\psi_1(C_{gt})\mid Z_{gt}}(y\mid z')\,
F_{\psi_2(C_{gt})\mid Z_{gt}}(y'\mid z')
\right),
\end{align*}
\normalsize
where
$F_{\psi_1\left(C_{g t}\right), \psi_2\left(C_{g t}\right) \mid Z_{g t}}\left(y, y^{\prime} \mid z^{\prime}\right):=\operatorname{Pr}\left(\psi_1\left(C_{g t}\right) \leq y, \psi_2\left(C_{g t}\right) \leq y^{\prime} \mid Z_{g t}=z^{\prime}\right)$.
\end{lemma}

The next lemma is in the spirit of \citet[Lemma A.12(i)]{AI2006}. It provides an equivalent representation of the bounds for $F_{Y_{11}(0)}$ in terms of the distribution of the transformed variable $\overline{k}_{\alpha}(C_{10})$ and $\underline{k}_{\alpha}(C_{10})$. In contrast to \citet{AI2006}, the result is formulated independently of the primitive discrete CiC structure and allows for underreporting. 

We remark that Assumption~\ref{AE0_TIES}(ii) in Lemma~\ref{BOUNDEQV} may be replaced by the weaker conditions
$\operatorname{supp}(C_{10})\subseteq \operatorname{supp}(U_{00})$ and 
$\operatorname{supp}(C_{10})\subseteq \operatorname{supp}(L_{00,\alpha})$.
\medskip

\begin{lemma}[Equivalent Representation of Bounds for $F_{Y_{11}(0)}$]\label{BOUNDEQV}
Suppose Assumption~\ref{AE0_TIES} holds.
Then, for each $\alpha\in\mathcal{A}$ and $y\in\mathcal{J}$,
\begin{align*}
L_{11,\alpha}^{(0)}\left(y\right)
&=
\frac{
\mathrm{Pr}\left(\overline{k}_{\alpha}\left(C_{10}\right)\le y\right)
-
\min\left\{\mathrm{Pr}\left(\overline{k}_{\alpha}\left(C_{10}\right)\le y\right),\alpha\right\}
}{
1-
\min\left\{\mathrm{Pr}\left(\overline{k}_{\alpha}\left(C_{10}\right)\le y\right),\alpha\right\}
},\\
U_{11,\alpha}^{(0)}\left(y\right)
&=
\inf_{z\in\mathcal{Z}_{10}}
\mathrm{Pr}\left(\underline{k}_{\alpha}\left(C_{10}\right)\le y\mid Z_{10}=z\right),
\end{align*}
with $L_{11,\alpha}^{(0)}$ and $U_{11,\alpha}^{(0)}$ defined in equation~(\ref{final2BOUND2}), and $\underline{k}_{\alpha}$ and $\overline{k}_{\alpha}$ in equation~(\ref{klowup}), respectively.
\end{lemma}

\medskip

To analyze the asymptotic behavior of the bound estimators in equation~(\ref{ESTLUBOUND}), we first study their infeasible analogues obtained by treating $\underline{k}_{\alpha}$ and $\overline{k}_{\alpha}$ as fixed population transformations:
\begin{align*}
\begin{aligned}
\widetilde{L}_{11,\alpha}^{(0)}(y)
&:=
\frac{
\frac{1}{N_{10}}\sum_{i=1}^{N_{10}}
\mathbb{I}\left\{{\overline{k}}_{\alpha}(C_{i,10})\le y\right\}
-
\min\!\left\{
\frac{1}{N_{10}}\sum_{i=1}^{N_{10}}
\mathbb{I}\left\{{\overline{k}}_{\alpha}(C_{i,10})\le y\right\},
\alpha
\right\}
}{
1-
\min\!\left\{
\frac{1}{N_{10}}\sum_{i=1}^{N_{10}}
\mathbb{I}\left\{{\overline{k}}_{\alpha}(C_{i,10})\le y\right\},
\alpha
\right\}
}
\\
\widetilde{U}_{11,\alpha}^{(0)}(y)
&:=
\min_{z\in \mathcal{Z}_{10}}
\frac{
\sum_{i=1}^{N_{10}}
\mathbb{I}\left\{{\underline{k}}_{\alpha}(C_{i,10})\le y\right\}
\mathbb{I}\left\{Z_{i,10}=z\right\}
}{
\sum_{i=1}^{N_{10}}
\mathbb{I}\left\{Z_{i,10}=z\right\}
}
\end{aligned},\quad y\in \mathcal{J}.
\end{align*}
 Using the notation introduced at the beginning of this subsection, they can be written as
 \begin{align*}
\begin{aligned}
\widetilde{L}_{11,\alpha}^{(0)}(y)
&=
\frac{
\widehat{F}_{\overline{k}_{\alpha}(C_{10})}(y)
-
\min\!\left\{
\widehat{F}_{\overline{k}_{\alpha}(C_{10})}(y),
\alpha
\right\}
}{
1-
\min\!\left\{
\widehat{F}_{\overline{k}_{\alpha}(C_{10})}(y),
\alpha
\right\}
},\qquad 
\widetilde{U}_{11,\alpha}^{(0)}(y)
=
\min_{z\in\mathcal{Z}_{10}}
\widehat{F}_{\underline{k}_{\alpha}(C_{10})\mid Z_{10}}(y\mid z).
\end{aligned}
\end{align*}

\medskip

\begin{lemma}[Convergence of Infeasible Bound Estimators for $F_{Y_{11}(0)}$]\label{BOUNDDIST}
Suppose Assumptions~\ref{AE0}--\ref{AE0_TIES} hold. Then, for each $\alpha\in\mathcal{A}$,
\[
\sqrt{N}
\begin{pmatrix}
\widetilde{L}_{11,\alpha}^{(0)}-L_{11,\alpha}^{(0)}
\\
\widetilde{U}_{11,\alpha}^{(0)}-U_{11,\alpha}^{(0)}
\end{pmatrix}
\rightsquigarrow
\begin{pmatrix}
G_{L,11,\alpha}^{(0)}
\\
G_{U,11,\alpha}^{(0)}
\end{pmatrix}\quad
\text{in }
\ell^{\infty}\!\left(\mathcal{J},\mathbb{R}^{2}\right),
\]
with
\[
\begin{pmatrix}
G_{L,11,\alpha}^{(0)}
\\
G_{U,11,\alpha}^{(0)}
\end{pmatrix}:=
\boldsymbol{\phi}^{(0)'}_{\alpha,\left(\theta_{L,\alpha}^{(0)},\theta_{U,\alpha}^{(0)}\right)}
\!\left(
G_{\overline{k}_{\alpha}(C_{10})},
G_{\underline{k}_{\alpha}(C_{10})\mid Z_{10}}
\right),
\]
$\boldsymbol{\phi}^{(0)'}_{\alpha,\left(\theta_{L,\alpha}^{(0)},\theta_{U,\alpha}^{(0)}\right)}$ the Hadamard directional derivative of $\boldsymbol{\phi}_{\alpha}^{(0)}$ at $\left(\theta_{L,\alpha}^{(0)},\theta_{U,\alpha}^{(0)}\right)$, $\theta_{L,\alpha}^{(0)}$, $\theta_{U,\alpha}^{(0)}$, and $\boldsymbol{\phi}_{\alpha}^{(0)}$ defined in equations~(\ref{eq:phi_def})--(\ref{eq:phi_defs}), and
\(
\left(
G_{\overline{k}_{\alpha}(C_{10})},
G_{\underline{k}_{\alpha}(C_{10})\mid Z_{10}}
\right)
\)
the weak limit in Lemma~\ref{LEMDIST}, applied with
$\left(g,t\right)=\left(1,0\right)$,
$\psi_1=\overline{k}_{\alpha}$, and
$\psi_2=\underline{k}_{\alpha}$.
Moreover,
\[
\left[
\boldsymbol{\phi}^{(0)'}_{\alpha,\left(\theta_{L,\alpha}^{(0)},\theta_{U,\alpha}^{(0)}\right)}(h_1,h_2)
\right](y)
=
\begin{pmatrix}
\dfrac{1}{1-\alpha}
h_1(y)
\left(
\mathbb{I}\left\{\theta_{L,\alpha}^{(0)}(y)>\alpha\right\}
+
\mathbb{I}\left\{\theta_{L,\alpha}^{(0)}(y)=\alpha\right\}
\mathbb{I}\left\{h_1(y)>0\right\}
\right)
\\
\min_{z\in\Psi_{\alpha}^{(0)}(y)} h_2(y,z)
\end{pmatrix},
\]
for all $h_1\in \ell^{\infty}\!\left(\mathcal{J},\mathbb{R}\right)$ and
$h_2\in \ell^{\infty}\!\left(\mathcal{J}\times\mathcal{Z}_{10},\mathbb{R}\right)$,
with $\Psi_{\alpha}^{(0)}(y)
:=
\operatorname{arg\,inf}_{z\in\mathcal{Z}_{10}}
\theta_{U,\alpha}^{(0)}(y,z)$.
\end{lemma}

\medskip

The final lemma is in the spirit of \citet[Lemma A.13]{AI2006}. It shows that the estimated
transformation functions coincide with their population counterparts on the finite set
$\mathcal{J}$ with probability approaching one. This property, which is akin to the sample median as discussed in Footnote~\ref{Median} in the main text, will be used below to replace
the population mappings $\underline{k}_{\alpha}$ and $\overline{k}_{\alpha}$ by their estimators
without affecting first-order asymptotics. In particular, for any deterministic sequence
$r_N \to \infty$ as $N\to \infty$,
\[
r_N \max_{y\in\mathcal{J}}
\left|\widehat{\overline{k}}_\alpha(y)-\overline{k}_\alpha(y)\right|
=o_{\mathrm{Pr}}(1),
\]
and the same holds for $\widehat{\underline{k}}_{\alpha}$. Importantly, it implies that
$$
\operatorname{Pr}\left(\widehat{\overline{k}}_\alpha=\overline{k}_\alpha \text { and } \widehat{\underline{k}}_\alpha=\underline{k}_\alpha \text { on } \mathcal{J}\right) \rightarrow 1 .
$$

\medskip

\begin{lemma}[Uniform Consistency of $\widehat{\overline{k}}_{\alpha}$ and $\widehat{\underline{k}}_{\alpha}$]\label{LEMPEE}
Suppose that Assumptions~\ref{AE0}--\ref{AE0_TIES} hold. Then, for each $\alpha\in\mathcal{A}$,
\begin{align*}
	\Pr\left(\left\{
\max_{y\in\mathcal{J}}
\left|
\widehat{\overline{k}}_{\alpha}(y)-\overline{k}_{\alpha}(y)
\right|
=0
\right\}\right)\to 1,\qquad
\Pr\left(\left\{
\max_{y\in\mathcal{J}}
\left|
\widehat{\underline{k}}_{\alpha}(y)-\underline{k}_{\alpha}(y)
\right|
=0
\right\}\right)\to 1.
\end{align*}
\end{lemma}

\subsection{Proofs of Main Results}

\begin{proof}[Proof of Proposition~\ref{PROPLIMIT}]
Fix $\alpha\in\mathcal{A}$, and define
\[
\mathcal{E}_{N,\alpha}
:=
\left\{
\max_{y\in\mathcal{J}}
\left|
\widehat{\overline{k}}_{\alpha}(y)-\overline{k}_{\alpha}(y)
\right|
=0
\right\}
\cap
\left\{
\max_{y\in\mathcal{J}}
\left|
\widehat{\underline{k}}_{\alpha}(y)-\underline{k}_{\alpha}(y)
\right|
=0
\right\}.
\]
By Lemma \ref{LEMPEE},
\[
\Pr\!\left(\mathcal{E}_{N,\alpha}\right)\to 1.
\]
On $\mathcal{E}_{N,\alpha}$, since $\mathcal{J}$ is finite and $C_{i,10}\in\mathcal{J}$ for all $i=1,\ldots,N_{10}$, we have
\[
\widehat{\overline{k}}_{\alpha}(C_{i,10})
=
\overline{k}_{\alpha}(C_{i,10}),
\qquad
\widehat{\underline{k}}_{\alpha}(C_{i,10})
=
\underline{k}_{\alpha}(C_{i,10}),
\qquad i=1,\ldots,N_{10}.
\]
Hence,
\[
\widehat{L}_{11,\alpha}^{(0)}(y)
=
\widetilde{L}_{11,\alpha}^{(0)}(y),
\qquad
\widehat{U}_{11,\alpha}^{(0)}(y)
=
\widetilde{U}_{11,\alpha}^{(0)}(y),
\qquad y\in\mathcal{J},
\]
so
\[
\sqrt{N}
\begin{pmatrix}
\widehat{L}_{11,\alpha}^{(0)}
-
\widetilde{L}_{11,\alpha}^{(0)}
\\
\widehat{U}_{11,\alpha}^{(0)}
-
\widetilde{U}_{11,\alpha}^{(0)}
\end{pmatrix}
=o_{\Pr}(1)
\quad
\text{in }
\ell^{\infty}\!\left(\mathcal{J},\mathbb{R}^{2}\right).
\]
Moreover, Lemma \ref{BOUNDDIST} yields
\[
\sqrt{N}
\begin{pmatrix}
\widetilde{L}_{11,\alpha}^{(0)}
-
L_{11,\alpha}^{(0)}
\\
\widetilde{U}_{11,\alpha}^{(0)}
-
U_{11,\alpha}^{(0)}
\end{pmatrix}
\rightsquigarrow
\begin{pmatrix}
G_{L,11,\alpha}^{(0)}
\\
G_{U,11,\alpha}^{(0)}
\end{pmatrix}
\quad
\text{in }
\ell^{\infty}\!\left(\mathcal{J},\mathbb{R}^{2}\right).
\]
The result then follows by Slutsky's theorem.
\end{proof}

\begin{proof}[Proof of Proposition~\ref{PROPZGAMMA}]
Fix $\alpha\in\mathcal{A}$. For each $y\in\mathcal{J}$ and $z\in\mathcal{Z}_{10}$, define
the infeasible empirical distribution functions and their bootstrap counterparts by
\small\begin{align*}
\widetilde{\theta}_{L,\alpha}^{(0)}(y)
&:=
\widehat{F}_{\overline{k}_{\alpha}(C_{10})}(y)
:=
\frac{1}{N_{10}}
\sum_{i=1}^{N_{10}}
\mathbb{I}\!\left\{
\overline{k}_{\alpha}(C_{i,10}) \le y
\right\},
\\
\widetilde{\theta}_{L,\alpha}^{(0),*}(y)
&:=
\widehat{F}^{\,*}_{\overline{k}_{\alpha}(C_{10})}(y)
:=
\frac{1}{N_{10}}
\sum_{i=1}^{N_{10}}
\mathbb{I}\!\left\{
\overline{k}_{\alpha}(C_{i,10}^{*}) \le y
\right\},
\\
\widetilde{\theta}_{U,\alpha}^{(0)}(y,z)
&:=
\widehat{F}_{\underline{k}_{\alpha}(C_{10})\mid Z_{10}}(y\mid z)
:=
\frac{
\frac{1}{N_{10}}
\sum_{i=1}^{N_{10}}
\mathbb{I}\!\left\{
\underline{k}_{\alpha}(C_{i,10}) \le y
\right\}
\mathbb{I}\!\left\{
Z_{i,10}=z
\right\}
}{
\frac{1}{N_{10}}
\sum_{i=1}^{N_{10}}
\mathbb{I}\!\left\{
Z_{i,10}=z
\right\}},
\\
\widetilde{\theta}_{U,\alpha}^{(0),*}(y,z)
&:=
\widehat{F}^{\,*}_{\underline{k}_{\alpha}(C_{10})\mid Z_{10}}(y\mid z)
:=
\frac{
\frac{1}{N_{10}}
\sum_{i=1}^{N_{10}}
\mathbb{I}\!\left\{
\underline{k}_{\alpha}(C_{i,10}^{*}) \le y
\right\}
\mathbb{I}\!\left\{
Z_{i,10}^{*}=z
\right\}
}{
\frac{1}{N_{10}}
\sum_{i=1}^{N_{10}}
\mathbb{I}\!\left\{
Z_{i,10}^{*}=z
\right\}}.
\end{align*}\normalsize
Because $\mathcal{J}$ and $\mathcal{Z}_{10}$ are finite, the relevant indicator classes are finite and therefore $\Pr$-Donsker.  
Hence, by Theorem 3.6.1 in \citet{VanderVaart1996}, using the same argument as in the proof of Lemma \ref{LEMDIST}, and since $N_{10}/N \xrightarrow{\Pr} \pi_{10}>0$, it follows that
\[
\sqrt{N}
\left(
\widetilde{\theta}_{L,\alpha}^{(0),*}-\widetilde{\theta}_{L,\alpha}^{(0)},
\;
\widetilde{\theta}_{U,\alpha}^{(0),*}-\widetilde{\theta}_{U,\alpha}^{(0)}
\right)
\stackrel{d^{*}}{\rightsquigarrow}
\left(
G_{\overline{k}_{\alpha}(C_{10})},
\;
G_{\underline{k}_{\alpha}(C_{10})\mid Z_{10}}
\right)
\quad
\text{in }
\ell^{\infty}\!\left(\mathcal{J},\mathbb{R}\right)
\times
\ell^{\infty}\!\left(\mathcal{J}\times\mathcal{Z}_{10},\mathbb{R}\right)
\]
in probability.\footnote{For a bootstrap statistic $T_N^b$ and a random element $D$, we write
$
T_N^b \xrightarrow{d^*} D
$
in probability if, conditional on the sample with probability approaching one,
$T_N^b$ converges weakly to $D$ under the bootstrap measure $\operatorname{Pr}^{*}$.} 
Moreover, by Lemma \ref{LEMPEE}, with probability approaching one,
 $\widehat{\overline{k}}_{\alpha}(y)=\overline{k}_{\alpha}(y)$ and $\widehat{\underline{k}}_{\alpha}(y)=\underline{k}_{\alpha}(y)$ for all $y\in \mathcal{J}$. The feasible and infeasible empirical functions, both for the sample and the bootstrap draws, therefore coincide with probability approaching one. Hence, by the same argument as in Proposition~\ref{PROPLIMIT},
\[
\sqrt{N}
\left(
\widehat{\theta}_{L,\alpha}^{(0),*}-\widehat{\theta}_{L,\alpha}^{(0)},
\;
\widehat{\theta}_{U,\alpha}^{(0),*}-\widehat{\theta}_{U,\alpha}^{(0)}
\right)
\stackrel{d^{*}}{\rightsquigarrow}
\left(
G_{\overline{k}_{\alpha}(C_{10})},
\;
G_{\underline{k}_{\alpha}(C_{10})\mid Z_{10}}
\right)
\quad
\text{in }
\ell^{\infty}\!\left(\mathcal{J},\mathbb{R}\right)
\times
\ell^{\infty}\!\left(\mathcal{J}\times\mathcal{Z}_{10},\mathbb{R}\right)
\]
in probability. By Lemma \ref{BOUNDDIST}, $\boldsymbol{\phi}_{\alpha}^{(0)}$ is
Hadamard directionally differentiable at
$\left(\theta_{L,\alpha}^{(0)},\theta_{U,\alpha}^{(0)}\right)$ with
\[
\boldsymbol{\phi}^{(0)'}_{\alpha,\left(\theta_{L,\alpha}^{(0)},\theta_{U,\alpha}^{(0)}\right)}
\!\left(
G_{\overline{k}_{\alpha}(C_{10})},
G_{\underline{k}_{\alpha}(C_{10})\mid Z_{10}}
\right)
=
\begin{pmatrix}
G_{L,11,\alpha}^{(0)}
\\
G_{U,11,\alpha}^{(0)}
\end{pmatrix}.
\]
Hence, Theorem 3.1 in \citet{HL2018}, together with
$\epsilon_{N}\to0$ and $\sqrt{N}\epsilon_{N}\to\infty$, implies that
\[
\begin{pmatrix}
T_{1,\alpha}^{(0),*}
\\
T_{2,\alpha}^{(0),*}
\end{pmatrix}
=
\widehat{\boldsymbol{\phi}}_{\alpha}^{(0)'}\!\left(
\sqrt{N}\left(\widehat{\theta}_{L,\alpha}^{(0),*}-\widehat{\theta}_{L,\alpha}^{(0)}\right),
\sqrt{N}\left(\widehat{\theta}_{U,\alpha}^{(0),*}-\widehat{\theta}_{U,\alpha}^{(0)}\right)
\right)
\stackrel{d^{*}}{\rightsquigarrow}
\begin{pmatrix}
G_{L,11,\alpha}^{(0)}
\\
G_{U,11,\alpha}^{(0)}
\end{pmatrix}
\quad
\text{in }
\ell^{\infty}\!\left(\mathcal{J},\mathbb{R}^{2}\right)
\]
in probability.

Because $\sigma(y)>0$ for all $y\in\mathcal{J}\setminus\{J\}$, each term in $m$ defined in equation~(\ref{EQmfunction}) is a rescaled supremum functional. By \citet[Lemma~C5]{MP2020}, the supremum functional is Lipschitz continuous with respect to the sup-norm, and the maximum of finitely many Lipschitz functions is again Lipschitz. Hence, by the continuous mapping theorem for Lipschitz maps \citep[Proposition~10.7]{K2008},
\[
M^{(0),*}_{\alpha}
=
m\!\left(\begin{pmatrix}
T_{1,\alpha}^{(0),*}
\\
T_{2,\alpha}^{(0),*}
\end{pmatrix}
\right)
\stackrel{d^{*}}{\rightsquigarrow}
m\!\left(
\begin{pmatrix}
G_{L,11,\alpha}^{(0)}
\\
G_{U,11,\alpha}^{(0)}
\end{pmatrix}
\right)
=
M_{\alpha}^{(0)}
\]
in probability.
Since the distribution function of $M_{\alpha}^{(0)}$ is continuous and strictly increasing in a neighborhood of $z_{\alpha}^{(0)}\left(1-\gamma\right)$, Corollary~3.2 of \citet{FS2015} yields
\[
\widehat{z}_{\alpha}^{(0)}\left(1-\gamma\right)
\xrightarrow{\mathrm{Pr}}
z_{\alpha}^{(0)}\left(1-\gamma\right),
\]
where $z_{\alpha}^{(0)}\left(1-\gamma\right)$ is defined in Proposition~\ref{PROPZGAMMA}. Moreover, by Proposition~\ref{PROPLIMIT} and another application of the continuous mapping theorem,
\[
m\!\left(
\sqrt{N}
\begin{pmatrix}
\widehat{L}_{11,\alpha}^{(0)}-L_{11,\alpha}^{(0)}
\\
\widehat{U}_{11,\alpha}^{(0)}-U_{11,\alpha}^{(0)}
\end{pmatrix}
\right)
\rightsquigarrow
M_{\alpha}^{(0)}.
\]
Finally, observe that, since $L_{11, \alpha}^{(0)}(y), U_{11, \alpha}^{(0)}(y) \in[0,1]$ for all $y \in \mathcal{J}\setminus\{J\}$,
\small$$
\left[
L_{11,\alpha}^{(0)}(y),
U_{11,\alpha}^{(0)}(y)
\right]
\subseteq
\mathrm{CS}_{\alpha}^{(0)}(y;1-\gamma)
\text{ for all } y\in\mathcal{J}\setminus\{J\}
\iff
m\!\left(
\sqrt{N}
\begin{pmatrix}
\widehat{L}_{11,\alpha}^{(0)}-L_{11,\alpha}^{(0)}
\\
\widehat{U}_{11,\alpha}^{(0)}-U_{11,\alpha}^{(0)}
\end{pmatrix}
\right)
\le
\widehat{z}_{\alpha}^{(0)}\left(1-\gamma\right).
$$\normalsize
At $y=J$, the inclusion holds automatically because $L_{11,\alpha}^{(0)}(J)=U_{11,\alpha}^{(0)}(J)=1$ and $\mathrm{CS}_{\alpha}^{(0)}(J;1-\gamma)=\{1\}$. 
Combining these convergence results with the equivalence above, Slutsky’s theorem, and continuity of the distribution function of $M_{\alpha}^{(0)}$ at $z_{\alpha}^{(0)}\left(1-\gamma\right)$ yields
\small
\begin{align*}
\lim_{N\to\infty}
\mathrm{Pr}\!\left(
\left[
L_{11,\alpha}^{(0)}(y),
U_{11,\alpha}^{(0)}(y)
\right]
\subseteq
\mathrm{CS}_{\alpha}^{(0)}(y;1-\gamma)
\ \text{for all } y\in\mathcal{J}
\right)
=
\mathrm{Pr}\!\left(
M_{\alpha}^{(0)}\le z_{\alpha}^{(0)}\left(1-\gamma\right)
\right)
=
1-\gamma .
\end{align*}
\normalsize
\end{proof}

\begin{proof}[Proof of Corollary~\ref{COR:CIINFERENCE}]
Fix $\alpha\in\mathcal{A}$ and $\gamma\in\left(0,1\right)$. Let
\small$$A_{N, \alpha}:=\left\{\left[L_{11, \alpha}^{(1)}(y), U_{11}^{(1)}(y)\right] \subseteq \mathrm{CS}_\alpha^{(1)}(y ; 1-\gamma / 2) \text{ and } \left[L_{11, \alpha}^{(0)}(y), U_{11, \alpha}^{(0)}(y)\right] \subseteq \mathrm{CS}_\alpha^{(0)}(y ; 1-\gamma / 2)\  \forall y \in \mathcal{J}\right\}.$$\normalsize
By Proposition~\ref{PROPZGAMMA}, applied with $\gamma/2$ in place of $\gamma$,  and by the analogous result for the factual confidence band, we have
$$\liminf _{N \rightarrow \infty} \operatorname{Pr}\left(A_{N, \alpha}\right) \geq 1-\gamma$$
by Bonferroni. On the event $A_{N,\alpha}$, Theorem~1 of \citet{CFMW2020}, applied separately to the lower and upper CDF bound functions, implies that for all $\tau\in(0,1)$,
$$\left[U_{11}^{(1),-1}(\tau), L_{11, \alpha}^{(1),-1}(\tau)\right] \subseteq \mathrm{CS}_\alpha^{(1),-1}(\tau ; 1-\gamma / 2),\quad \left[U_{11, \alpha}^{(0),-1}(\tau), L_{11, \alpha}^{(0),-1}(\tau)\right] \subseteq \mathrm{CS}_\alpha^{(0),-1}(\tau ; 1-\gamma / 2) .$$
Since the pointwise Minkowski difference is monotone under set inclusion,\footnote{For nonempty subsets $A_1,A_2,B_1,B_2\subseteq \mathbb{R}$ with $A_1\subseteq A_2$ and $B_1\subseteq B_2$, 
we have $
A_1\ominus B_1 \subseteq A_2\ominus B_2.
$}
it follows that, for all $\tau\in(0,1)$,
\[
\left[L_{\mathrm{QTT},\alpha}(\tau),U_{\mathrm{QTT},\alpha}(\tau)\right]
=
\left[U_{11}^{(1),-1}(\tau),L_{11,\alpha}^{(1),-1}(\tau)\right]
\ominus
\left[U_{11,\alpha}^{(0),-1}(\tau),L_{11,\alpha}^{(0),-1}(\tau)\right]
\]
is contained in
\[
\mathrm{CS}_{\alpha}^{(1),-1}(\tau;1-\gamma/2)
\ominus
\mathrm{CS}_{\alpha}^{(0),-1}(\tau;1-\gamma/2)
=
\mathrm{CS}_{\mathrm{QTT},\alpha}(\tau;1-\gamma).
\]
Hence $A_{N,\alpha}$ implies the desired QTT coverage event, and the result follows.\end{proof}

\subsection{Proofs of Auxiliary Results}

\begin{proof}[Proof of Lemma \ref{LEMDIST}]
Fix $\left(g,t\right)\in\{0,1\}\times\{0,1\}$. 
For each $y\in\mathcal{J}$ and $\left(y',z\right)\in\mathcal{J}\times\mathcal{Z}_{gt}$ define
\small\begin{align*}
\widehat{p}_{1,y}
&:=
\frac{1}{N}\sum_{i=1}^N
\mathbb{I}\!\left\{G_i=g,T_i=t\right\}
\mathbb{I}\!\left\{\psi_1(C_i)\le y\right\},
\qquad \widehat{p}_{2,y',z}
:=
\frac{1}{N}\sum_{i=1}^N
\mathbb{I}\!\left\{G_i=g,T_i=t\right\}
\mathbb{I}\!\left\{\psi_2(C_i)\le y'\right\}
\mathbb{I}\!\left\{Z_i=z\right\},\\
\widehat{q}_z&:=\frac{1}{N} \sum_{i=1}^N \mathbb{I}\left\{G_i=g, T_i=t\right\} \mathbb{I}\left\{Z_i=z\right\},\qquad
\widehat{\pi}_{g t}:=\frac{1}{N} \sum_{i=1}^N \mathbb{I}\left\{G_i=g, T_i=t\right\}, \\
p_{1,y}
&:= 
\pi_{gt}\,
F_{\psi_1(C_{gt})}(y),
\qquad
p_{2,y',z}
:= 
\pi_{gt}\,
\mathrm{Pr}\!\left(\psi_2(C_{gt})\le y',Z_{gt}=z\right),
\qquad q_z
:=
\pi_{gt}\mathrm{Pr}\!\left(Z_{gt}=z\right).
\end{align*}\normalsize
Collect these quantities into the vectors
\begin{align*}
\widehat{\theta}
:=&
\left(
\left(\widehat{p}_{1,y}\right)_{y\in\mathcal{J}},
\left(\widehat{p}_{2,y',z}\right)_{\left(y',z\right)\in\mathcal{J}\times\mathcal{Z}_{gt}},
\left(\widehat{q}_z\right)_{z\in\mathcal{Z}_{gt}},
\widehat{\pi}_{gt}
\right),\\
\theta
:=&
\left(
\left(p_{1,y}\right)_{y\in\mathcal{J}},
\left(p_{2,y',z}\right)_{\left(y',z\right)\in\mathcal{J}\times\mathcal{Z}_{gt}},
\left(q_z\right)_{z\in\mathcal{Z}_{gt}},
\pi_{gt}
\right),
\end{align*}
By the multivariate CLT, $\sqrt{N}\left(\widehat{\theta}-\theta\right)$ converges to a mean-zero Gaussian vector.   
Since $\pi_{gt}>0$ is maintained and, by Assumption~\ref{AE0A}, $q_z>0$ for all $z\in\mathcal{Z}_{gt}$, the map
$\varphi(\theta)=\left(\left(p_{1, y} / \pi_{gt}\right)_{y \in \mathcal{J}},\left(p_{2, y, z} / q_z\right)_{(y, z) \in \mathcal{J} \times \mathcal{Z}_{g t}}\right)$ is continuously differentiable at $\theta$. The delta method then yields the asserted joint weak convergence:
\small$$
\sqrt{N}
\left(
\widehat{F}_{\psi_1(C_{gt})}-F_{\psi_1(C_{gt})},
\widehat{F}_{\psi_2(C_{gt})\mid Z_{gt}}-F_{\psi_2(C_{gt})\mid Z_{gt}}
\right)
\rightsquigarrow
\left(
G_{\psi_1(C_{gt})},
G_{\psi_2(C_{gt})\mid Z_{gt}}
\right)
$$\normalsize
in $\ell^{\infty}\!\left(\mathcal{J},\mathbb{R}\right)
\times
\ell^{\infty}\!\left(\mathcal{J}\times\mathcal{Z}_{gt},\mathbb{R}\right)$, where $\left(G_{\psi_1(C_{gt})},G_{\psi_2(C_{gt})\mid Z_{gt}}\right)$ is a mean-zero Gaussian
process. 

To characterize the covariance structure of the limiting process, some standard algebra yields 
\small
\begin{align*}
	&\quad \ \sqrt{N}
\left(
\left(\widehat F_{\psi_1(C_{gt})}(y)-F_{\psi_1(C_{gt})}(y)\right)_{y\in\mathcal J},
\;
\left(\widehat F_{\psi_2(C_{gt})\mid Z_{gt}}(y'\mid z)-F_{\psi_2(C_{gt})\mid Z_{gt}}(y'\mid z)\right)_{(y',z)\in\mathcal J\times\mathcal Z_{gt}}
\right)\\
&=
\frac{1}{\sqrt N}\sum_{i=1}^N
\Big(
\left(\operatorname{IF}_{1,y,i}\right)_{y\in\mathcal J},
\;
\left(\operatorname{IF}_{2,y',z,i}\right)_{(y',z)\in\mathcal J\times\mathcal Z_{gt}}
\Big)
+o_{\mathrm{Pr}}(1),
\end{align*}
\normalsize
where, for each $y\in\mathcal{J}$ and $\left(y',z\right)\in\mathcal{J}\times\mathcal{Z}_{gt}$,
\begin{align*}
\operatorname{IF}_{1,y,i}
&:=
\frac{1}{\pi_{gt}}\,
\mathbb{I}\!\left\{G_i=g,T_i=t\right\}
\left(
\mathbb{I}\!\left\{\psi_1(C_i)\le y\right\}
-
F_{\psi_1(C_{gt})}(y)
\right),
\\
\operatorname{IF}_{2,y',z,i}
&:=
\frac{1}{\pi_{gt}}\,
\frac{\mathbb{I}\!\left\{G_i=g,T_i=t\right\}\mathbb{I}\!\left\{Z_i=z\right\}}
{\Pr\!\left(Z_{gt}=z\right)}
\left(
\mathbb{I}\!\left\{\psi_2(C_i)\le y'\right\}
-
F_{\psi_2(C_{gt})\mid Z_{gt}}(y'\mid z)
\right).
\end{align*}
The covariance kernels follow by direct calculation of the influence-function covariances.
\end{proof}

\begin{proof}[Proof of Lemma \ref{BOUNDEQV}]
Fix $\alpha\in\mathcal{A}\subseteq [\alpha^*,1)$ and $y\in\mathcal{J}$.
We begin by observing that\begin{align}\label{BOUNDEQV_SUPP}
	\begin{aligned}
		\overline{k}_{\alpha}\left(y\right)
&=
L_{01,\alpha}^{-1}\!\left(U_{00}\left(y\right)\right)\in\operatorname{supp}\left(L_{01,\alpha}\right)\\
\underline{k}_{\alpha}\left(y\right)
&=U_{01}^{-1}\!\left(L_{00,\alpha}\left(y-1\right)\right)\in\operatorname{supp}\left(U_{01}\right).
	\end{aligned}
	\end{align}
To see the first inclusion, observe that since $\operatorname{supp}\left(L_{01, \alpha}\right) \subseteq\mathcal{J}$ is finite,  by definition of the generalized inverse,
\begin{align*}
L_{01, \alpha}^{-1}(q)=\min \left\{t \in \operatorname{supp}\left(L_{01, \alpha}\right) \mid L_{01, \alpha}(t) \geq q\right\} \in \operatorname{supp}\left(L_{01, \alpha}\right),\quad q\in [0,1].
\end{align*}
The second inclusion follows analogously by applying the same argument to $U_{01}^{-1}$.\medskip

\noindent\underline{Lower bound.}  If $y<\underline{y}_{L,01}$ or $y>\overline{y}_{L,01}$, then by definition $L_{11,\alpha}^{(0)}(y)=0$ in the former case and $L_{11,\alpha}^{(0)}(y)=1$ in the latter. Moreover, \eqref{BOUNDEQV_SUPP} implies $\mathrm{Pr}\!\left(\overline{k}_{\alpha}(C_{10})\le y\right)=0$ if $y<\underline{y}_{L,01}$ and $\mathrm{Pr}\!\left(\overline{k}_{\alpha}(C_{10})\le y\right)=1$ if $y>\overline{y}_{L,01}$. Hence, the desired representation holds.

Now, suppose $\underline{y}_{L,01}\le y\le \overline{y}_{L,01}$.
By the definitions of $L_{11,\alpha}^{(0)}$ and $L_{10,\alpha}$ with $\alpha<1$,
\footnotesize\begin{align*}
	L_{11,\alpha}^{(0)}\left(y\right)
&=
L_{10,\alpha}\!\left(U_{00}^{(-1)}\!\left(L_{01,\alpha}\left(y\right)\right)\right)
=
\frac{
\mathrm{Pr}\!\left(C_{10}\le U_{00}^{(-1)}\!\left(L_{01,\alpha}\left(y\right)\right)\right)-\min\left\{\mathrm{Pr}\!\left(C_{10}\le U_{00}^{(-1)}\!\left(L_{01,\alpha}\left(y\right)\right)\right),\alpha\right\}
}{
1-\min\left\{\mathrm{Pr}\!\left(C_{10}\le U_{00}^{(-1)}\!\left(L_{01,\alpha}\left(y\right)\right)\right),\alpha\right\}}.
\end{align*}\normalsize
Thus, for the desired representation, it suffices to show that
\begin{align}\label{BOUNDEQV_EQ1}
\mathrm{Pr}\!\left(\overline{k}_{\alpha}\!\left(C_{10}\right)\le y\right)
=
\mathrm{Pr}\!\left(C_{10}\le U_{00}^{(-1)}\!\left(L_{01,\alpha}(y)\right)\right).
\end{align}
Let $c\in\operatorname{supp}\left(C_{10}\right)$. By Assumption~\ref{AE0_TIES}(ii), $\operatorname{supp}\left(C_{10}\right)\subseteq \operatorname{supp}\left(U_{00}\right)$, so $c\in \operatorname{supp}\left(U_{00}\right)$. Then $U_{00}(c)\in\left(0,1\right]$. Applying Lemma~\ref{lem:inv_mon}(iv) with $F=L_{01,\alpha}$, $q=U_{00}(c)$, and $x=y$, together with Lemma~\ref{lem:inv_mon}(v) with $F=U_{00}$, $x=c$, and $q=L_{01,\alpha}(y)$, yields
\begin{align*}
\overline{k}_{\alpha}(c)\le y
&\iff
L_{01,\alpha}^{-1}\!\left(U_{00}(c)\right)\le y
\iff
U_{00}(c)\le L_{01,\alpha}(y)
\iff
c\le U_{00}^{(-1)}\!\left(L_{01,\alpha}(y)\right).
\end{align*}
Since $C_{10}\in\operatorname{supp}\left(C_{10}\right)$ a.s., \eqref{BOUNDEQV_EQ1} follows, hence the desired representation.

\noindent\underline{Upper bound.} 
If $y<\underline{y}_{U,01}$ or $y>\overline{y}_{U,01}$, then by definition $U_{11,\alpha}^{(0)}(y)=0$ in the former case and $U_{11,\alpha}^{(0)}(y)=1$ in the latter. Moreover, \eqref{BOUNDEQV_SUPP} implies $\mathrm{Pr}\!\left(\underline{k}_{\alpha}(C_{10})\le y\right)=0$ if $y<\underline{y}_{U,01}$ and $\mathrm{Pr}\!\left(\underline{k}_{\alpha}(C_{10})\le y\right)=1$ if $y>\overline{y}_{U,01}$, and consequently $\mathrm{Pr}\!\left(\underline{k}_{\alpha}(C_{10})\le y \mid Z_{10}=z\right)=0$ or $1$, respectively, for all $z\in\mathcal{Z}_{10}$. Hence, the desired representation holds.

Now, suppose $\underline{y}_{U,01}\le y\leq  \overline{y}_{U,01}$. 
By the definitions of $U_{11,\alpha}^{(0)}$ and $U_{10}$,
$$
U_{11,\alpha}^{(0)}\!\left(y\right)
=
U_{10}\!\left(L_{00,\alpha}^{-1}\!\left(U_{01}\!\left(y\right)\right)\right)
=
\inf_{z\in \mathcal{Z}_{10}}
\mathrm{Pr}\!\left(C_{10}\le L_{00,\alpha}^{-1}\!\left(U_{01}\!\left(y\right)\right)\mid Z_{10}=z\right).
$$
Thus, for the desired representation, it suffices to show that, for each $z\in \mathcal{Z}_{10}$,
\begin{align}\label{BOUNDEQV_EQ2}
\mathrm{Pr}\!\left(\underline{k}_{\alpha}\!\left(C_{10}\right)\le y\mid Z_{10}=z\right)
=
\mathrm{Pr}\!\left(C_{10}\le L_{00,\alpha}^{-1}\!\left(U_{01}\!\left(y\right)\right)\mid Z_{10}=z\right).
\end{align}
Since $C_{10}\in\operatorname{supp}\left(C_{10}\right)$ a.s., it suffices to show that for each $c\in\operatorname{supp}\left(C_{10}\right)$,
\small\begin{align*}
\underline{k}_{\alpha}\!\left(c\right)\le y
\iff
U_{01}^{-1}\!\left(L_{00,\alpha}\left(c-1\right)\right)\le y
\iff
L_{00,\alpha}\left(c-1\right)\le U_{01}\left(y\right)
\iff
c\le L_{00,\alpha}^{-1}\!\left(U_{01}\!\left(y\right)\right).
\end{align*}\normalsize

The first equivalence follows from the definition of $\underline{k}_{\alpha}$. 
For the second equivalence, note that if $L_{00,\alpha}(c-1)=0$, then both sides hold since $U_{01}^{-1}(0)=\underline{y}_{U,01}\le y$ and $0\le U_{01}(y)$. If $L_{00,\alpha}(c-1)\in(0,1]$, then Lemma~\ref{lem:inv_mon}(iv), applied with $F=U_{01}$, $q=L_{00,\alpha}(c-1)$, and $x=y$, yields the equivalence.

For the last equivalence, note that $0<U_{01}(y)\le 1$. If $U_{01}(y)=1$, then $c\le L_{00,\alpha}^{-1}(1)=L_{00,\alpha}^{-1}(U_{01}(y))$ and $L_{00,\alpha}(c-1)\le 1$ hold trivially. If $U_{01}(y)\in(0,1)$, then Assumption~\ref{AE0_TIES}(i) implies that $L_{00,\alpha}(c-1)\le U_{01}(y)$ is equivalent to $L_{00,\alpha}(c-1)<U_{01}(y)$. Applying Lemma~\ref{lem:inv_mon}(iv) in contrapositive form with $F=L_{00,\alpha}$, $q=U_{01}(y)$, and $x=c-1$, we obtain
\[
L_{00,\alpha}(c-1)<U_{01}(y)
\iff
L_{00,\alpha}^{-1}\!\left(U_{01}(y)\right)>c-1
\iff
L_{00,\alpha}^{-1}\!\left(U_{01}(y)\right)\ge c,
\]
where the last step uses that $L_{00,\alpha}^{-1}$ takes values in $\mathcal{J}$. This establishes \eqref{BOUNDEQV_EQ2}.
\end{proof}

\begin{proof}[Proof of Lemma \ref{BOUNDDIST}]
Let $\alpha\in\mathcal{A}\subseteq [\alpha^*,1)$. For each $y\in\mathcal{J}$ and $z\in\mathcal{Z}_{10}$, define
$$
\widetilde{\theta}_{L,\alpha}^{(0)}(y)
:=
\widehat{F}_{\overline{k}_{\alpha}(C_{10})}(y),
\qquad
\widetilde{\theta}_{U,\alpha}^{(0)}(y,z)
:=
\widehat{F}_{\underline{k}_{\alpha}(C_{10})\mid Z_{10}}(y\mid z),
$$
Moreover, by Lemma~\ref{BOUNDEQV} and the definition of
$\left(\theta_{L,\alpha}^{(0)},\theta_{U,\alpha}^{(0)}\right)$ in
equation~(\ref{eq:phi_def}), we have, for all $y\in\mathcal{J}$,
\begin{align*}
\begin{pmatrix}
L_{11,\alpha}^{(0)}(y)
\\
U_{11,\alpha}^{(0)}(y)
\end{pmatrix}
=
\left[
\boldsymbol{\phi}_{\alpha}^{(0)}\!\left(\theta_{L,\alpha}^{(0)},\theta_{U,\alpha}^{(0)}\right)
\right](y),
\qquad
\begin{pmatrix}
\widetilde{L}_{11,\alpha}^{(0)}(y)
\\
\widetilde{U}_{11,\alpha}^{(0)}(y)
\end{pmatrix}
=
\left[
\boldsymbol{\phi}_{\alpha}^{(0)}\!\left(\widetilde{\theta}_{L,\alpha}^{(0)},\widetilde{\theta}_{U,\alpha}^{(0)}\right)
\right](y).
\end{align*}

\medskip
\noindent\underline{Lower bound map.}
First, for any $p\in[0,1]$,
\begin{align*}
\frac{p-\min\left\{p,\alpha\right\}}{1-\min\left\{p,\alpha\right\}}
&=
\mathbb{I}\left\{p\le \alpha\right\}\frac{p-p}{1-p}
+
\mathbb{I}\left\{p>\alpha\right\}\frac{p-\alpha}{1-\alpha}
=
\frac{1}{1-\alpha}
\max\left\{p-\alpha,0\right\}.
\end{align*}
Hence, by equation~(\ref{eq:phi_defs}), the first component of $\boldsymbol{\phi}_{\alpha}^{(0)}$, denoted by $\phi_{1,\alpha}^{(0)}$, can be written as
\[
\left[\phi_{1,\alpha}^{(0)}(\theta)\right](y)
=
\frac{1}{1-\alpha}
\max\{\theta(y)-\alpha,0\}.
\]
For fixed $y$, the map
$\theta \mapsto \left[\phi_{1,\alpha}^{(0)}(\theta)\right](y)$
is a composition of
$\theta \mapsto \theta(y)$,
$u \mapsto u-\alpha$,
$v \mapsto \max\{v,0\}$,
and $w \mapsto w/(1-\alpha)$.
The first, second, and fourth maps are linear and hence Hadamard differentiable.
The map $v \mapsto \max\{v,0\}$, denoted by $m$, is Hadamard directionally differentiable with derivative
$m_{v_0}^{\prime}(\dot{v})=\dot{v}\left(\mathbb{I}\left\{0<v_0\right\} +\mathbb{I}\left\{v_0=0\right\}  \mathbb{I}\{0<\dot{v}\}\right)$, where $m_{\vartheta}^{\prime}(\nu)$ denotes its directional derivative at $\vartheta$ in direction $\nu$ (see, e.g., the directional derivative result in \citealp[p.~90]{MP2020}, applied to $\left(a_1, a_2\right)=(v, 0)$).

By the chain rule for Hadamard directionally differentiable maps
\citep[Lemma~C2]{MP2020},
$\phi_{1,\alpha}^{(0)}$ is Hadamard directionally differentiable at $\theta_{L,\alpha}^{(0)}$
tangentially to
$\ell^{\infty}\!\left(\mathcal{J},\mathbb{R}\right)$,
with derivative
\small\begin{align*}
\left[\phi_{1,\alpha,\theta_{L,\alpha}^{(0)}}^{(0)'}(h)\right](y)
&=
\frac{1}{1-\alpha}
m'_{\theta_{L,\alpha}^{(0)}(y)-\alpha}\!\left(h(y)\right)=
\frac{1}{1-\alpha}
h(y)
\left(
\mathbb{I}\left\{\theta_{L,\alpha}^{(0)}(y)>\alpha\right\}
+
\mathbb{I}\left\{\theta_{L,\alpha}^{(0)}(y)=\alpha\right\}
\mathbb{I}\left\{h(y)>0\right\}
\right),
\end{align*}\normalsize
for all $h \in \ell^{\infty}\left(\mathcal{J} , \mathbb{R}\right)$.

\noindent\underline{Upper bound map.}
Since $\mathcal{Z}_{10}$ is finite,
the infimum in equation~(\ref{eq:phi_defs}) is a finite minimum.
By \citet[Lemma~C4]{MP2020} and the identity
$\inf A = -\sup(-A)$,
the second component of $\boldsymbol{\phi}_{\alpha}^{(0)}$, denoted by $\phi_{2,\alpha}^{(0)}$, is Hadamard directionally differentiable at $\theta_{U,\alpha}^{(0)}$
tangentially to
$\ell^{\infty}\!\left(\mathcal{J}\times\mathcal{Z}_{10},\mathbb{R}\right)$,
with derivative
\[
\left[\phi_{2,\alpha,\theta_{U,\alpha}^{(0)}}^{(0)'}(h)\right](y)
=
\min_{z\in\Psi_{\alpha}^{(0)}(y)} h(y,z),
\]
for all $h \in \ell^{\infty}\left(\mathcal{J} \times \mathcal{Z}_{10}, \mathbb{R}\right)$, where $\Psi_{\alpha}^{(0)}(y)
:=
\operatorname{arg\,inf}_{z\in\mathcal{Z}_{10}}
\theta_{U,\alpha}^{(0)}(y,z)$.

\medskip
\noindent\underline{Combined map.} Since $\phi_{1,\alpha}^{(0)}$ and $\phi_{2,\alpha}^{(0)}$ are Hadamard directionally differentiable
at $\theta_{L,\alpha}^{(0)}$ and $\theta_{U,\alpha}^{(0)}$, respectively, the map
$\boldsymbol{\phi}_{\alpha}^{(0)}$ is Hadamard directionally differentiable at
$\left(\theta_{L,\alpha}^{(0)},\theta_{U,\alpha}^{(0)}\right)$ tangentially to 
\(
\ell^{\infty}\!\left(\mathcal{J},\mathbb{R}\right)
\times
\ell^{\infty}\!\left(\mathcal{J}\times\mathcal{Z}_{10},\mathbb{R}\right),
\)
with derivative
\[
\boldsymbol{\phi}^{(0)'}_{\alpha,\left(\theta_{L,\alpha}^{(0)},\theta_{U,\alpha}^{(0)}\right)}(h_1,h_2)
=\begin{pmatrix}
\phi_{1,\alpha,\theta_{L,\alpha}^{(0)}}^{(0)'}(h_1)\\
\phi_{2,\alpha,\theta_{U,\alpha}^{(0)}}^{(0)'}(h_2)
\end{pmatrix}.
\]
By Lemma~\ref{LEMDIST}, applied with $(g,t)=(1,0)$, $\psi_1=\overline{k}_{\alpha}$, and $\psi_2=\underline{k}_{\alpha}$,
\[
\sqrt{N}
\left(
\widetilde{\theta}_{L,\alpha}^{(0)}-\theta_{L,\alpha}^{(0)},
\;
\widetilde{\theta}_{U,\alpha}^{(0)}-\theta_{U,\alpha}^{(0)}
\right)
\rightsquigarrow
\left(
G_{\overline{k}_{\alpha}(C_{10})},
\;
G_{\underline{k}_{\alpha}(C_{10})\mid Z_{10}}
\right)
\quad
\text{in }
\ell^{\infty}\!\left(\mathcal{J},\mathbb{R}\right)
\times
\ell^{\infty}\!\left(\mathcal{J}\times\mathcal{Z}_{10},\mathbb{R}\right).
\]Applying the functional delta method for Hadamard directionally differentiable functions \citep[Theorem~2.1]{FS2019} yields
\begin{align*}
\sqrt{N}
\begin{pmatrix}
\widetilde{L}_{11,\alpha}^{(0)}-L_{11,\alpha}^{(0)}
\\
\widetilde{U}_{11,\alpha}^{(0)}-U_{11,\alpha}^{(0)}
\end{pmatrix}
&=
\sqrt{N}
\left(
\boldsymbol{\phi}_{\alpha}^{(0)}\!\left(\widetilde{\theta}_{L,\alpha}^{(0)},\widetilde{\theta}_{U,\alpha}^{(0)}\right)
-
\boldsymbol{\phi}_{\alpha}^{(0)}\!\left(\theta_{L,\alpha}^{(0)},\theta_{U,\alpha}^{(0)}\right)
\right)
\\
&\rightsquigarrow
\boldsymbol{\phi}^{(0)'}_{\alpha,\left(\theta_{L,\alpha}^{(0)},\theta_{U,\alpha}^{(0)}\right)}
\!\left(
G_{\overline{k}_{\alpha}(C_{10})},
G_{\underline{k}_{\alpha}(C_{10})\mid Z_{10}}
\right)
=
\begin{pmatrix}
G_{L,11,\alpha}^{(0)}
\\
G_{U,11,\alpha}^{(0)}
\end{pmatrix}\quad
\text{in }
\ell^{\infty}\!\left(\mathcal{J},\mathbb{R}^{2}\right).
\end{align*}
\end{proof}

\begin{proof}[Proof of Lemma \ref{LEMPEE}]
Let $\alpha\in\mathcal{A}\subseteq [\alpha^*,1)$. The proof parallels Lemma A.13 of \citet{AI2006}. We establish the first statement for $\widehat{\overline{k}}_{\alpha}$. The argument for the second statement is analogous after replacing $\left(L_{01,\alpha},U_{00}\right)$ with $\left(U_{01},L_{00,\alpha}\right)$ and adjusting $\nu_{\alpha}$ accordingly (with $y=0$ treated separately instead of $y=J$).

First, by Lemma~\ref{LEMDIST}, applied with $\left(g,t\right)=\left(0,1\right)$ and $\left(g,t\right)=\left(0,0\right)$ and with $\psi_1=\psi_2=\mathrm{id}$, together with the continuous mapping theorem applied to
$
f \mapsto
\frac{f-\min\{f,\alpha\}}{1-\min\{f,\alpha\}}$ and 
$f \mapsto \inf_{z\in\mathcal{Z}_{00}} f(\cdot,z),$ 
we obtain
\[
\max_{y\in\mathcal{J}}
\left|
\widehat{L}_{01,\alpha}\left(y\right)-L_{01,\alpha}\left(y\right)
\right|
\xrightarrow{\mathrm{Pr}}0,
\qquad
\max_{y\in\mathcal{J}}
\left|
\widehat{U}_{00}\left(y\right)-U_{00}\left(y\right)
\right|
\xrightarrow{\mathrm{Pr}}0.
\]

Define
$
\nu_{\alpha}
:=
\min_{\substack{y\in\left\{-1\right\}\cup\mathcal{J},\\ m\in\mathcal{J}\setminus\{J\}}}
\left|L_{01,\alpha}\left(y\right)-U_{00}\left(m\right)\right|.$
By Assumption~\ref{AE0_TIES}(ii), $\operatorname{supp}\left(U_{00}\right)=\mathcal{J}$, so $0<U_{00}(m)<1$ for $m<J$. Combined with Assumption~\ref{AE0_TIES}(i) and $L_{01,\alpha}(-1)=0$, this implies $\nu_\alpha>0$. Also let $\delta_\alpha:=1-L_{01,\alpha}\!\left(\overline{k}_{\alpha}(J)-1\right)>0$. Define the event
\[
\mathcal{E}_{N,\alpha}
:=
\left\{
\max_{y\in\mathcal{J}}
\left|
\widehat{L}_{01,\alpha}(y)-L_{01,\alpha}(y)
\right|
\le \min\left\{\frac{\nu_\alpha}{3},\frac{\delta_\alpha}{2}\right\}
\right\}
\cap
\left\{
\max_{y\in\mathcal{J}}
\left|
\widehat{U}_{00}(y)-U_{00}(y)
\right|
\le \frac{\nu_\alpha}{3}
\right\}.
\]
Then $\mathrm{Pr}\left(\mathcal{E}_{N,\alpha}\right)\to 1$. It suffices to show that on $\mathcal{E}_{N,\alpha}$,
\[
\widehat{\overline{k}}_{\alpha}(y)
=
\overline{k}_{\alpha}(y),
\qquad y\in\mathcal{J}.
\]
If this holds, then
$
\mathcal{E}_{N, \alpha} \subseteq\left\{\max _{y \in\mathcal{J}}\left|\widehat{\overline{k}}_\alpha(y)-\overline{k}_\alpha(y)\right|=0\right\} 
$, hence
$
\operatorname{Pr}\left(\max _{y \in\mathcal{J}}\left|\widehat{\overline{k}}_\alpha(y)-\overline{k}_\alpha(y)\right| = 0\right)  \geq \Pr(\mathcal{E}_{N,\alpha})\rightarrow 1 .
$

Suppose $y\in\mathcal{J}\setminus \{J\}$.    By definition of the generalized inverse,
$$L_{01, \alpha}\left(\overline{k}_\alpha(y)-1\right)<U_{00}(y) < L_{01, \alpha}\left(\overline{k}_\alpha(y)\right)
$$
where the second inequality is strict by Assumption~6(i). On $\mathcal{E}_{N,\alpha}$, using $L_{01, \alpha}\left(\overline{k}_\alpha(y)\right)-{U}_{00}\left(y\right)\ge \nu_{\alpha}$ and ${U}_{00}\left(y\right)-L_{01, \alpha}\left(\overline{k}_\alpha(y)-1\right)\ge \nu_{\alpha}$ by definition of $\nu_{\alpha}$, simple algebra yields
$$\widehat{L}_{01, \alpha}\left(\overline{k}_\alpha(y)-1\right)<\widehat{U}_{00}(y) < \widehat{L}_{01, \alpha}\left(\overline{k}_\alpha(y)\right),
$$
so $\widehat{L}_{01, \alpha}^{-1}\left(\widehat{U}_{00}(y)\right)=\overline{k}_\alpha(y)$, that is, $\widehat{\overline{k}}_\alpha(y)=\overline{k}_\alpha(y)$.

For $y=J$, since $U_{00}(J)=\widehat{U}_{00}(J)=1$, we have $\overline{k}_\alpha(J)= L_{01, \alpha}^{-1}(1)$ and $\widehat{\overline{k}}_\alpha(J)=\widehat{L}_{01, \alpha}^{-1}(1)$. On $\mathcal{E}_{N,\alpha}$, using $\max_{y\in\mathcal{J}}
\left|
\widehat{L}_{01,\alpha}(y)-L_{01,\alpha}(y)
\right|
\le \delta_\alpha/2
$, simple algebra yields $\widehat{L}_{01,\alpha}(\overline{k}_{\alpha}(J)-1)<1$. Moreover, $L_{01, \alpha}\left(\overline{k}_\alpha(J)\right)=1$ implies $F_{C_{01}}\left(\overline{k}_\alpha(J)\right)=1$, hence $\widehat{L}_{01, \alpha}\left(\overline{k}_\alpha(J)\right)=1$. Therefore, by definition of the generalized inverse,
 $\widehat{\overline{k}}_\alpha(J)=\widehat{L}_{01,\alpha}^{-1}(1)=\overline{k}_\alpha(J)$.
\end{proof}

\section{Additional Covariates}\label{sec:AddCov}

In this section, we briefly outline how the analysis extends to additional exogenous covariates. We focus on the case where the covariate vector $\mathbf{X}_{gt}$ has finite support in each $(g,t)$ cell. 

Let
$
\mathcal{X}
:=
\bigcap_{(g,t)\in\{0,1\}\times\{0,1\}}
\operatorname{supp}\!\left(\mathbf{X}_{gt}\right).
$
The extension proceeds by imposing Assumptions~\ref{AE0}--\ref{AE0_TIES} conditional on $\mathbf{X}=\mathbf{x}$ for each $\mathbf{x}\in\mathcal{X}$. In particular, we assume that, conditional on $\mathbf{X}=\mathbf{x}$, each group-time cell has positive probability, that $\operatorname{supp}(Z_{gt}\mid \mathbf{X}_{gt}=\mathbf{x})$ is finite, and that the regularity conditions in Assumption~\ref{AE0_TIES} continue to hold.

\begin{corollary}[Extension with Discrete Covariates]\label{PROPCOV}
Suppose the conditional extension described above holds. Then, for each $\mathbf{x}\in\mathcal{X}$, the conclusions of Propositions~\ref{PROPLIMIT}--\ref{PROPZGAMMA} and Corollary~\ref{COR:CIINFERENCE} continue to hold after replacing all unconditional objects by their conditional analogues given $\mathbf{X}=\mathbf{x}$.
\end{corollary}

\begin{proof}[Proof of Corollary \ref{PROPCOV}]
Fix $\mathbf{x}\in\mathcal{X}$. Conditional on $\mathbf{X}=\mathbf{x}$, the problem is identical to the unconditional setup, with cell-specific conditional distributions and finite support $\operatorname{supp}\!\left(Z_{gt}\mid \mathbf{X}_{gt}=\mathbf{x}\right)$. Because $\mathbf{X}$ has finite support and
\[
\Pr\!\left(G=g,T=t,\mathbf{X}=\mathbf{x}\right)>0
\]
for all $(g,t)\in\{0,1\}\times\{0,1\}$, the effective sample size in each $(g,t,\mathbf{x})$ cell is of order $N$. The arguments used in Sections~\ref{sec:proofbounds}--\ref{sec:proofestinf} therefore apply conditionally, with only notational changes. We omit the repeated details.
\end{proof}

\medskip

\begin{remark}[Continuous Covariates]
When $\mathbf{X}_{gt}$ or $Z_{gt}$ contains continuous components, conditioning on exact covariate cells is no longer practical. Two alternative approaches may be used in this case.

First, the covariates may be discretized on a finite grid, after which the cellwise estimators described above apply directly. While straightforward to implement, this approach may become impractical when the researcher includes many additional covariates.

As a more tractable alternative, one may instead estimate the conditional distribution functions in a first step using, for example, distribution regression \citep{CFM2013,CFMW2020}. This follows a common strategy in partially identified models, where identification is established nonparametrically while implementation proceeds via flexible parametric or semiparametric first-step estimators; see, e.g., \citet{MPZ2024}. Specifically, for each $y\in\mathcal{J}\setminus\{J\}$ and $(g,t)$ one may run a binary response regression with dependent variable $\mathbb{I}\{C_{gt}\le y\}$:
\[
\Pr\!\left(C_{gt}\le y\mid Z_{gt}=z,\mathbf{X}_{gt}=\mathbf{x}\right)
=
\Lambda\!\left(B(z,\mathbf{x})^{\prime}\theta_{gt}(y)\right),
\]
where $B(z,\mathbf{x})$ is a vector of transformations of $(z,\mathbf{x})$,  $\theta_{gt}(y)$ is a parameter vector, and $\Lambda$ is a known link function such as logit or probit. The fitted conditional distribution functions can then be plugged into the formulas for the bounds. Provided the first-step estimators satisfy suitable regularity conditions, the same functional delta method and bootstrap strategy as in Section~\ref{sec:estimation} should apply. A full theoretical analysis of this extension is left for future work.
\end{remark}

\section{Semiparametric Model}\label{sec:parametricmodel}

{While \citet{GKM2023} clarify the relationship between CiC and copula stability in continuous settings, their analysis does not directly extend to the discrete CiC model considered here. We establish an equivalence result that characterizes the discrete CiC model as a threshold-crossing representation for a latent outcome variable with suitable location and scale restrictions across groups and time. This characterization is used below to construct and estimate a semiparametric model for consumption and reporting decisions with covariates.

\begin{proposition}[Equivalence of Discrete CiC Models]\label{CIC_equiv_general} Suppose that
\(\mathrm{supp}\!\left(Y(0)\mid T=t\right)=\mathcal{J}\) for all \(t\in \{0,1\}\).
 Then Assumption~\ref{A2} holds with $\mathrm{Var}\left[U\mid G=g\right]<+\infty$ for all $g\in \{0,1\}$ if and only if there exists a real-valued random variable $V$  and real-valued parameters $\{\eta_{gt},\lambda_{gt},\kappa_{t,j}\}_{(g,t,j)\in \{0,1\}\times \{0,1\}\times \mathcal{J}\setminus \{J\}}$, with $\lambda_{gt}>0$ for all $(g,t)\in \{0,1\}\times \{0,1\}$, such that
 \begin{enumerate}[label=(\arabic*)]
	\item $V\mid G=g$ is continuously distributed  with $\mathrm{Var}\left[V\mid G=g\right]<+\infty$ for all $g\in \{0,1\}$,
	\item $V\perp T\mid G$,
	\item $\mathrm{supp}\left(\eta_{GT}+\lambda_{GT}V\mid G=1,T=0 \right)\subseteq \mathrm{supp}\left(\eta_{GT}+\lambda_{GT}V \mid G=0,T=0\right)$,
	\item $\displaystyle\eta_{11}-\frac{\lambda_{11}}{\lambda_{10}} \eta_{10}=\eta_{01}-\frac{\lambda_{01}}{\lambda_{00}} \eta_{00}$ and $\displaystyle \frac{\lambda_{11}}{\lambda_{10}}=\frac{\lambda_{01}}{\lambda_{00}},$
\item $\kappa_{t,0}=0$, and if $J\geq 2$, $\kappa_{t,1}=1$
      and $\kappa_{t,j-1}<\kappa_{t,j}$ for all $(t, j) \in\{0,1\} \times\{1, \ldots, J-1\}$,
\end{enumerate}
and
$$Y(0)
=\begin{cases}
0, & \text{if } \eta_{GT}+\lambda_{GT}V \le \kappa_{T,0},\\
j, & \text{if } \kappa_{T,j-1} < \eta_{GT}+\lambda_{GT}V \le \kappa_{T,j},\ j=1,\dots,J-1,\\
J, & \text{if } \kappa_{T,J-1} < \eta_{GT}+\lambda_{GT}V,
\end{cases}
\quad\text{a.s.},
$$where $\eta_{GT}:=\sum_{(g,t)} \eta_{gt}\mathbb{I}\{G=g,T=t\}$ and similarly for $\lambda_{GT}$, and $\kappa_{T,j}:=\sum_{t} \kappa_{t,j}\mathbb{I}\{T=t\}$.
\end{proposition}
Proposition \ref{CIC_equiv_general} highlights that the discrete CiC model can be transformed into an alternative threshold-crossing model without affecting \(Y(0)\) almost surely. We additionally impose $\mathrm{Var}\!\left[U\mid G=g\right]<+\infty$, a mild technical condition ensuring that the equivalent threshold-crossing representation admits a well-defined scale parameter, while $\mathrm{supp}\left(Y\left(0\right)\mid T=t\right)=\mathcal{J}$ guarantees that the associated thresholds are well defined and strictly ordered. Note that these extra conditions are only required for the equivalence result of Proposition~\ref{CIC_equiv_general}, and not used in any other part of the paper.

Here, $\eta_{gt}$ may be interpreted as the location (e.g., mean) parameter of a corresponding latent variable, say $Y^{\ast}(0):=\eta_{GT}+\lambda_{GT}V$, for group $G=g$ and period $T=t$, while $\lambda_{gt}$ captures the corresponding scale. The discrete outcome \(Y(0)\in\mathcal{J}\) is generated by comparing this latent index to time-specific thresholds \(\kappa_{t,j}\). Part (4) of Proposition~\ref{CIC_equiv_general} imposes structured restrictions on these parameters. The location parameters \(\eta_{gt}\) satisfy a weighted ``parallel trends''-type assumption, while the scale parameters \(\lambda_{gt}\) satisfy a proportional ``parallel trends''-type assumption. Specifically, the change in the latent mean for the treated group, $\eta_{11}-\eta_{10}$, is linked to the corresponding change for the control group, $\eta_{01}-\eta_{00}$, after adjustment by the scaling parameters, which act as tailored weighting factors. The weights adjust for components of common time trends that cannot be removed by simple within-group mean differencing.
 If time trends affect exclusively the location parameters, so that \(\lambda_{g0}=\lambda_{g1}\) for all \(g\), these restrictions reduce to the conventional parallel trends assumption and  nest standard ordered Logit or Probit DiD specifications.
}

\subsection{Point Identification}\label{ssec:point}

In this subsection, we impose additional structure and consider a semiparametric model that jointly captures the consumption and reporting decisions. Specifically, we extend the threshold-crossing representation of the discrete CiC model in Proposition~\ref{CIC_equiv_general} by allowing the latent index to depend on observed covariates, and proceed analogously for the reporting decision. While this framework is more restrictive than the partial-identification approach, it yields point identification of the counterfactual distribution and, as shown below, of treatment parameters related to both consumption and reporting behavior.

To formalize reporting behavior, let \(R(d)\), \(d \in \{0,1\}\), denote the potential \emph{reporting intention}, taking values in the same finite set \(\mathcal J = \{0,1,\ldots,J\}\). We assume \(J \ge 2\) throughout this section.\footnote{In the binary case \(J = 1\), the consumption and reporting equations reduce to single-index threshold models that are invariant to positive rescalings of the latent index. As a result, the location and scale parameters are not separately identified. Propositions~\ref{PROP3}--\ref{PROP7} therefore exclude \(J = 1\).} We interpret $R(d)$ as the highest consumption level that an individual is willing to report truthfully under treatment state $d$. The reported potential outcome is then
\begin{equation}\label{EQOBSERVABILITY2}
C(d)=\min\{Y(d),R(d)\}.
\end{equation}
Thus, respondents report their true consumption $Y(d)$ if and only if $Y(d)\le R(d)$. For instance, if $Y(0)=2$ and $R(0)=1$, then in the absence of treatment the respondent reports consumption at level $1$ even though the true level is $2$. This formulation nests the familiar ``false-zero'' case \citep[see, e.g.,][]{BHSZ2018,GHSZ2018,NDT2019}, but is more flexible because it allows underreporting to any lower outcome level. The definition in \eqref{EQOBSERVABILITY2} also bears similarities to the binary selection framework in \citet{KSV2015}. Identification is more challenging here, however, because the reporting decision is not separately observed. Unlike standard selection models with an observed selection indicator, the reporting intention $R(d)$ can only be partially inferred from the joint realization of consumption and reporting behavior \citep[see also][]{P1980}. Finally, reporting behavior is allowed to respond to the treatment status $d$. In our empirical application, legalization may alter the perceived stigma associated with consumption and thereby affect individuals' willingness to report truthfully.

In the spirit of Proposition~\ref{CIC_equiv_general}, we next specify a semiparametric threshold-crossing model for consumption and reporting intention in the absence of treatment. Let \(\mathbf X_{gt}\) and \(\mathbf Z_{gt}\) denote observed covariate vectors entering the consumption and reporting equations, respectively. The two vectors may share common components. Exclusion restrictions will be imposed below to identify the consumption and reporting equations separately. Write \(\mathcal X_{gt}:=\operatorname{supp}(\mathbf X_{gt})\), \(\mathcal Z_{gt}:=\operatorname{supp}(\mathbf Z_{gt})\), and define the augmented covariates \(\overline{\mathbf X}_{gt}:=(1,\mathbf X_{gt}')'\) and \(\overline{\mathbf Z}_{gt}:=(1,\mathbf Z_{gt}')'\), with corresponding supports \(\overline{\mathcal X}_{gt}:=\operatorname{supp}(\overline{\mathbf X}_{gt})\) and \(\overline{\mathcal Z}_{gt}:=\operatorname{supp}(\overline{\mathbf Z}_{gt})\). 

\begin{assumption}[Semiparametric Threshold-Crossing Model]\label{S1}
For each \((g,t)\in\{0,1\}\times\{0,1\}\), the potential outcomes satisfy
\begin{align*}
Y_{gt}(0)
&=
\begin{cases}
0, & \text{if } Y^*_{gt}(0)\le \kappa_{t,0}^{(0)},\\
j, & \text{if } \kappa_{t,j-1}^{(0)}<Y^*_{gt}(0)\le \kappa_{t,j}^{(0)},\quad j=1,\ldots,J-1,\\
J, & \text{if } \kappa_{t,J-1}^{(0)}<Y^*_{gt}(0),
\end{cases}\\
R_{gt}(0)
&=
\begin{cases}
0, & \text{if } R^*_{gt}(0)\le \iota_{t,0}^{(0)},\\
j, & \text{if } \iota_{t,j-1}^{(0)}<R^*_{gt}(0)\le \iota_{t,j}^{(0)},\quad j=1,\ldots,J-1,\\
J, & \text{if } \iota_{t,J-1}^{(0)}<R^*_{gt}(0),
\end{cases}
\end{align*}
where
\begin{align*}
Y^*_{gt}(0)
&=
\overline{\mathbf X}_{gt}'\overline{\boldsymbol{\eta}}_{gt}^{(0)}
+\lambda_{gt}^{(0)}\varepsilon_{gt}^{(0)},\\
R^*_{gt}(0)
&=
\overline{\mathbf Z}_{gt}'\overline{\boldsymbol{\pi}}_{gt}^{(0)}
+\zeta_{gt}^{(0)}\nu_{gt}^{(0)},
\end{align*}
with \(\lambda_{gt}^{(0)}>0\) and \(\zeta_{gt}^{(0)}>0\). The thresholds satisfy \(\kappa_{t,0}^{(0)}=0\), \(\kappa_{t,1}^{(0)}=1\), and \(\kappa_{t,j-1}^{(0)}<\kappa_{t,j}^{(0)}\) for \(j=1,\ldots,J-1\), and analogously \(\iota_{t,0}^{(0)}=0\), \(\iota_{t,1}^{(0)}=1\), and \(\iota_{t,j-1}^{(0)}<\iota_{t,j}^{(0)}\) for \(j=1,\ldots,J-1\).
\end{assumption}

Assumption~\ref{S1} specifies a semiparametric model for the latent outcomes $Y^*_{gt}(0)$ and $R^*_{gt}(0)$, respectively. The additive structure is used in the identification argument of Proposition~\ref{PROP3} below, but we conjecture that weaker, non-additive  structures could also be adopted at the cost of stronger continuity restrictions on the observable covariates and suitable monotonicity restrictions on the corresponding non-additive function \citep[see, e.g.][]{Matzkin2003}.

For later use, decompose $\overline{\boldsymbol{\eta}}_{gt}^{(0)}
=
\left(\eta_{c,gt}^{(0)},\eta_{1,gt}^{(0)},\boldsymbol{\eta}_{-1,gt}^{(0)\prime}\right)'$, $\overline{\boldsymbol{\pi}}_{gt}^{(0)}
=
\left(\pi_{c,gt}^{(0)},\pi_{1,gt}^{(0)},\boldsymbol{\pi}_{-1,gt}^{(0)\prime}\right)'$, $
\overline{\mathbf X}_{gt}
=
\left(1,X_{1,gt},\mathbf X_{-1,gt}'\right)'$ and $\overline{\mathbf Z}_{gt}
=
\left(1,Z_{1,gt},\mathbf Z_{-1,gt}'\right)'.$

We now impose the semiparametric analogue of the restrictions on the latent location and scale parameters across groups and time in the threshold-crossing representation of the discrete CiC model in Proposition~\ref{CIC_equiv_general}.

\begin{assumption}[CiC-type Parallel Trends]\label{S2}
Let \(\overline{\mathcal X}:=\bigcap_{(g,t)\in\{0,1\}\times\{0,1\}}\overline{\mathcal X}_{gt}\) and \(\overline{\mathcal Z}:=\bigcap_{(g,t)\in\{0,1\}\times\{0,1\}}\overline{\mathcal Z}_{gt}\).
\begin{enumerate}[label=(\roman*)]
\item For every \(\overline{\mathbf x}\in\overline{\mathcal X}\),
\[
\overline{\mathbf x}'\overline{\boldsymbol{\eta}}_{11}^{(0)}
-
\frac{\lambda_{11}^{(0)}}{\lambda_{10}^{(0)}}
\overline{\mathbf x}'\overline{\boldsymbol{\eta}}_{10}^{(0)}
=
\overline{\mathbf x}'\overline{\boldsymbol{\eta}}_{01}^{(0)}
-
\frac{\lambda_{01}^{(0)}}{\lambda_{00}^{(0)}}
\overline{\mathbf x}'\overline{\boldsymbol{\eta}}_{00}^{(0)},
\qquad
\frac{\lambda_{11}^{(0)}}{\lambda_{10}^{(0)}}
=
\frac{\lambda_{01}^{(0)}}{\lambda_{00}^{(0)}}.
\]
\item For every \(\overline{\mathbf z}\in\overline{\mathcal Z}\),
\[
\overline{\mathbf z}'\overline{\boldsymbol{\pi}}_{11}^{(0)}
-
\frac{\zeta_{11}^{(0)}}{\zeta_{10}^{(0)}}
\overline{\mathbf z}'\overline{\boldsymbol{\pi}}_{10}^{(0)}
=
\overline{\mathbf z}'\overline{\boldsymbol{\pi}}_{01}^{(0)}
-
\frac{\zeta_{01}^{(0)}}{\zeta_{00}^{(0)}}
\overline{\mathbf z}'\overline{\boldsymbol{\pi}}_{00}^{(0)},
\qquad
\frac{\zeta_{11}^{(0)}}{\zeta_{10}^{(0)}}
=
\frac{\zeta_{01}^{(0)}}{\zeta_{00}^{(0)}}.
\]
\end{enumerate}
\end{assumption}

Assumption~\ref{S2} imposes, for each fixed covariate value, a weighted ``parallel trends''-type restriction on the latent location parameters and a proportional ``parallel trends''-type restriction on the latent scale parameters, separately for consumption and reporting intention.

The observation rule \eqref{EQOBSERVABILITY2}, together with the threshold-crossing specification and the dependence structure introduced below, implies that, for each \(j=1,\ldots,J\),
\footnotesize\begin{align}
\begin{aligned}
\Pr\!\left(C_{gt}(0)\ge j \mid \mathbf X_{gt},\mathbf Z_{gt}\right)
&=
\Pr\!\left(Y_{gt}(0)\ge j,\ R_{gt}(0)\ge j
\mid \mathbf X_{gt},\mathbf Z_{gt}\right)\\
&=
\Pr\!\left(
Y^*_{gt}(0)>\kappa_{t,j-1}^{(0)},\
R^*_{gt}(0)>\iota_{t,j-1}^{(0)}
\,\middle|\,
\mathbf X_{gt},\mathbf Z_{gt}
\right)\\
&=
\overline{C}_{gt}^{(0)}\!\left(
F_{-\varepsilon_{gt}^{(0)}}\!\left(
\frac{\overline{\mathbf X}_{gt}'\overline{\boldsymbol{\eta}}_{gt}^{(0)}
-\kappa_{t,j-1}^{(0)}}{\lambda_{gt}^{(0)}}
\right),
F_{-\nu_{gt}^{(0)}}\!\left(
\frac{\overline{\mathbf Z}_{gt}'\overline{\boldsymbol{\pi}}_{gt}^{(0)}
-\iota_{t,j-1}^{(0)}}{\zeta_{gt}^{(0)}}
\right)
\right),
\end{aligned}\label{EQUPPERTAIL}
\end{align}\normalsize
where \(F_{-\varepsilon_{gt}^{(0)}}\) and \(F_{-\nu_{gt}^{(0)}}\) denote the CDFs of \(-\varepsilon_{gt}^{(0)}\) and \(-\nu_{gt}^{(0)}\), respectively, and \(\overline{C}_{gt}^{(0)}\) denotes the copula of \(( -\varepsilon_{gt}^{(0)},-\nu_{gt}^{(0)})\). As in related bivariate latent-index models, the copula captures the dependence between the unobserved components of consumption and reporting \citep[cf.][]{CFL2025,HV2017}. The key difference here is that reporting intention is itself latent. Accordingly, the second margin in \eqref{EQUPPERTAIL} is not directly identified from observables, so standard results for models with fully observed binary outcomes do not apply directly.\footnote{In fact, since only the composite outcome $C(0) =
\min\{Y (0),R(0)\}$ is observed, even imposing a parametric copula does not by itself
restore the variation needed to separately disentangle latent consumption, latent reporting,
and their dependence.}

To identify the parameters in \eqref{EQUPPERTAIL}, we impose restrictions on the distributions of the latent unobservables, on their dependence structure, and on excluded variation. 

\begin{assumption}[Unobservables]\label{S3}
For each \((g,t)\in\{0,1\}\times\{0,1\}\):
\begin{enumerate}[label=(\roman*)]
\item
\((\varepsilon_{gt}^{(0)},\nu_{gt}^{(0)})\) is jointly independent of \((\mathbf X_{gt}',\mathbf Z_{gt}')'\). Their marginal CDFs \(F_{-\varepsilon_{gt}^{(0)}}\) and \(F_{-\nu_{gt}^{(0)}}\) are strictly increasing and real analytic, with zero median and unit variance. In addition, for each \(g\in\{0,1\}\),
\[
F_{-\varepsilon_{g0}^{(0)}}=F_{-\varepsilon_{g1}^{(0)}}
\qquad\text{and}\qquad
F_{-\nu_{g0}^{(0)}}=F_{-\nu_{g1}^{(0)}}.
\]
\item
The copula \(\overline{C}_{gt}^{(0)}\) in \eqref{EQUPPERTAIL} is strictly increasing and real analytic in each coordinate.
\end{enumerate}
\end{assumption}

\begin{assumption}[Exclusion Restriction and Covariates]\label{S4}
\begin{enumerate}[label=(\roman*)]
\item For each \((g,t)\in\{0,1\}\times\{0,1\}\):
\begin{enumerate}[label=(\alph*)]
\item \(X_{1,gt}\) is excluded from \(\mathbf Z_{gt}\).
\item \(Z_{1,gt}\) is excluded from \(\mathbf X_{gt}\).
\end{enumerate}
\item For each \((g,t)\neq(1,1)\), \(\eta_{1,gt}^{(0)}\neq 0\) and \(\pi_{1,gt}^{(0)}\neq 0\), and for each  \(\overline{\mathbf w}\in \operatorname{supp}\left(\overline{\mathbf W}_{gt}\right)\), there exists an open rectangle \(I_1\times I_2
\subseteq
\operatorname{supp}\left(
\left(X_{1,gt},Z_{1,gt}\right)
\mid
\overline{\mathbf W}_{gt}=\overline{\mathbf w}
\right)
\), where
$
\overline{\mathbf W}_{gt}
:=
\left(1,\mathbf X_{-1,gt}',\mathbf Z_{-1,gt}'\right)'.
$
\item
For each \((g,t)\in\{0,1\}\times\{0,1\}\), \(\overline{\mathcal X}_{gt}\) and \(\overline{\mathcal Z}_{gt}\) are not contained in proper linear subspaces of \(\mathbb R^{d_{X_{gt}}+1}\) and \(\mathbb R^{d_{Z_{gt}}+1}\), respectively, where \(d_{X_{gt}}\) and \(d_{Z_{gt}}\) denote the dimensions of \(\mathbf X_{gt}\) and \(\mathbf Z_{gt}\).
\end{enumerate}
\end{assumption}

Assumption~\ref{S3}(i) imposes covariate independence of the latent unobservables, while still allowing unrestricted dependence between \(\varepsilon_{gt}^{(0)}\) and \(\nu_{gt}^{(0)}\). This is important in the present context, since individuals with systematically high latent consumption may also be more or less likely to misreport than others. In that respect, our setup differs from the literature on binary choice models with two-sided misreporting, which typically imposes conditional independence between the unobservables governing the outcome and reporting decisions \citep[see][]{HASM1998,L2000}. Assumption~\ref{S3}(i) also imposes within-group time invariance of the marginal distributions of the unobservables. This is the semiparametric counterpart of the time-stability restriction underlying the discrete CiC representation in Proposition~\ref{CIC_equiv_general}. The zero-median and unit-variance conditions normalize the location and scale of \(\varepsilon_{gt}^{(0)}\) and \(\nu_{gt}^{(0)}\).\footnote{
As an alternative to the zero median, normalization based on zero mean could also be imposed. 
The proof adapts immediately.}

Assumption~\ref{S3}(ii) imposes a smoothness condition through real analyticity. Its role is to deliver identification by extrapolation: if two real analytic functions coincide on a neighborhood, then they coincide everywhere. In our setting, this replaces the large-support arguments commonly used in selection or triangular models \citep[e.g.,][]{C1986,Lewbel2007,Corradi2023}. Related extrapolation arguments also appear in microeconometrics and industrial organization \citep[see][]{AB2017,PF2023,IW2022}. Moreover, the real analytic property is satisfied by many standard copula families, including the Gaussian, Frank, Clayton, Plackett, Joe, and Gumbel families.

Assumption~\ref{S4} provides the excluded variation needed to isolate one latent index from the other. Parts (i)(a) and (i)(b) impose exclusion restrictions for the consumption and reporting equations, respectively. As will be shown, identification of each equation requires only one such exclusion. Part (ii) requires that the excluded variables enter their respective latent indices with nonzero coefficients and exhibit local continuous variation in their conditional support. Part (iii) is a standard nondegeneracy condition on the covariate support, ruling out collinearity.

We obtain the following identification result.

\begin{proposition}[Identification of Counterfactual Parameters]\label{PROP3}
Suppose Assumptions~\ref{S1}--\ref{S4} hold, and
\(
\mathcal X_{11}\subseteq \bigcap_{(g,t)\neq(1,1)}\mathcal X_{gt}
\)
and
\(
\mathcal Z_{11}\subseteq \bigcap_{(g,t)\neq(1,1)}\mathcal Z_{gt}.
\)
Then:
\begin{enumerate}[label=(\roman*)]
\item Under Assumption~\ref{S4}(i)(b), for each \((g,t)\in\{0,1\}\times\{0,1\}\), the parameters \(\overline{\boldsymbol{\eta}}_{gt}^{(0)}\), \(\lambda_{gt}^{(0)}\), \(\{\kappa_{t,j}^{(0)}\}_{j=2}^{J-1}\), and the marginal distribution \(F_{-\varepsilon_{gt}^{(0)}}\) are identified. In particular,
\[
\overline{\boldsymbol{\eta}}_{11}^{(0)}
=
\overline{\boldsymbol{\eta}}_{01}^{(0)}
+
\frac{\overline{\boldsymbol{\eta}}_{10}^{(0)}
-\overline{\boldsymbol{\eta}}_{00}^{(0)}}{\lambda_{00}^{(0)}/\lambda_{01}^{(0)}},
\qquad
\lambda_{11}^{(0)}
=
\frac{\lambda_{01}^{(0)}\lambda_{10}^{(0)}}{\lambda_{00}^{(0)}}.
\]
\item Under Assumption~\ref{S4}(i)(a), for each \((g,t)\in\{0,1\}\times\{0,1\}\), the parameters \(\overline{\boldsymbol{\pi}}_{gt}^{(0)}\), \(\zeta_{gt}^{(0)}\), \(\{\iota_{t,j}^{(0)}\}_{j=2}^{J-1}\), and the marginal distribution \(F_{-\nu_{gt}^{(0)}}\) are identified. In particular,
\[
\overline{\boldsymbol{\pi}}_{11}^{(0)}
=
\overline{\boldsymbol{\pi}}_{01}^{(0)}
+
\frac{\overline{\boldsymbol{\pi}}_{10}^{(0)}
-\overline{\boldsymbol{\pi}}_{00}^{(0)}}{\zeta_{00}^{(0)}/\zeta_{01}^{(0)}},
\qquad
\zeta_{11}^{(0)}
=
\frac{\zeta_{01}^{(0)}\zeta_{10}^{(0)}}{\zeta_{00}^{(0)}}.
\]
\end{enumerate}
\end{proposition}

To identify treatment effects on consumption and reporting intention in the treated group after treatment, we now specify the corresponding structure for the cell \((g,t)=(1,1)\) under treatment.

\begin{assumption}[Treated-Group Post-Treatment Structure]\label{S5}
For the treated group in the post-treatment period,
\begin{align*}
Y_{11}(1)
&=
\begin{cases}
0, & \text{if } Y^*_{11}(1)\le \kappa_{1,0}^{(1)},\\
j, & \text{if } \kappa_{1,j-1}^{(1)}<Y^*_{11}(1)\le \kappa_{1,j}^{(1)},\quad j=1,\ldots,J-1,\\
J, & \text{if } \kappa_{1,J-1}^{(1)}<Y^*_{11}(1),
\end{cases}\\
R_{11}(1)
&=
\begin{cases}
0, & \text{if } R^*_{11}(1)\le \iota_{1,0}^{(1)},\\
j, & \text{if } \iota_{1,j-1}^{(1)}<R^*_{11}(1)\le \iota_{1,j}^{(1)},\quad j=1,\ldots,J-1,\\
J, & \text{if } \iota_{1,J-1}^{(1)}<R^*_{11}(1),
\end{cases}
\end{align*}
where
\begin{align*}
	Y^*_{11}(1)
&=
\overline{\mathbf X}_{11}'\overline{\boldsymbol{\eta}}_{11}^{(1)}
+\lambda_{11}^{(1)}\varepsilon_{11}^{(1)},\\
R^*_{11}(1)
&=
\overline{\mathbf Z}_{11}'\overline{\boldsymbol{\pi}}_{11}^{(1)}
+\zeta_{11}^{(1)}\nu_{11}^{(1)}.
\end{align*}
with \(\lambda_{11}^{(1)}>0\) and \(\zeta_{11}^{(1)}>0\). The thresholds satisfy \(\kappa_{1,0}^{(1)}=0\), \(\kappa_{1,1}^{(1)}=1\), and \(\kappa_{1,j-1}^{(1)}<\kappa_{1,j}^{(1)}\) for \(j=1,\ldots,J-1\), and analogously \(\iota_{1,0}^{(1)}=0\), \(\iota_{1,1}^{(1)}=1\), and \(\iota_{1,j-1}^{(1)}<\iota_{1,j}^{(1)}\) for \(j=1,\ldots,J-1\).

In addition,
\begin{enumerate}[label=(\roman*)]
\item
\((\varepsilon_{11}^{(1)},\nu_{11}^{(1)})\) is jointly independent of \((\mathbf X_{11}',\mathbf Z_{11}')'\). Their marginal CDFs \(F_{-\varepsilon_{11}^{(1)}}\) and \(F_{-\nu_{11}^{(1)}}\) are strictly increasing and real analytic, with zero median and unit variance.
\item
The copula \(\overline{C}_{11}^{(1)}\) of \(( -\varepsilon_{11}^{(1)},-\nu_{11}^{(1)})\) is strictly increasing and real analytic in each coordinate.
\item \(\eta_{1,11}^{(1)}\neq 0\) and \(\pi_{1,11}^{(1)}\neq 0\), and for each \(\overline{\mathbf w}\in \operatorname{supp}\left(\overline{\mathbf W}_{11}\right)\), there exists an open rectangle \(I_1 \times I_2 \subseteq \operatorname{supp}\left(\left(X_{1,11}, Z_{1,11}\right) \mid \overline{\mathbf W}_{11}=\overline{\mathbf w}\right)\), where \(\overline{\mathbf W}_{11} := \left(1,\mathbf X_{-1,11}',\mathbf Z_{-1,11}'\right)'.\)
\end{enumerate}
\end{assumption}

Assumption~\ref{S5} is the treatment analogue of Assumptions~\ref{S1}, \ref{S3}, and \ref{S4} for the treated group in the post-treatment period.

For \(j\in\mathcal J\), \(\mathbf x\in\mathcal X_{11}\), and \(\mathbf z\in\mathcal Z_{11}\), define the conditional DTTs for consumption and reporting intention by
\begin{align}
\begin{aligned}
\tau(j\mid \mathbf{x})
&:=
\Pr\!\left(Y_{11}(1)=j\mid \mathbf X_{11}=\mathbf x\right)
-
\Pr\!\left(Y_{11}(0)=j\mid \mathbf X_{11}=\mathbf x\right),\\
\tau^{r}(j\mid \mathbf{z})
&:=
\Pr\!\left(R_{11}(1)=j\mid \mathbf Z_{11}=\mathbf z\right)
-
\Pr\!\left(R_{11}(0)=j\mid \mathbf Z_{11}=\mathbf z\right).
\end{aligned}
\label{EQCOND_DTT}
\end{align}
The corresponding marginalized versions of the conditional DTTs in \eqref{EQCOND_DTT} are
\[
\tau(j):=\int_{\mathcal X_{11}}\tau(j\mid \mathbf x)\,dF_{\mathbf X_{11}}(\mathbf x),
\qquad
\tau^{r}(j):=\int_{\mathcal Z_{11}}\tau^{r}(j\mid \mathbf z)\,dF_{\mathbf Z_{11}}(\mathbf z).
\]

We then obtain the following identification result.

\begin{proposition}[Identification of DTTs]\label{PROP4}
Suppose Assumptions~\ref{S1}--\ref{S5} hold, and
\(
\mathcal X_{11}\subseteq \bigcap_{(g,t)\neq(1,1)}\mathcal X_{gt}
\)
and
\(
\mathcal Z_{11}\subseteq \bigcap_{(g,t)\neq(1,1)}\mathcal Z_{gt}.
\)
Also, let \(\kappa_{1,-1}^{(d)}:=-\infty\), \(\kappa_{1,J}^{(d)}:=+\infty\), and \(\iota_{1,-1}^{(d)}:=-\infty\), \(\iota_{1,J}^{(d)}:=+\infty\) for \(d\in\{0,1\}\). Then:
\begin{enumerate}[label=(\roman*)]
\item Under Assumption~\ref{S4}(i)(b), the parameters \(\overline{\boldsymbol{\eta}}_{11}^{(1)}\), \(\lambda_{11}^{(1)}\), \(\{\kappa_{1,j}^{(1)}\}_{j=0}^{J-1}\), and the marginal distribution \(F_{-\varepsilon_{11}^{(1)}}\) are identified. Consequently, for each \(j\in\mathcal J\) and \(\mathbf x\in\mathcal X_{11}\), $\tau(j\mid \mathbf{x})$ and $\tau(j)$ are identified by
\begin{align*}
\tau(j\mid \mathbf{x})
=
&\left[
F_{-\varepsilon_{11}^{(1)}}\!\left(
\frac{\overline{\mathbf x}'\overline{\boldsymbol{\eta}}_{11}^{(1)}
-\kappa_{1,j-1}^{(1)}}
{\lambda_{11}^{(1)}}
\right)
-
F_{-\varepsilon_{11}^{(1)}}\!\left(
\frac{\overline{\mathbf x}'\overline{\boldsymbol{\eta}}_{11}^{(1)}
-\kappa_{1,j}^{(1)}}
{\lambda_{11}^{(1)}}
\right)
\right]\\
&-
\left[
F_{-\varepsilon_{11}^{(0)}}\!\left(
\frac{\overline{\mathbf x}'\overline{\boldsymbol{\eta}}_{11}^{(0)}
-\kappa_{1,j-1}^{(0)}}
{\lambda_{11}^{(0)}}
\right)
-
F_{-\varepsilon_{11}^{(0)}}\!\left(
\frac{\overline{\mathbf x}'\overline{\boldsymbol{\eta}}_{11}^{(0)}
-\kappa_{1,j}^{(0)}}
{\lambda_{11}^{(0)}}
\right)
\right]
\end{align*}
and $
\tau(j)
=
\int_{\mathcal X_{11}}\tau(j\mid \mathbf x)\,dF_{\mathbf X_{11}}(\mathbf x).
$
\item Under Assumption~\ref{S4}(i)(a), the parameters \(\overline{\boldsymbol{\pi}}_{11}^{(1)}\), \(\zeta_{11}^{(1)}\), \(\{\iota_{1,j}^{(1)}\}_{j=0}^{J-1}\), and the marginal distribution \(F_{-\nu_{11}^{(1)}}\) are identified. Consequently, for each \(j\in\mathcal J\) and \(\mathbf z\in\mathcal Z_{11}\), $\tau^{r}(j\mid \mathbf{z})$ and $\tau^{r}(j)$ are identified by
\begin{align*}
\tau^{r}(j\mid \mathbf{z})
=
&\left[
F_{-\nu_{11}^{(1)}}\!\left(
\frac{\overline{\mathbf z}'\overline{\boldsymbol{\pi}}_{11}^{(1)}
-\iota_{1,j-1}^{(1)}}
{\zeta_{11}^{(1)}}
\right)
-
F_{-\nu_{11}^{(1)}}\!\left(
\frac{\overline{\mathbf z}'\overline{\boldsymbol{\pi}}_{11}^{(1)}
-\iota_{1,j}^{(1)}}
{\zeta_{11}^{(1)}}
\right)
\right]\\
&-
\left[
F_{-\nu_{11}^{(0)}}\!\left(
\frac{\overline{\mathbf z}'\overline{\boldsymbol{\pi}}_{11}^{(0)}
-\iota_{1,j-1}^{(0)}}
{\zeta_{11}^{(0)}}
\right)
-
F_{-\nu_{11}^{(0)}}\!\left(
\frac{\overline{\mathbf z}'\overline{\boldsymbol{\pi}}_{11}^{(0)}
-\iota_{1,j}^{(0)}}
{\zeta_{11}^{(0)}}
\right)
\right]
\end{align*}
and
$
\tau^{r}(j)
=
\int_{\mathcal Z_{11}}\tau^{r}(j\mid \mathbf z)\,dF_{\mathbf Z_{11}}(\mathbf z).
$
\end{enumerate}
\end{proposition}

\subsection{Parametric Estimation}\label{ssec:estimation}

Propositions~\ref{PROP3}--\ref{PROP4} are nonconstructive because the identification argument relies on an ill-conditioned continuation of the marginal distribution and copula functions using the real analytic property. Instead, to regularize the problem, we impose a
parametric specification to obtain tractable estimators. That is, we strengthen the semiparametric model in
Section~\ref{ssec:point} by fixing the marginal distributions of the
latent unobservables and imposing a parametric copula. Specifically, for all
$(g,t)\in\{0,1\}\times\{0,1\}$ and $d=g\cdot t$, we set
\[
F_{-\varepsilon_{gt}^{(d)}}=\Phi,
\qquad
F_{-\nu_{gt}^{(d)}}=\Phi,
\]
where \(\Phi\) denotes the CDF of the standard
normal distribution, and assume that the copula \(\overline{C}_{gt}^{(d)}\)
belongs to a parametric family \(\overline{C}(\cdot,\cdot;\rho)\), where
\(\rho\in\mathcal R\) and \(\mathcal R\) denotes the natural open
parameter domain of the copula family, e.g., \(\mathcal R=(-1,1)\)
for the Gaussian copula, \(\mathcal R=(0,\infty)\) for the Clayton
copula, and \(\mathcal R=\mathbb R\setminus\{0\}\) for the Frank copula. To discriminate between these different parametric choices of the copula function, researchers may in practice apply the likelihood ratio test suggested by \citet{V1989}.\footnote{As pointed out by an anonymous referee, the same holds true for the choice of the marginal CDF  $\Phi$.}
Finally, note that more flexible parameterizations, allowing for heterogeneity in the marginal
distributions or in the copula across groups, time periods, or treatment
states, can be accommodated at the cost of additional parameters. 

For $J\geq 2$, $(g,t)\in\{0,1\}\times\{0,1\}$, and $d=g\cdot t$, let
\[
\boldsymbol{\vartheta}_{gt}^{(d)}
:=
\bigl(
\overline{\boldsymbol{\eta}}_{gt}^{(d)\prime},
\lambda_{gt}^{(d)},
\overline{\boldsymbol{\pi}}_{gt}^{(d)\prime},
\zeta_{gt}^{(d)}
\bigr)^{\prime},\quad \boldsymbol{\kappa}_{t}^{(d)}
:=
(\kappa_{t,2}^{(d)},\ldots,\kappa_{t,J-1}^{(d)})^{\prime},
\quad
\boldsymbol{\iota}_{t}^{(d)}
:=
(\iota_{t,2}^{(d)},\ldots,\iota_{t,J-1}^{(d)})^{\prime}.
\]
Collect all parameters in
\[
\boldsymbol{\theta}
:=
\bigl(
\boldsymbol{\vartheta}_{00}^{(0)\prime},
\boldsymbol{\vartheta}_{01}^{(0)\prime},
\boldsymbol{\vartheta}_{10}^{(0)\prime},
\boldsymbol{\vartheta}_{11}^{(1)\prime},
\boldsymbol{\kappa}_{0}^{(0)\prime},
\boldsymbol{\kappa}_{1}^{(0)\prime},
\boldsymbol{\kappa}_{1}^{(1)\prime},
\boldsymbol{\iota}_{0}^{(0)\prime},
\boldsymbol{\iota}_{1}^{(0)\prime},
\boldsymbol{\iota}_{1}^{(1)\prime},
\rho
\bigr)^{\prime}
\in \boldsymbol{\Theta},
\]
where \(\boldsymbol{\Theta}\) is a subset of the space of admissible parameters $ \{\boldsymbol{\theta}\in \mathbb{R}^{d_{\theta}}\mid \lambda_{g t}^{(d)}>0, \zeta_{g t}^{(d)}>0,\kappa_{t,j-1}^{(d)}<\kappa_{t,j}^{(d)},\iota_{t,j-1}^{(d)}<\iota_{t,j}^{(d)},\rho \in \mathcal{R}\}$, with \(d_{\theta}:=\dim(\boldsymbol{\theta})\). Let $\boldsymbol{\theta}_0 \in \boldsymbol{\Theta}$ denote the true parameter value.
For $\boldsymbol{\theta}\in\boldsymbol{\Theta}$, define
\begin{align*}
q_{gt,j}^{(d)}(\mathbf x,\mathbf z;\boldsymbol\theta)
:=
\overline C\!\left(
\Phi\!\left(
\frac{\overline{\mathbf x}'\overline{\boldsymbol\eta}_{gt}^{(d)}-\kappa_{t,j-1}^{(d)}}{\lambda_{gt}^{(d)}}
\right),
\Phi\!\left(
\frac{\overline{\mathbf z}'\overline{\boldsymbol\pi}_{gt}^{(d)}-\iota_{t,j-1}^{(d)}}{\zeta_{gt}^{(d)}}
\right);
\rho
\right),\quad j\in\{1,\ldots,J\},
\end{align*}
with
$
q_{gt,0}^{(d)}(\cdot,\cdot;\boldsymbol\theta):=1
$ and
$q_{gt,J+1}^{(d)}(\cdot,\cdot;\boldsymbol\theta):=0.
$
The corresponding probabilities are
\[
p_{gt,j}^{(d)}(\mathbf x,\mathbf z;\boldsymbol\theta)
:=
q_{gt,j}^{(d)}(\mathbf x,\mathbf z;\boldsymbol\theta)
-
q_{gt,j+1}^{(d)}(\mathbf x,\mathbf z;\boldsymbol\theta).
\]

Let the sample $\{(C_i,\mathbf X_i',\mathbf Z_i',G_i,T_i)'\}_{i=1}^{N}$ be i.i.d.\ draws from the population. As in the main text, we re-index these observations as $\{(C_{i,gt},\mathbf X_{i,gt}',\mathbf Z_{i,gt}')'\}_{i=1}^{N_{gt}}$. Also recall that we have maintained \(\pi_{gt}=\mathrm{Pr}\!\left(G=g,T=t\right)>0\) for all \((g,t)\). The conditional likelihood contribution of observation $i$,
given $(\mathbf X_i,\mathbf Z_i,G_i,T_i)$, is
\[
L_i(\boldsymbol\theta):=
\prod_{g\in\{0,1\}}\prod_{t\in\{0,1\}}\prod_{j\in\mathcal J}
\left[
p_{gt,j}^{(g\cdot t)}(\mathbf X_i,\mathbf Z_i;\boldsymbol\theta)
\right]^{\mathbb{I}\{G_i=g,T_i=t,C_i=j\}}.
\]
With the log-likelihood function $\ell_i(\boldsymbol\theta):=\log L_i(\boldsymbol\theta)$,
the maximum-likelihood estimator (MLE) is
\[
\widehat{\boldsymbol\theta}
:=
\arg\max_{\boldsymbol\theta\in\boldsymbol\Theta}
\frac{1}{N}\sum_{i=1}^{N}\ell_i(\boldsymbol\theta).
\]

Following Propositions~\ref{PROP3}--\ref{PROP4}, we construct plug-in estimators
for the counterfactual parameters and DTTs. 
To this end, define the untreated counterfactual parameters for the treatment group by
\begin{align*}
	\overline{\boldsymbol{\eta}}_{11}^{(0)}(\boldsymbol{\theta})
:=
\overline{\boldsymbol{\eta}}_{01}^{(0)}
+
\frac{\lambda_{01}^{(0)}}{\lambda_{00}^{(0)}}
\left(
\overline{\boldsymbol{\eta}}_{10}^{(0)}
-
\overline{\boldsymbol{\eta}}_{00}^{(0)}
\right),
\qquad
\lambda_{11}^{(0)}(\boldsymbol{\theta})
:=
\frac{\lambda_{01}^{(0)}\lambda_{10}^{(0)}}{\lambda_{00}^{(0)}},\\
\overline{\boldsymbol{\pi}}_{11}^{(0)}(\boldsymbol{\theta})
:=
\overline{\boldsymbol{\pi}}_{01}^{(0)}
+
\frac{\zeta_{01}^{(0)}}{\zeta_{00}^{(0)}}
\left(
\overline{\boldsymbol{\pi}}_{10}^{(0)}
-
\overline{\boldsymbol{\pi}}_{00}^{(0)}
\right),
\qquad
\zeta_{11}^{(0)}(\boldsymbol{\theta})
:=
\frac{\zeta_{01}^{(0)}\zeta_{10}^{(0)}}{\zeta_{00}^{(0)}}.
\end{align*}
For \(j\in\mathcal J\), \(\mathbf x\in\mathcal X_{11}\), and
\(\mathbf z\in\mathcal Z_{11}\), let
\small
\begin{align*}
\tau(j\mid \mathbf x;\boldsymbol{\theta})
:=
&\left[
\Phi\!\left(
\frac{
\overline{\mathbf x}'\overline{\boldsymbol\eta}_{11}^{(1)}
-
\kappa_{1,j-1}^{(1)}
}{
\lambda_{11}^{(1)}
}
\right)
-
\Phi\!\left(
\frac{
\overline{\mathbf x}'\overline{\boldsymbol\eta}_{11}^{(1)}
-
\kappa_{1,j}^{(1)}
}{
\lambda_{11}^{(1)}
}
\right)
\right] -
\left[
\Phi\!\left(
\frac{
\overline{\mathbf x}'\overline{\boldsymbol\eta}_{11}^{(0)}(\boldsymbol{\theta})
-
\kappa_{1,j-1}^{(0)}
}{
\lambda_{11}^{(0)}(\boldsymbol{\theta})
}
\right)
-
\Phi\!\left(
\frac{
\overline{\mathbf x}'\overline{\boldsymbol\eta}_{11}^{(0)}(\boldsymbol{\theta})
-
\kappa_{1,j}^{(0)}
}{
\lambda_{11}^{(0)}(\boldsymbol{\theta})
}
\right)
\right],
\\
\tau^{r}(j\mid \mathbf z;\boldsymbol{\theta})
:=
&\left[
\Phi\!\left(
\frac{
\overline{\mathbf z}'\overline{\boldsymbol\pi}_{11}^{(1)}
-
\iota_{1,j-1}^{(1)}
}{
\zeta_{11}^{(1)}
}
\right)
-
\Phi\!\left(
\frac{
\overline{\mathbf z}'\overline{\boldsymbol\pi}_{11}^{(1)}
-
\iota_{1,j}^{(1)}
}{
\zeta_{11}^{(1)}
}
\right)
\right] -
\left[
\Phi\!\left(
\frac{
\overline{\mathbf z}'\overline{\boldsymbol\pi}_{11}^{(0)}(\boldsymbol{\theta})
-
\iota_{1,j-1}^{(0)}
}{
\zeta_{11}^{(0)}(\boldsymbol{\theta})
}
\right)
-
\Phi\!\left(
\frac{
\overline{\mathbf z}'\overline{\boldsymbol\pi}_{11}^{(0)}(\boldsymbol{\theta})
-
\iota_{1,j}^{(0)}
}{
\zeta_{11}^{(0)}(\boldsymbol{\theta})
}
\right)
\right].
\end{align*}
\normalsize
Then, under correct specification,
\[
\tau(j\mid\mathbf x)=\tau(j\mid\mathbf x;\boldsymbol{\theta}_0),
\qquad
\tau^{r}(j\mid\mathbf z)=\tau^{r}(j\mid\mathbf z;\boldsymbol{\theta}_0),
\]
and the corresponding plug-in estimators are given by
\[
\widehat{\tau}(j\mid\mathbf x)
:=
\tau(j\mid\mathbf x;\widehat{\boldsymbol{\theta}}),
\qquad
\widehat{\tau}^{r}(j\mid\mathbf z)
:=
\tau^{r}(j\mid\mathbf z;\widehat{\boldsymbol{\theta}}).
\]
The marginalized DTT estimators are
\begin{align}
\widehat{\tau}(j)
&:=
\int_{\mathcal X_{11}}\widehat{\tau}(j\mid\mathbf x)\,
d\widehat{F}_{\mathbf X_{11}}(\mathbf x)
=
\frac{1}{N_{11}}\sum_{i=1}^{N}\mathbb{I}\{G_i=1,T_i=1\}
\widehat{\tau}(j\mid \mathbf X_i),\label{MARGINALC}\\
\widehat{\tau}^{r}(j)
&:=
\int_{\mathcal Z_{11}}\widehat{\tau}^{r}(j\mid\mathbf z)\,
d\widehat{F}_{\mathbf Z_{11}}(\mathbf z)
=
\frac{1}{N_{11}}\sum_{i=1}^{N}\mathbb{I}\{G_i=1,T_i=1\}
\widehat{\tau}^{r}(j\mid \mathbf Z_i),\label{MARGINALR}
\end{align}
where \(N_{11}:=\sum_{i=1}^{N}\mathbb{I}\{G_i=1,T_i=1\}\), and
\(\widehat{F}_{\mathbf X_{11}}\) and \(\widehat{F}_{\mathbf Z_{11}}\) denote the
empirical distribution functions \citep[see, e.g.,][]{CFM2013}.

We impose the following conditions:

\begin{assumption}[Sampling]\label{AE1}
$\{(C_i,\mathbf X_i',\mathbf Z_i',G_i,T_i)'\}_{i=1}^{N}$ are i.i.d. copies of $(C,\mathbf X',\mathbf Z',G,T)$.
\end{assumption}

\begin{assumption}[Correct Specification]\label{AE2}
$\boldsymbol{\Theta}$ is compact, $\boldsymbol{\theta}_{0}\in\operatorname{int}\left(\boldsymbol{\Theta}\right)$, and 
$
p_{gt,j}^{(d)}\left(\mathbf x,\mathbf z;\boldsymbol{\theta}_{0}\right)
=
\Pr\left(
C_i=j
\,\middle|\,
\mathbf X_i=\mathbf x,\mathbf Z_i=\mathbf z,G_i=g,T_i=t
\right)
$
for all $j\in\mathcal J$, 
$(g,t)\in\{0,1\}\times\{0,1\}$ with $d=g\cdot t$,  and $\left(\mathbf x,\mathbf z\right)\in
\operatorname{supp}\left(\left(\mathbf X_{gt},\mathbf Z_{gt}\right)\right)$.
\end{assumption}
%
%
%

\begin{assumption}[Regularity]\label{AE3} There exists a neighborhood $\mathcal N$ of $\boldsymbol{\theta}_0$ such that

\begin{enumerate}[label=(\roman*)]
\item $\mathrm E\left[
\sup_{\boldsymbol{\theta}\in\boldsymbol{\Theta}}
\left|
\ell_i\left(\boldsymbol{\theta}\right)
\right|
\right]
<
\infty$.
\item $\overline{C}(u, v ; \rho)$ is strictly increasing in $\rho$, and  twice continuously differentiable  in $(u, v, \rho)$ on $(0,1)^2 \times \mathcal{R}$.

\item $
\mathrm{E}\left[\sup_{\boldsymbol{\theta}\in\mathcal N}
\left\|
\nabla_{\boldsymbol{\theta}}
p_{gt,j}^{(g\cdot t)}\left(\mathbf X_{i,gt}
,\mathbf Z_{i,gt};\boldsymbol{\theta}\right)
\right\|
\right]<
\infty$ and $\mathrm{E}\left[\sup_{\boldsymbol{\theta}\in\mathcal N}
\left\|
\nabla_{\boldsymbol{\theta}\boldsymbol{\theta}}^2
p_{gt,j}^{(g\cdot t)}\left(\mathbf X_{i,gt}
,\mathbf Z_{i,gt};\boldsymbol{\theta}\right)
\right\|
\right]<
\infty$ for all $(g,t)\in \{0,1\}\times \{0,1\}$.

\item
$
\mathbf J_0
:=
\mathrm E\left[
\nabla_{\boldsymbol{\theta}}\ell_i\left(\boldsymbol{\theta}_{0}\right)
\nabla_{\boldsymbol{\theta}}\ell_i\left(\boldsymbol{\theta}_{0}\right)^\prime
\right]
$
exists and is nonsingular.

\item $
\mathrm E\left[
\sup_{\boldsymbol{\theta}\in\mathcal N}
\left\|
\nabla_{\boldsymbol{\theta}\boldsymbol{\theta}}^{2}
\ell_i\left(\boldsymbol{\theta}\right)
\right\|
\right]
<
\infty.
$
\end{enumerate}
\end{assumption}

Assumption~\ref{AE1} extends the i.i.d.\ sampling condition to the observed vector \(\left(C,\mathbf X',\mathbf Z',G,T\right)'\). Assumption~\ref{AE2} imposes correct specification by requiring that the model-implied conditional probabilities \(p_{gt,j}^{(d)}\left(\mathbf x,\mathbf z;\boldsymbol{\theta}_0\right)\) coincide with the corresponding conditional probabilities of the observed outcome in the population.

Assumption~\ref{AE3} collects a set of standard regularity conditions for maximum-likelihood estimation \citep{NM1994}. Assumption~\ref{AE3}(ii) adds to the identification assumptions a strict monotonicity and smoothness requirement with respect to the copula parameter \(\rho\). This condition is weak and is satisfied by many commonly used parametric copula families, including the Frank, Gaussian, and Clayton copulas. Assumptions~\ref{AE3}(i), \ref{AE3}(iii), and \ref{AE3}(v) are automatically satisfied if, for each \((g,t)\), the joint support of \(\left(\mathbf X_{gt},\mathbf Z_{gt}\right)\) is compact.

The following result gives the asymptotic distribution of the
estimators for the conditional and unconditional DTTs.

\begin{proposition}[Convergence of DTT Estimators]\label{PROP7}
Suppose the conditions of Proposition \ref{PROP4} and Assumptions~\ref{AE1}--\ref{AE3} hold.
\begin{enumerate}[label=(\roman*)]

\item For any \(j\in\mathcal J\) and \(\mathbf x\in\mathcal X_{11}\),
\[
\sqrt{N}\left(\widehat{\tau}\left(j\mid\mathbf x\right)-\tau\left(j\mid\mathbf x\right)\right)
\rightsquigarrow
N\left(
0,
\nabla_{\boldsymbol{\theta}}\tau\left(j\mid\mathbf x;\boldsymbol{\theta}_0\right)^\prime
\mathbf J_0^{-1}
\nabla_{\boldsymbol{\theta}}\tau\left(j\mid\mathbf x;\boldsymbol{\theta}_0\right)
\right),
\]
If, in addition,
$
\mathrm E\left[
\sup_{\boldsymbol{\theta}\in\mathcal N}
\left\|
\nabla_{\boldsymbol{\theta}}
\tau\left(j\mid \mathbf X_{11};\boldsymbol{\theta}\right)
\right\|
\right]
<\infty,
$ then
$$\sqrt{N}(\widehat{\tau}(j)-\tau(j)) \rightsquigarrow N\left(0, \overline{\nabla}_{\boldsymbol{\theta}} \tau_j^{\prime} \mathbf{J}_0^{-1} \overline{\nabla}_{\boldsymbol{\theta}} \tau_j+\frac{1}{\pi_{11}} \operatorname{Var}\left(\tau\left(j \mid \mathbf{X}_{11} ; \boldsymbol{\theta}_0\right)\right)\right),$$
where $\overline{\nabla}_{\boldsymbol{\theta}} \tau_j:=\int_{\mathcal{X}_{11}} \nabla_{\boldsymbol{\theta}} \tau\left(j \mid \mathbf{x} ; \boldsymbol{\theta}_0\right) d F_{\mathbf{X}_{11}}(\mathbf{x}).$

\item For any \(j\in\mathcal J\) and \(\mathbf z\in\mathcal Z_{11}\),
\[
\sqrt{N}\left(\widehat{\tau}^{r}\left(j\mid\mathbf z\right)-\tau^{r}\left(j\mid\mathbf z\right)\right)
\rightsquigarrow
N\left(
0,
\nabla_{\boldsymbol{\theta}}\tau^{r}\left(j\mid\mathbf z;\boldsymbol{\theta}_0\right)^\prime
\mathbf J_0^{-1}
\nabla_{\boldsymbol{\theta}}\tau^{r}\left(j\mid\mathbf z;\boldsymbol{\theta}_0\right)
\right).
\]
If, in addition,
$
\mathrm E\left[
\sup_{\boldsymbol{\theta}\in\mathcal N}
\left\|
\nabla_{\boldsymbol{\theta}}
\tau^r\left(j\mid \mathbf Z_{11};\boldsymbol{\theta}\right)
\right\|
\right]
<\infty,
$ then
$$\sqrt{N}(\widehat{\tau}^r(j)-\tau^r(j)) \rightsquigarrow N\left(0, \overline{\nabla}_{\boldsymbol{\theta}} \tau_j^{r\prime} \mathbf{J}_0^{-1} \overline{\nabla}_{\boldsymbol{\theta}} \tau^r_j+\frac{1}{\pi_{11}} \operatorname{Var}\left(\tau^r\left(j \mid \mathbf{Z}_{11} ; \boldsymbol{\theta}_0\right)\right)\right),$$
where $\overline{\nabla}_{\boldsymbol{\theta}} \tau_j^r:=\int_{\mathcal{Z}_{11}} \nabla_{\boldsymbol{\theta}} \tau^r\left(j \mid \mathbf{z} ; \boldsymbol{\theta}_0\right) d F_{\mathbf{Z}_{11}}(\mathbf{z}).$
\end{enumerate}
\end{proposition}

\subsection{Empirical Application}\label{sec:empirical_parametric}
In this section, we re-visit the empirical application from the main text using the parametric model from the previous section. As covariates,  we follow \citet{HWB2022}, and not only include individual-level covariates like gender, age, ethnicity, and parental education, but also certain state-level control variables such as unemployment rate and median income, fixed at pre-treatment levels.\footnote{State-level unemployment rates as well as the state median income were constructed from data of the Bureau of Labor Statistics (\texttt{https://www.bls.gov/}).} Moreover, we employ state-level marijuana prices from 2015 as an instrument for the consumption (\texttt{state\_price\_m}), and the ``survey cooperation'' measure (\texttt{trust}) for the reporting.\footnote{Marijuana price data source: \\\texttt{https://oxfordtreatment.com/substance-abuse/marijuana/average-cost-of-marijuana/} (last accessed: 12-10-2023).} Here, marijuana prices are presumed to exclusively influence consumption decisions. The plausibility of the exclusion restriction relies on the assumption that consumption behavior is sensitive to prices, while reporting behavior is not, given that we also control for other state-level variables. 

In Table \ref{Empirical1}, we present our initial findings for benchmark models without underreporting and without additional control variables. We compare three specifications: a binary-response Probit DiD (\texttt{pdid}), an ordered-response Probit DiD (\texttt{pdid\_ordered}), and a generalized DiD specification for ordered outcomes (\texttt{gdid}). Here, the two Probit DiD benchmarks impose the standard parallel-trends restriction on the latent index, whereas \texttt{gdid} allows for the more general CiC-type restrictions on latent location and scale, analogous to those in Proposition~\ref{PROP3}. Our first set of results for \texttt{gdid} suggests that $\tau(j)$ is not statistically different from zero for $j\in\{0,1,2\}$ at any conventional levels. These results align with the estimates from the more restrictive ordered and binary Probit DiD benchmarks.

\begin{table}
\begin{center}
  \caption{Unconditional DTTs (No Misreporting)} 
  \label{Empirical1}
\footnotesize
\begin{tabular}[t]{l|r|r|r|r|r|r}
\hline\hline
  & pdid & pdid (sd) & pdid\_ordered & pdid\_ordered (sd)& gdid & gdid (sd)\\
\hline
\hline
\hspace{1em}$\tau(0)$ & -0.0071 & 0.0056 & -0.0067 & 0.0056 & -0.0071  & 0.0056 \\
\hline
\hspace{1em}$\tau(1)$ & 0.0071  & 0.0056  & 0.0027  & 0.0045  & 0.0047  & 0.0040  \\
\hline
\hspace{1em}$\tau(2)$ &  NA     & NA      & 0.0039     & 0.0033      &  0.0025     & 0.0042      \\
\hline 
\hline  
\end{tabular}\end{center}

\footnotesize Notes: Data between 2015-2018 are used. The table contains the estimates of the DTTs $\tau(j)$, $j\in\{0,1,2\}$ for three different models estimated \textbf{without} additional controls. Specifically, the results are for the Probit DiD with binary outcomes (\texttt{pdid}), the Probit DiD with ordered outcomes (\texttt{pdid\_ordered}), and a generalized DiD specification for ordered outcomes (\texttt{gdid}). For the Probit DiD with binary outcomes, $\tau(1)$ indicates the effect for users with outcomes $\geq$ 1. Standard errors are computed via the Delta method and can be found in the corresponding columns next to the coefficient estimates.\normalsize
\end{table}

Next, we move to specifications without misreporting but including covariates to account for differences across groups in terms of observed characteristics. We present estimates of  $\tau(j)$ in Table \ref{Empirical2} (first panel). Given nearly identical variance parameters across groups and time in unreported parameter estimates for Table \ref{Empirical1}, we constrain $\lambda_{gt}$ to be constant in this specification.\footnote{Unreported results (available upon request) show that allowing for flexible variance parameters does not alter the estimates qualitatively.} Notably, the estimates of $\widehat{\tau}(j)$ now appear to be larger in absolute terms, with a significant decrease in non-consumption probability by around 2 percentage points, and significant increases in consumption probabilities by approximately 1 percentage point. Compared to the unconditional results, conditioning on additional regressors seems to amplify the effect sizes in absolute terms. 

In the final step, we estimate models with covariates that incorporate misreporting as described in Section~\ref{ssec:estimation} (see second panel of Table \ref{Empirical2}). As before, we only present estimates of the marginalized treatment effects $\tau(j)$ and $\tau^{r}(j)$ on consumption and misreporting, respectively, as defined in  (\ref{MARGINALC}) and (\ref{MARGINALR}), respectively. Unreported results for the slope coefficient estimates (available from the authors upon request) show that the slope estimates for the consumption decision closely mirror those from the specification without misreporting. Both of the instruments are found to be significant at the 5 or 10\% level across groups and time (except for the treatment group after treatment).\footnote{The positive and significant effects for marijuana prices align with other studies \citep[e.g.][]{GHSZ2018}, potentially reflecting drug quality and availability effects, in the absence of quality as a control variable.} Notably, we see more pronounced effect sizes for the DTTs after accounting for misreporting (increased by more than 50\% for all levels), with all $\widehat{\tau}(j)$ being significant at the 5\% level. No significant treatment effects are found on the reporting behavior. These findings are robust to replacing the Frank copula with the asymmetric Clayton copula.

\begin{table}
\begin{center}
  \caption{DTTs (Marginalized)} 
  \label{Empirical2}
\footnotesize
\begin{tabular}[t]{l|r|r}
\multicolumn{3}{c}{\textbf{Without Misreporting}}\\
\hline\hline
  & gdid & gdid (sd)\\\hline
\hline
\hspace{1em}$\tau(0)$ & -0.0200 & 0.0082\\
\hline
\hspace{1em}$\tau(1)$ & 0.0088 & 0.0036\\
\hline
\hspace{1em}$\tau(2)$ & 0.0113 & 0.0047\\
\hline
\hline\end{tabular}\\\bigskip

\begin{tabular}[t]{l|r|r|r|r}
\multicolumn{5}{c}{\textbf{With Misreporting}}\\
\hline\hline
  & gdid\_frank & gdid\_frank (sd) & gdid\_clayton & gdid\_clayton (sd)\\
\hline
\hline
\hspace{1em}$\tau(0)$ & -0.0327 & 0.0137 & -0.0327 & 0.0141\\
\hline
\hspace{1em}$\tau(1)$ & 0.0138 & 0.0057 & 0.0138 & 0.0058\\
\hline
\hspace{1em}$\tau(2)$ & 0.0189 & 0.0081 & 0.0189 & 0.0083\\
\hline
\hspace{1em}$\tau^{r}(0)$ & 0.1199 & 0.1213 & 0.1242 & 0.1442\\
\hline
\hspace{1em}$\tau^{r}(1)$ & 0.0125 & 0.0116 & 0.0179 & 0.0268\\
\hline
\hspace{1em}$\tau^{r}(2)$ & -0.1324 & 0.1307 & -0.1421 & 0.1698\\
\hline
\end{tabular}\end{center}

\footnotesize Data between 2015--2018 are used. The table contains the estimates of the DTTs $\tau(j)$ (and $\tau^r(j)$), $j\in\{0,1,2\}$, for three different models estimated \textbf{with} additional controls $\mathbf{X}_{gt}$ (and $\mathbf{Z}_{gt}$). The first panel of results (``Without Misreporting'') displays estimates of treatment effects $\tau(j)$, $j\in\{0,1,2\}$, for the model estimated without misreporting. These treatment effect estimates are marginalized with respect to the distribution of $\mathbf{X}_{11}$. The second panel of results (``With Misreporting'') displays estimates of treatment effects $\tau(j)$ and $\tau^{r}(j)$, $j\in\{0,1,2\}$, for the model with misreporting. The estimates of $\tau(j)$ are marginalized with respect to the distribution of $\mathbf{X}_{11}$, whereas the estimates of $\tau^{r}(j)$ are marginalized with respect to the distribution of $\mathbf{Z}_{11}$. The first two columns display results for a Frank Copula specification, the last two columns show corresponding results for a Clayton Copula specification. Standard errors are computed via the Delta method and can be found in the corresponding columns next to the coefficient estimates.\normalsize

\end{table}

In summary, the empirical findings of this section reveal nuanced effects of recreational marijuana legalization on 8th-grade high-school students' short-term consumption and reporting behavior. Our initial analysis without misreporting and covariates showed no significant changes in consumption probabilities. However, after introducing covariates, we observed a more pronounced impact, with a significant decrease in non-consumption probability and a significant increase in consumption probabilities. These effect sizes were further amplified and remain significant after accounting for misreporting, while no significant treatment effects are found on the reporting behavior. The results are also in line with the nonparametric bounds analysis of the main paper in that also in the latter case results did not allow to rule out negative (positive) treatment effects in the lower (upper) part of the distribution when controlling for underreporting at the 10\% significance level. This was in contrast to the case without misreporting, where negative (or positive) effects could be rejected at the lower end of the distribution. Nevertheless, it should be noted that, as a limitation of our study, the categorization of the outcome variable focuses on usage frequency and does not capture the doses consumed per occasion, potentially overlooking variations in consumption intensity.

\subsection{Proofs}

\begin{proof}[Proof of Proposition \ref{CIC_equiv_general}]

We prove the equivalence in two directions. 
For the only-if part, starting from Assumption~\ref{A2}, we first show that $h$ can be taken left-continuous, then rewrite it in threshold-crossing form, and finally construct $(\eta,\lambda,V)$ to obtain the representation in Proposition~\ref{CIC_equiv_general}.
For the if part, we prove the converse implication by first decomposing $(\eta,\lambda)$ and then constructing a corresponding pair $(U,h)$.

\underline{\textbf{Only-if part:}}

\underline{\textit{Step 1. \(h(\cdot,t)\) can be taken left-continuous.}}
Define the left-continuous modification of \(h(\cdot,t)\) by
\[
h_-(u,t) := \lim_{x\uparrow u} h(x,t), \qquad u\in\mathbb{R},\ t\in\{0,1\}.
\]
Because one-sided limits exist for any nondecreasing function, \(h_-\) is well defined. 
By construction, \(h_-\) is nondecreasing and left-continuous in \(u\) for each \(t\). 
For each \(t\in\{0,1\}\), let 
$
D_t := \{u\in\mathbb{R} \mid h_-(u,t)\ne h(u,t)\},
$
that is, the set of left-discontinuity points of \(h\).
Since any monotone function has at most countably many discontinuities, \(D_t\) is at most countable. 
Moreover, since $U \mid G=g$ is continuously distributed and $U \perp T \mid G$, it follows that $U \mid T=t$ is also continuously distributed. Hence,
$
\mathrm{Pr}(U\in D_t\mid T=t)=0,
$
and therefore
\begin{align*}
&\quad \ \mathrm{Pr}\!\left(h(U,T)=h_-(U,T)\right)
=\sum_{t\in\{0,1\}}\mathrm{Pr}\!\left(h(U,t)=h_-(U,t)\mid T=t\right)\mathrm{Pr}(T=t)\\
&=\sum_{t\in\{0,1\}}\mathrm{Pr}\!\left(U\notin D_t\mid T=t\right)\mathrm{Pr}(T=t)=1.
\end{align*}
Because $h_-$ is nondecreasing and \(h(U,T)=h_-(U,T)\) a.s., \(h\) can be replaced by its left-continuous modification \(h_-\) without loss of generality. Hence, we assume left-continuity of $h$ in the subsequent part.

\underline{\textit{Step 2. $h(\cdot,t)$ admits a threshold-crossing structure.}} 
Define
$$
	S_{t,j}:=\left\{u\in\mathbb{R}\mid h(u,t)\le j\right\},\quad \gamma_{t,j}:=\sup S_{t,j},\quad t\in\{0,1\},\ j\in\mathcal{J}\setminus \{J\}.
$$
For each $t\in\{0,1\}$ and $j\in\mathcal{J}\setminus \{J\}$, the set $S_{t,j}$ is nonempty, since
$\mathrm{supp}\!\left(Y(0)\mid T=t\right)=\mathcal{J}$ implies
$\mathrm{Pr}\!\left(h\left(U,t\right)=j\mid T=t\right)>0$, and hence there exists
$u\in\mathbb{R}$ such that $h\left(u,t\right)\le j$. Moreover, $\gamma_{t,j}$ is finite. Indeed, since $j\le J-1$, there exists $u'\in\mathbb{R}$ such that
$h\left(u',t\right)=j+1$, and monotonicity of $u\mapsto h\left(u,t\right)$ implies
$S_{t,j}\subseteq\left(-\infty,u'\right]$. Thus $S_{t,j}$ is bounded above and
$\gamma_{t,j}<\infty$.

We note that 
\begin{align}\label{proof:CIC_equiv_general1}
	S_{t,j}=\left(-\infty,\gamma_{t,j}\right],\qquad t\in\{0,1\},\ j\in\mathcal{J}\setminus \{J\}.
\end{align}
To see this, fix $t\in\{0,1\}$ and $j\in\mathcal{J}\setminus \{J\}$, and let \(y\in\left(-\infty,\gamma_{t,j}\right]\). 
If \(y<\gamma_{t,j}\), then by the property of the supremum, there exists 
\(u\in S_{t,j}\) with \(y<u\). Monotonicity of \(u\mapsto h(u,t)\) yields
\(h(y,t)\le h(u,t)\le j\),
hence \(y\in S_{t,j}\).  
If \(y=\gamma_{t,j}\), again by the property of the supremum, there exists a sequence \(\left(u_n\right)\) in \(S_{t,j}\) with 
\(u_n\uparrow \gamma_{t,j}\). Left-continuity of \(u\mapsto h(u,t)\) yields
\(h(y,t)=h(\gamma_{t,j},t)=\lim_{n\to\infty}h(u_n,t)\le j\),
so again \(y\in S_{t,j}\).  
Thus \(\left(-\infty,\gamma_{t,j}\right]\subseteq S_{t,j}\). The reverse inclusion 
\(S_{t,j}\subseteq\left(-\infty,\gamma_{t,j}\right]\) is immediate from the definition 
of \(\gamma_{t,j}\), so (\ref{proof:CIC_equiv_general1}) follows.

Moreover, we note
\begin{align}\label{proof:CIC_equiv_general2}
	\gamma_{t,0}<\gamma_{t,1}<\dots<\gamma_{t,J-1}<+\infty,\qquad t\in\{0,1\}.
\end{align}
To see this, fix $t\in\{0,1\}$ and $j\in\{0,1,\dots,J-2\}$, and assume, for contradiction, that $\gamma_{t,j}\geq\gamma_{t,j+1}$. Then
$
	\mathrm{Pr}\!\left(U\le \gamma_{t,j}\mid T=t\right)\geq \mathrm{Pr}\!\left(U\le \gamma_{t,j+1}\mid T=t\right).
$
However, using  
\(\mathrm{supp}\!\left(Y(0)\mid T=t\right)=\mathcal{J}\), $S_{t,j}=\left\{u\in\mathbb{R}\mid h(u,t)\le j\right\}$, and (\ref{proof:CIC_equiv_general1}),  we have
\begin{align*}
&\mathrm{Pr}\!\left(U\le \gamma_{t,j}\mid T=t\right)
=\mathrm{Pr}\!\left(U\in S_{t,j}\mid T=t\right)
=\mathrm{Pr}\!\left(h(U,t)\le j\mid T=t\right)\\
&=\sum_{k=0}^{j}\mathrm{Pr}\!\left(Y(0)=k\mid T=t\right)
<\sum_{k=0}^{j+1}\mathrm{Pr}\!\left(Y(0)=k\mid T=t\right)\\
&=\mathrm{Pr}\!\left(h(U,t)\le j+1\mid T=t\right)
=\mathrm{Pr}\!\left(U\le \gamma_{t,j+1}\mid T=t\right),
\end{align*}
a contradiction.

Define the threshold-crossing function
\begin{align}\label{proof:CIC_equiv_general3}
	\widetilde{h}(u,t)
:=
\begin{cases}
0, & \text{if } u \le \gamma_{t,0},\\
j, & \text{if } \gamma_{t,j-1}<u\le \gamma_{t,j},\ j=1,\dots,J-1,\\
J, & \text{if } \gamma_{t,J-1}<u,
\end{cases}
\qquad u\in\mathbb{R},\ t\in \{0,1\}.
\end{align}
By construction, $\widetilde{h}(\cdot,t)$ is nondecreasing and left-continuous.  
For each $j\in\mathcal{J}\setminus \{J\}$, we have
\[
\{u\in\mathbb{R}\mid \widetilde{h}(u,t)\le j\}
=
(-\infty,\gamma_{t,j}]
=
S_{t,j}
=
\{u\in\mathbb{R}\mid h(u,t)\le j\},
\]
where the second equality is from (\ref{proof:CIC_equiv_general1}).
Hence, for each $u\in\mathbb{R}$ (i.e., pointwise),
\[
\widetilde{h}(u,t)\le j
\quad\iff \quad
h(u,t)\le j,
\qquad t\in\{0,1\},\ j\in\mathcal{J}\setminus \{J\}.
\]
Therefore, 
$
\left\{\widetilde{h}(U,t)\le j\right\}
=
\left\{h(U,t)\le j\right\}
$ for all $t\in\{0,1\}$ and $j\in\mathcal{J}\setminus \{J\}$. In particular,
\[
\left\{h(U,t)\le j<\widetilde{h}(U,t)\right\}
=\emptyset,
\qquad
\left\{\widetilde{h}(U,t)\le j<h(U,t)\right\}
=\emptyset.
\]

Now, fix $t\in\{0,1\}$. Since $Y(0)=h(U,T)$ a.s.\ and 
$\mathrm{supp}\!\left(Y(0)\mid T=t\right)=\mathcal{J}$, we have
$
\mathrm{Pr}\!\left(h(U,t)\in\mathcal{J}\mid T=t\right)=1.
$
Moreover, by construction, $\widetilde{h}(U,t)\in\mathcal{J}$ everywhere.
Hence, conditional on $\{T=t\}$, both $h(U,t)$ and $\widetilde{h}(U,t)$ take values in
$\mathcal{J}$ a.s., and therefore,
\small$$
\mathrm{Pr}\!\left(h(U,t)\neq \widetilde{h}(U,t)\mid T=t\right)
=
\mathrm{Pr}\!\left(
\bigcup_{j=0}^{J-1}
\left(
\left\{h(U,t)\le j<\widetilde{h}(U,t)\right\}
\cup
\left\{\widetilde{h}(U,t)\le j<h(U,t)\right\}
\right) \mid T=t
\right)
=0.
$$\normalsize
Consequently,
\[
\mathrm{Pr}\!\left(h(U,T)=\widetilde{h}(U,T)\right)
=
\sum_{t\in\{0,1\}}
\mathrm{Pr}\!\left(h(U,t)=\widetilde{h}(U,t)\mid T=t\right)\mathrm{Pr}(T=t)
=1.
\]
That is, $h(U,T)=\widetilde{h}(U,T)$ a.s., so $h$ can always be replaced by its threshold-crossing representation $\widetilde{h}$ without loss of generality.

Finally, to express this threshold structure in a standardized form, define, for each $t\in \{0,1\}$, 
\begin{align*}
\alpha_t&:=\begin{cases}
-\gamma_{t,0}, & J=1,\\[2pt]
-\dfrac{\gamma_{t,0}}{\gamma_{t,1}-\gamma_{t,0}}, & J>1,
\end{cases}
\quad
\beta_t:=\begin{cases}
1, & J=1,\\[2pt]
\dfrac{1}{\gamma_{t,1}-\gamma_{t,0}}, & J>1,
\end{cases}
\in(0,+\infty),
\\
\kappa_{t,j}&:=\begin{cases}
\gamma_{t,j}-\gamma_{t,0}, & J=1,\\[2pt]
\dfrac{\gamma_{t,j}-\gamma_{t,0}}{\gamma_{t,1}-\gamma_{t,0}}, & J>1,
\end{cases}
\qquad j=0,1,\dots,J-1.
\end{align*}
By construction, $\kappa_{t,0}=0$. When $J\ge 2$, we also have $\kappa_{t,1}=1$, and by \eqref{proof:CIC_equiv_general2}, the thresholds satisfy $\kappa_{t,j-1}<\kappa_{t,j}$ for all $j\ge 1$. Hence, the threshold-crossing function (\ref{proof:CIC_equiv_general3}) can be written into
\[
\widetilde{h}(u,t)
=
\begin{cases}
0, & \text{if } \alpha_t+\beta_t u \le \kappa_{t,0},\\
j, & \text{if } \kappa_{t,j-1}<\alpha_t+\beta_t u \le \kappa_{t,j},\ j=1,\dots,J-1,\\
J, & \text{if } \kappa_{t,J-1}<\alpha_t+\beta_t u,
\end{cases}
\qquad u\in\mathbb{R},\ t\in \{0,1\}.
\]
In what follows, we assume this representation of $h$.

\underline{\textit{Step 3. Existence of $V$  and $(\eta,\lambda)$.}}  For $g \in\{0,1\}$, let
$$\mu_g:=\mathrm{E}[U \mid G=g], \quad \sigma_g:=\sqrt{\operatorname{Var}[U \mid G=g]} \in (0,+\infty),
$$
where $\sigma_g\in (0,+\infty)$ is due to continuity of $U\mid G=g$ and $\mathrm{Var}\left[U\mid G=g\right]<\infty$. Define the random variables
$$
V:=\frac{U-\mu_G}{\sigma_G}, \quad
\mu_{G }:=\sum_{g\in \{0,1\}} \mu_{g} \mathbb{I}\{G=g\}, \quad
\sigma_{G }:=\sum_{g\in \{0,1\}} \sigma_{g} \mathbb{I}\{G=g\}.
$$
$V \mid G=g$ is continuously distributed because $U\mid G=g$ is continuously distributed, and it satisfies $\operatorname{E}\left[V \mid G=g\right]=0$ and $\operatorname{Var}\left[V \mid G=g\right]=1$ by construction. Besides,  $V \perp T \mid G $ because $U\perp T \mid G$. Set
$$
\eta_{g t}:=\alpha_t+\beta_t \mu_g, \quad \lambda_{g t}:=\beta_t \sigma_g \in (0,+\infty),
$$
where $\beta_t\in (0,+\infty)$ by construction from Step 2.
Then, with $\alpha_T := \sum_{t\in \{0,1\}} \alpha_t\mathbb{I}\{T=t\}$ and $\beta_T := \sum_{t\in \{0,1\}} \beta_t\mathbb{I}\{T=t\}$, we have
$$
\eta_{G T}+\lambda_{G T} V=\left(\alpha_T+\beta_T \mu_G\right)+\left(\beta_T \sigma_G\right) V=\alpha_T+\beta_T U \quad a.s.
$$
Moreover,  $U\perp T \mid G$, $\operatorname{supp}(U \mid G=1) \subseteq \operatorname{supp}(U \mid G=0)$ and $\beta_0>0$ immediately imply $\operatorname{supp}(\alpha_T+\beta_T U \mid G=1,T=0) \subseteq \operatorname{supp}(\alpha_T+\beta_T U \mid G=0,T=0)$, hence
$$
\operatorname{supp}\left(\eta_{G T}+\lambda_{G T} V \mid G=1, T=0\right) \subseteq \operatorname{supp}\left(\eta_{G T}+\lambda_{G T} V \mid G=0, T=0\right).
$$
Using Step 2 and defining $\kappa_{T,j} := \sum_{t\in \{0,1\}} \kappa_{t,j}\mathbb{I}\{T=t\}$, we obtain the desired representation
$$
Y(0)=\left\{\begin{array}{ll}
0, & \text { if } \eta_{G T}+\lambda_{G T} V \leq \kappa_{T,0} \\
j, & \text { if } \kappa_{T,j-1}<\eta_{G T}+\lambda_{G T} V \leq \kappa_{T,j},\ j=1,\dots,J-1 \\
J, & \text { if } \kappa_{T,J-1}<\eta_{G T}+\lambda_{G T} V
\end{array}\right. \text { a.s. }
$$

It remains to verify condition (4) in Proposition \ref{CIC_equiv_general}. We can write
$
\frac{\lambda_{11}}{\lambda_{10}}=\frac{\beta_1 \sigma_1}{\beta_0 \sigma_1}=\frac{\beta_1}{\beta_0}=\frac{\beta_1 \sigma_0}{\beta_0 \sigma_0}=\frac{\lambda_{01}}{\lambda_{00}} .
$
Moreover,
$
\eta_{11}-\frac{\lambda_{11}}{\lambda_{10}} \eta_{10}=\left(\alpha_1+\beta_1 \mu_1\right)-\frac{\beta_1}{\beta_0}\left(\alpha_0+\beta_0 \mu_1\right)=\alpha_1-\frac{\beta_1}{\beta_0} \alpha_0,
$
and similarly,
$
\eta_{01}-\frac{\lambda_{01}}{\lambda_{00}} \eta_{00}=\left(\alpha_1+\beta_1 \mu_0\right)-\frac{\beta_1}{\beta_0}\left(\alpha_0+\beta_0 \mu_0\right)=\alpha_1-\frac{\beta_1}{\beta_0} \alpha_0 .
$
Hence,
$$
\eta_{11}-\frac{\lambda_{11}}{\lambda_{10}} \eta_{10}=\eta_{01}-\frac{\lambda_{01}}{\lambda_{00}} \eta_{00}.
$$

\underline{\textbf{If part:}}

\underline{\textit{Step 1. Decomposing $\eta$ and $\lambda$.}} We show that there exist parameters $\left(\alpha_t, \beta_t, \mu_g, \sigma_g\right)$ with $\beta_t,\sigma_g\in (0,+\infty)$ such that
\begin{align}\label{proof:CIC_equiv_general4}
\eta_{g t}=\alpha_t+\beta_t \mu_g, \quad \lambda_{g t}=\beta_t \sigma_g
\end{align}
for all $(g, t) \in\{0,1\} \times\{0,1\}$.
First, write
\begin{align*}
\lambda_{g t}=b_0+\left(b_1-b_0\right) g+\left(b_2-b_0\right) t+\left(b_3-b_2-b_1+b_0\right) g t,
\end{align*}
where $b_0:=\lambda_{00}, b_1:=\lambda_{10}, b_2:=\lambda_{01}$, and $b_3:=\lambda_{11}$.
Under $\lambda_{11} / \lambda_{10}=\lambda_{01} / \lambda_{00}$ in condition (4) of Proposition \ref{CIC_equiv_general}, we have
$
b_3=\frac{b_1 b_2}{b_0}.
$
Substituting $b_3$ back, we obtain
$$
\lambda_{gt}
= b_0 + (b_1 - b_0)g + (b_2 - b_0)t 
   + \left(\frac{b_1 b_2}{b_0} - b_1 - b_2 + b_0\right)gt = \underbrace{\left(1 + \frac{b_2 - b_0}{b_0}t\right)}_{=:\,\beta_t}
   \;\underbrace{\left(b_0 + (b_1 - b_0)g\right)}_{=:\,\sigma_g}.
$$
That is,
$
	\beta_0=1,\beta_1=\frac{\lambda_{01}}{\lambda_{00}}, \sigma_0=\lambda_{00}$ and $ \sigma_1=\lambda_{10}.
$
Such $\beta_t$ and $\sigma_g$ are one particular choice demonstrating their existence, while alternative definitions are equally valid. Because each $\lambda_{g t}\in (0,+\infty)$, it follows that $\beta_t,\sigma_g\in (0,+\infty)$. 

Next, write
\begin{align*}
	\eta_{gt}=a_0+(a_1-a_0)g+(a_2-a_0)t+(a_3-a_2-a_1+a_0)gt,
\end{align*}
where $a_0:=\eta_{00}, a_1:=\eta_{10}, a_2:=\eta_{01}$ and $a_3:=\eta_{11}$.  Under \(\eta_{11}-\frac{\lambda_{11}}{\lambda_{10}} \eta_{10}=\eta_{01}-\frac{\lambda_{01}}{\lambda_{00}} \eta_{00} \) and $\lambda_{11} / \lambda_{10}=\lambda_{01} / \lambda_{00}$ in condition (4) of Proposition \ref{CIC_equiv_general}, it holds that
$
a_3=a_2+\frac{a_1-a_0}{b_0/b_2}.
$
With $\beta_t=\left(1+\frac{b_2-b_0}{b_0}t\right)$ from above, this immediately leads to:
\small$$
	\eta_{gt}=a_0+(a_1-a_0)g+(a_2-a_0)t+\left(\frac{a_1-a_0}{b_0/b_2}-a_1+a_0\right)gt=\underbrace{a_0+(a_2-a_0)t}_{=:\alpha_t}+\underbrace{(a_1-a_0)g}_{=:\mu_g}\underbrace{\left(1+\frac{b_2-b_0}{b_0}t\right)}_{\beta_t}.
$$\normalsize
That is, 
$
	\alpha_0=\eta_{00}, \alpha_1=\eta_{01}, \mu_0=0$ and $ \mu_1=\eta_{10}-\eta_{00}.
$

\underline{\textit{Step 2. Existence of $h$ and $U$.}} Define the random variable
$
U:=\mu_G+\sigma_G V,
$
where $\mu_G := \sum_{g\in \{0,1\}} \mu_g\mathbb{I}\{G=g\}$, and analogously for other quantities here and below. Define the function $h: \mathbb{R} \times\{0,1\} \rightarrow \mathbb{R}$,
$$
h(u, t)
:=\begin{cases}
0, & \text{if } \alpha_t + \beta_t u \le \kappa_{t,0}\\
j, & \text{if } \kappa_{t,j-1} < \alpha_t + \beta_t u \le \kappa_{t,j},\ j = 1, \dots, J-1\\
J, & \text{if } \kappa_{t,J-1} < \alpha_t + \beta_t u
\end{cases}.
$$
By (\ref{proof:CIC_equiv_general4}), we have
$
\alpha_T+\beta_T U=\left(\alpha_T+\beta_T \mu_G\right)+\left(\beta_T \sigma_G\right) V=\eta_{G T}+\lambda_{G T} V$ a.s.
Hence,  
\begin{align*}
		Y(0) &
=\begin{cases}
0, & \text{if } \eta_{GT}+\lambda_{GT}V \le \kappa_{T,0},\\
j, & \text{if } \kappa_{T,j-1} < \eta_{GT}+\lambda_{GT}V \le \kappa_{T,j},\ j=1,\dots,J-1,\\
J, & \text{if } \kappa_{T,J-1} < \eta_{GT}+\lambda_{GT}V,
\end{cases}\\
&=\begin{cases}
0, & \text{if } \alpha_T + \beta_T U \le \kappa_{T,0}\\
j, & \text{if } \kappa_{T,j-1} < \alpha_T + \beta_T U \le \kappa_{T,j},\ j = 1, \dots, J-1\\
J, & \text{if } \kappa_{T,J-1} < \alpha_T + \beta_T U
\end{cases}\\
&= h(U,T)      \quad\text{a.s.}
\end{align*}

It remains to verify the conditions in Assumption~\ref{A2}. By construction, $h(\cdot,t)$ is nondecreasing. Moreover, $U \mid G=g$ is continuously distributed with $\operatorname{Var}[U \mid G=g]<+\infty$ and satisfies $U \perp T \mid G$, because $V\mid G=g$ is continuously distributed with $\operatorname{Var}[V \mid G=g]<+\infty$ and $V\perp T \mid G$. Furthermore, since $\eta_{GT}+\lambda_{GT}V=\alpha_T+\beta_T U$ a.s., the condition
$
\operatorname{supp}\left(\eta_{G T}+\lambda_{G T} V \mid G=1, T=0\right) \subseteq \operatorname{supp}\left(\eta_{G T}+\lambda_{G T} V \mid G=0, T=0\right)
$
implies
\[
\operatorname{supp}\left(\alpha_T+\beta_T U \mid G=1, T=0\right)\subseteq \operatorname{supp}\left(\alpha_T+\beta_T U \mid G=0, T=0\right).
\]
Under $U \perp T \mid G$ and $\beta_0>0$, we obtain
\begin{align*}
\operatorname{supp}(U \mid G=1)
&=\operatorname{supp}(U \mid G=1, T=0)
=\frac{1}{\beta_0}\left(\operatorname{supp}\left(\alpha_0+\beta_0 U \mid G=1, T=0\right)-\alpha_0\right)\\
&\subseteq \frac{1}{\beta_0}\left(\operatorname{supp}\left(\alpha_0+\beta_0 U \mid G=0, T=0\right)-\alpha_0\right)\\
&=\operatorname{supp}(U \mid G=0, T=0)
=\operatorname{supp}(U \mid G=0).
\end{align*}
\end{proof}

\begin{proof}[Proof of Proposition \ref{PROP3}]
Denote
\(
\widetilde{\varepsilon}_{gt}^{(0)}:=\lambda_{gt}^{(0)}\varepsilon_{gt}^{(0)}
\)
and
\(
\widetilde{\nu}_{gt}^{(0)}:=\zeta_{gt}^{(0)}\nu_{gt}^{(0)},
\)
with corresponding CDFs
\[
F_{-\widetilde{\varepsilon}_{gt}^{(0)}}(v)
:=
F_{-\varepsilon_{gt}^{(0)}}\!\left(\frac{v}{\lambda_{gt}^{(0)}}\right),
\qquad
F_{-\widetilde{\nu}_{gt}^{(0)}}(v)
:=
F_{-\nu_{gt}^{(0)}}\!\left(\frac{v}{\zeta_{gt}^{(0)}}\right),
\qquad v\in\mathbb R.
\]
Then \eqref{EQUPPERTAIL} can be rewritten, for \(j\in \mathcal{J}\setminus\{0\}\), as
\begin{align*}
\Pr\!\left(C_{gt}(0)\ge j \mid \mathbf X_{gt},\mathbf Z_{gt}\right)
=
\overline{C}_{gt}^{(0)}\!\left(
F_{-\widetilde{\varepsilon}_{gt}^{(0)}}
\bigl(
\overline{\mathbf X}_{gt}'\overline{\boldsymbol{\eta}}_{gt}^{(0)}-\kappa_{t,j-1}^{(0)}
\bigr),
F_{-\widetilde{\nu}_{gt}^{(0)}}
\bigl(
\overline{\mathbf Z}_{gt}'\overline{\boldsymbol{\pi}}_{gt}^{(0)}-\iota_{t,j-1}^{(0)}
\bigr)
\right).
\end{align*}
We prove part (i). Part (ii) follows by the same argument after interchanging \((\mathbf X,\overline{\boldsymbol{\eta}},\lambda,\kappa)\) and \((\mathbf Z,\overline{\boldsymbol{\pi}},\zeta,\iota)\), using Assumption~\ref{S4}(i)(a) in place of Assumption~\ref{S4}(i)(b).

\underline{Identification of $\overline{\boldsymbol{\eta}}_{gt}^{(0)}$, \((g,t)\neq (1,1)\).}
First, let \((g,t)\neq(1,1)\). By Assumption~\ref{S3} and { Assumption~\ref{S4}(i)(b)--(ii)}, the coefficient \(\pi_{1,gt}^{(0)}\) is nonzero, and its sign is identified from
\begin{align*}
\operatorname{sgn}\!\left(
\mathrm{E}\!\left[
\frac{\partial}{\partial Z_{1,gt}}
\Pr\!\left(C_{gt}(0)\ge j \mid \overline{\mathbf W}_{gt},X_{1,gt},Z_{1,gt}\right)
\right]
\right)
\end{align*}
for some { \(j\in\mathcal J\setminus\{0\}\)}, where \(\operatorname{sgn}(x)=\mathbb{I}\{x>0\}-\mathbb{I}\{x<0\}\). Without loss of generality, we may suppose that \(\pi_{1,gt}^{(0)}>0\) for the rest of the proof. If \(\pi_{1,gt}^{(0)}<0\), the same argument applies with the limit \(z\to -\infty\) in place of \(z\to +\infty\).

Now, fix { \(j\in\mathcal J\setminus\{0\}\)}. Define, for any \(\mathbf{x}\in\mathbb{R}^{d_{X_{gt}}}\) and \(\mathbf{z}_{-1}\in\mathbb{R}^{d_{Z_{gt}}-1}\),
\[
c_{\mathbf x,\mathbf z_{-1},gt}(z)
:=
\overline{C}_{gt}^{(0)}\!\left(
F_{-\widetilde{\varepsilon}_{gt}^{(0)}}
\bigl(
\overline{\mathbf x}'\overline{\boldsymbol{\eta}}_{gt}^{(0)}-\kappa_{t,j-1}^{(0)}
\bigr),
F_{-\widetilde{\nu}_{gt}^{(0)}}
\bigl(
\pi_{1,gt}^{(0)}z+\overline{\mathbf z}_{-1}'\overline{\boldsymbol{\pi}}_{-1,gt}^{(0)}
-\iota_{t,j-1}^{(0)}
\bigr)
\right),
\quad z\in\mathbb R,
\]
where \(\overline{\mathbf{z}}_{-1}:=(1,\mathbf{z}_{-1}')'\) and \(\overline{\boldsymbol{\pi}}_{-1,gt}^{(0)}:=(\pi_{c, g t}^{(0)},\boldsymbol{\pi}_{-1,gt}^{(0)'})'\). Under Assumption~\ref{S3}, this function is real analytic in \(z\), since the elementary functions are real analytic and compositions of real analytic functions remain real analytic. We consider any alternative parameterization
\(
\left(
\overline{\boldsymbol{\eta}}_{gt}^{(0)*},\lambda_{gt}^{(0)*},
\overline{\boldsymbol{\pi}}_{gt}^{(0)*},\zeta_{gt}^{(0)*},\left\{\kappa_{t, j}^{(0) *}\right\}_{j=2}^{J-1},\left\{\iota_{t, j}^{(0) *}\right\}_{j=2}^{J-1}{ ,}
F_{-\widetilde{\varepsilon}_{gt}^{(0)}}^*,F_{-\widetilde{\nu}_{gt}^{(0)}}^*,
\overline{C}_{gt}^{(0)*}
\right)
\)
that satisfies the maintained assumptions and generates the same observable conditional tail probability. For \(j=1,2\), the relevant thresholds are { normalized} to zero and one, so starred and unstarred values coincide. The associated function \(c_{\mathbf x,\mathbf z_{-1},gt}^*(z)\) is also real analytic on \(\mathbb R\).

Fix $\overline{\mathbf{w}}=\left(1, \mathbf{x}_{-1}^{\prime}, \mathbf{z}_{-1}^{\prime}\right)^{\prime} \in \operatorname{supp}\left(\overline{\mathbf{W}}_{g t}\right)$.\footnote{This is to be understood as  $\overline{\mathbf{w}}=\left(1, \mathbf{x}_{-1}^{\prime}, \mathbf{z}_{-1}^{\prime}\right)^{\prime} \in \operatorname{supp}\left(\overline{\mathbf{W}}_{g t}\right)$ such that $c_{\mathbf{x}, \mathbf{z}_{-1}, g t}(z)=c_{\mathbf{x}, \mathbf{z}_{-1}, g t}^*(z) $ for $\mathrm{P}_{\left(X_{1, g t}, Z_{1, g t}\right) \mid \overline{\mathbf{W}}_{g t}=\overline{\mathbf{w}}}$-a.e. $(x, z)$, where $\mathbf{x}:=\left(x, \mathbf{x}_{-1}^{\prime}\right)^{\prime}$. Note that the previous equality holds for $\mathrm{P}_{\overline{\mathbf W}_{gt}}$-a.e. $\overline{\mathbf{w}}$
by observational equivalence.} By Assumption~\ref{S4}(ii), we have $
c_{\mathbf{x}, \mathbf{z}_{-1}, g t}(z)=c_{\mathbf{x}, \mathbf{z}_{-1}, g t}^*(z)
$ for almost every $(x, z)\in I_1\times I_2\subseteq
\operatorname{supp}\left(
\left(X_{1,gt},Z_{1,gt}\right)
\mid
\overline{\mathbf W}_{gt}=\overline{\mathbf w}\right)$, where $\mathbf{x}:=\left(x, \mathbf{x}_{-1}^{\prime}\right)^{\prime}$. Because both sides are continuous in $\left(x,z\right)$,  this equality extends to every $(x, z) \in I_1 \times I_2$, since otherwise failure of equality at a point would imply failure of equality on some nonempty open subset of $I_1\times I_2$ with positive conditional probability. Hence, for each fixed $x \in I_1$, the function
\begin{align*}
z \mapsto c_{\mathbf{x}, \mathbf{z}_{-1}, g t}(z)-c_{\mathbf{x}, \mathbf{z}_{-1}, g t}^*(z)
\end{align*}
is real analytic on $\mathbb{R}$ and zero on the open interval $I_2$. By the identity theorem for real analytic functions, the two functions therefore coincide everywhere on \(\mathbb{R}\). Taking limits as \(z\to\infty\), using \(\pi_{1,gt}^{(0)}>0\) {and \(\pi_{1,gt}^{(0)*}>0\)}, the continuity of \(F_{-\widetilde{\nu}_{gt}^{(0)}}\) {and \(F_{-\widetilde{\nu}_{gt}^{(0)}}^*\)}, and the copula property \(\overline{C}_{gt}^{(0)}(u,1)=\overline{C}_{gt}^{(0)*}(u,1)=u\), yields
$$
F_{-\widetilde{\varepsilon}_{gt}^{(0)}}
\!\left(
\overline{\mathbf{x}}'\overline{\boldsymbol{\eta}}_{gt}^{(0)}
-\kappa_{t,j-1}^{(0)}
\right)=\lim_{z\to\infty} c_{\mathbf{x},\mathbf{z}_{-1},gt}(z)=
\lim_{z\to\infty} c_{\mathbf{x},\mathbf{z}_{-1},gt}^*(z)
=
F_{-\widetilde{\varepsilon}_{gt}^{(0)}}^*
\!\left(
\overline{\mathbf{x}}'\overline{\boldsymbol{\eta}}_{gt}^{(0)*}
-{\kappa_{t,j-1}^{(0)*}}
\right).
$$
Because the equality holds for every \(x \in I_1\), the identity theorem for real analytic functions again implies that
\[
F_{-\widetilde{\varepsilon}_{gt}^{(0)}}
\!\left(
\eta_{1,gt}^{(0)}x+\overline{\mathbf{x}}_{-1}'\overline{\boldsymbol{\eta}}_{-1,gt}^{(0)}
-\kappa_{t,j-1}^{(0)}
\right)
=
F_{-\widetilde{\varepsilon}_{gt}^{(0)}}^*
\!\left(
\eta_{1,gt}^{(0)*}x+\overline{\mathbf{x}}_{-1}'\overline{\boldsymbol{\eta}}_{-1,gt}^{(0)*}
-{\kappa_{t,j-1}^{(0)*}}
\right)
\]
for all \(x\in \mathbb{R}\), where \(\overline{\mathbf{x}}_{-1}:=(1,\mathbf{x}_{-1}')'\), $\overline{\boldsymbol{\eta}}_{-1,gt}^{(0)}:=(\eta_{c, g t}^{(0)},\boldsymbol{\eta}_{-1,gt}^{(0)'})'$ and $\overline{\boldsymbol{\eta}}_{-1,gt}^{(0)*}:=(\eta_{c, g t}^{(0)*},\boldsymbol{\eta}_{-1,gt}^{(0)*'})'$. { We now apply this identity at the normalized thresholds $0$ and $1$.} Since both CDFs are strictly increasing with median zero under Assumption~\ref{S3}, evaluating at
\[
x_j^0
:=
\frac{\kappa_{t,j-1}^{(0)}
-\overline{\mathbf x}_{-1}'\overline{\boldsymbol{\eta}}_{-1,gt}^{(0)}}{\eta_{1,gt}^{(0)}},
\qquad j\in\{1,2\},
\]
yields
\[
\eta_{1,gt}^{(0)*}x_j^0
+\overline{\mathbf x}_{-1}'\overline{\boldsymbol{\eta}}_{-1,gt}^{(0)*}
-\kappa_{t,j-1}^{(0)}
=
0,
\qquad j\in\{1,2\}.
\]
Rearranging the terms and using \(\kappa_{t,0}^{(0)}=0\) and \(\kappa_{t,1}^{(0)}=1\), we obtain a system of linear equations in \(\bigl(\eta_{1,gt}^{(0)*},\,\overline{\mathbf{x}}_{-1}'\overline{\boldsymbol{\eta}}_{-1,gt}^{(0)*}\bigr)\):
\begin{align*}
\eta_{1,gt}^{(0)*}
\frac{-\overline{\mathbf{x}}_{-1}'\overline{\boldsymbol{\eta}}_{-1,gt}^{(0)}}{\eta_{1,gt}^{(0)}}
+
\overline{\mathbf{x}}_{-1}'\overline{\boldsymbol{\eta}}_{-1,gt}^{(0)*}
=0,\qquad 
\eta_{1,gt}^{(0)*}
\frac{1-\overline{\mathbf{x}}_{-1}'\overline{\boldsymbol{\eta}}_{-1,gt}^{(0)}}{\eta_{1,gt}^{(0)}}
+
\overline{\mathbf{x}}_{-1}'\overline{\boldsymbol{\eta}}_{-1,gt}^{(0)*}
=1.
\end{align*}
The system has full rank, since
$
\det
\begin{pmatrix}
-\frac{\overline{\mathbf{x}}_{-1}'\overline{\boldsymbol{\eta}}_{-1,gt}^{(0)}}{\eta_{1,gt}^{(0)}} & 1\\
\frac{1-\overline{\mathbf{x}}_{-1}'\overline{\boldsymbol{\eta}}_{-1,gt}^{(0)}}{\eta_{1,gt}^{(0)}} & 1
\end{pmatrix}
=
{ -\frac{1}{\eta_{1,gt}^{(0)}}}
\neq 0.
$
Hence, the solution is unique. Solving the system yields
\[
\eta_{1,gt}^{(0)*}=\eta_{1,gt}^{(0)},
\qquad
\overline{\mathbf{x}}_{-1}'\overline{\boldsymbol{\eta}}_{-1,gt}^{(0)*}
=
\overline{\mathbf{x}}_{-1}'\overline{\boldsymbol{\eta}}_{-1,gt}^{(0)}.
\]
Thus, \(\eta_{1,gt}^{(0)}\) is identified, and
$
\overline{\mathbf{x}}_{-1}'\overline{\boldsymbol{\eta}}_{-1,gt}^{(0)*}
=
\overline{\mathbf{x}}_{-1}'\overline{\boldsymbol{\eta}}_{-1,gt}^{(0)}
$
for almost every \(\mathbf x_{-1}\in \mathcal X_{-1,gt}\).
Since
$
\mathbf x_{-1}\mapsto
\overline{\mathbf{x}}_{-1}'
\left(
\overline{\boldsymbol{\eta}}_{-1,gt}^{(0)*}
-
\overline{\boldsymbol{\eta}}_{-1,gt}^{(0)}
\right)
$
is continuous, this equality again extends to every
\(\mathbf x_{-1}\in \mathcal X_{-1,gt}\). Because
\(\overline{\mathcal X}_{-1,gt}\) is not contained in a proper linear subspace by Assumption~\ref{S4}(iii), it follows that
$
\overline{\boldsymbol{\eta}}_{-1,gt}^{(0)*}
=
\overline{\boldsymbol{\eta}}_{-1,gt}^{(0)},
$
{ so} \(\overline{\boldsymbol{\eta}}_{-1,gt}^{(0)}\) is identified.

\underline{Identification of $\lambda_{gt}^{(0)}$ and $F_{-\varepsilon_{gt}^{(0)}}$, \((g,t)\neq (1,1)\).}
Given identification of \(\overline{\boldsymbol{\eta}}_{gt}^{(0)}\), we have
\[
F_{-\widetilde{\varepsilon}_{gt}^{(0)}}
\!\left(
\eta_{1,gt}^{(0)}x
+
\overline{\mathbf{x}}_{-1}'\overline{\boldsymbol{\eta}}_{-1,gt}^{(0)}
-
\kappa_{t,j-1}^{(0)}
\right)
=
F_{-\widetilde{\varepsilon}_{gt}^{(0)}}^*
\!\left(
\eta_{1,gt}^{(0)}x
+
\overline{\mathbf{x}}_{-1}'\overline{\boldsymbol{\eta}}_{-1,gt}^{(0)}
-
\kappa_{t,j-1}^{(0)}
\right),\quad j=1,2,
\]
for all \(x\in I_1\). Since both sides are real analytic in \(x\) and $I_1$ is an open interval, the identity theorem again implies that this equality holds for every \(x\in\mathbb R\). Since \(\eta_{1,gt}^{(0)}\neq 0\), we have
\[
F_{-\widetilde{\varepsilon}_{gt}^{(0)}}(v)
=
F_{-\widetilde{\varepsilon}_{gt}^{(0)}}^*(v),
\qquad v\in\mathbb R,
\]
so the marginal distribution \(F_{-\widetilde{\varepsilon}_{gt}^{(0)}}\) is identified. Since \(\widetilde{\varepsilon}_{gt}^{(0)}=\lambda_{gt}^{(0)}\varepsilon_{gt}^{(0)}\) and Assumption~\ref{S3}(i) normalizes \(\varepsilon_{gt}^{(0)}\) to have unit variance, \(\lambda_{gt}^{(0)}\) is identified from
\[
(\lambda_{gt}^{(0)})^2
=
\operatorname{Var}\!\left(\widetilde{\varepsilon}_{gt}^{(0)}\right).
\]
It then follows that
\[
F_{-\varepsilon_{gt}^{(0)}}(u)
=
F_{-\widetilde{\varepsilon}_{gt}^{(0)}}\!\left(\lambda_{gt}^{(0)}u\right),
\qquad u\in\mathbb R,
\]
so \(F_{-\varepsilon_{gt}^{(0)}}\) is identified as well.

\underline{Identification of \(\{\kappa_{t,j}^{(0)}\}_{j=2}^{J-1}\), \(t\in \{0,1\}\).}
Given identification of \(\overline{\boldsymbol{\eta}}_{gt}^{(0)}\) and \(F_{-\widetilde{\varepsilon}_{gt}^{(0)}}\), we have
\[
F_{-\widetilde{\varepsilon}_{gt}^{(0)}}
\!\left(
\eta_{1,gt}^{(0)}x
+
\overline{\mathbf{x}}_{-1}'\overline{\boldsymbol{\eta}}_{-1,gt}^{(0)}
-
\kappa_{t,j-1}^{(0)}
\right)
=
F_{-\widetilde{\varepsilon}_{gt}^{(0)}}
\!\left(
\eta_{1,gt}^{(0)}x
+
\overline{\mathbf{x}}_{-1}'\overline{\boldsymbol{\eta}}_{-1,gt}^{(0)}
-
\kappa_{t,j-1}^{(0)*}
\right),\quad j\in \mathcal{J}\setminus \{0,1,2\},
\]
for all \(x\in I_1\). Since \(F_{-\widetilde{\varepsilon}_{gt}^{(0)}}\) is strictly increasing, it follows that
\[
\kappa_{t,j-1}^{(0)}
=
\kappa_{t,j-1}^{(0)*},
\]
so 
\(
\{\kappa_{t,j}^{(0)}\}_{j=2}^{J-1}
\)
is identified.

\underline{Identification of $\overline{\boldsymbol{\eta}}_{11}^{(0)}$, $\lambda_{11}^{(0)}$, and $F_{-\varepsilon_{11}^{(0)}}$.}
Finally, consider \((g,t)=(1,1)\). Assumption~\ref{S2} yields
\[
\lambda_{11}^{(0)}
=
\frac{\lambda_{01}^{(0)}\lambda_{10}^{(0)}}{\lambda_{00}^{(0)}}
\quad \text{and} \quad
\overline{\mathbf x}'\overline{\boldsymbol{\eta}}_{11}^{(0)}
=
\overline{\mathbf x}'\overline{\boldsymbol{\eta}}_{01}^{(0)}
+
\frac{\lambda_{01}^{(0)}}{\lambda_{00}^{(0)}}
\overline{\mathbf x}'
\left(
\overline{\boldsymbol{\eta}}_{10}^{(0)}
-
\overline{\boldsymbol{\eta}}_{00}^{(0)}
\right)
\]
for all \(\overline{\mathbf x}\in\overline{\mathcal X}\). Because  $\overline{\mathcal X}=\bigcap_{(g,t)\in\{0,1\}\times\{0,1\}}\overline{\mathcal X}_{gt}$ and $\mathcal{X}_{11} \subseteq \bigcap_{(g, t) \neq(1,1)} \mathcal{X}_{g t},$ we have \(\overline{\mathcal{X}}=\overline{\mathcal X}_{11}\). Since \(\overline{\mathcal X}_{11}\) is not contained in a proper linear subspace by Assumption~\ref{S4}(iii), this implies
\[
\overline{\boldsymbol{\eta}}_{11}^{(0)}
=
\overline{\boldsymbol{\eta}}_{01}^{(0)}
+
\frac{\lambda_{01}^{(0)}}{\lambda_{00}^{(0)}}
\left(
\overline{\boldsymbol{\eta}}_{10}^{(0)}
-
\overline{\boldsymbol{\eta}}_{00}^{(0)}
\right).
\]
Assumption~\ref{S3}(i) immediately implies
\(
F_{-\varepsilon_{11}^{(0)}}
=
F_{-\varepsilon_{10}^{(0)}}.
\)
\end{proof}

\begin{proof}[Proof of Proposition \ref{PROP4}]
We prove part (i). Part (ii) follows analogously upon replacing
\((\mathbf X,\overline{\boldsymbol{\eta}},\lambda,\kappa,\varepsilon)\) by
\((\mathbf Z,\overline{\boldsymbol{\pi}},\zeta,\iota,\nu)\) and using
Assumption~\ref{S4}(i)(a) in place of Assumption~\ref{S4}(i)(b).

Under the support condition
\(
\mathcal X_{11}\subseteq \bigcap_{(g,t)\neq(1,1)}\mathcal X_{gt},
\)
Proposition~\ref{PROP3}(i) identifies
\(\overline{\boldsymbol{\eta}}_{11}^{(0)}\),
\(\lambda_{11}^{(0)}\),
\(\{\kappa_{1,j}^{(0)}\}_{j=2}^{J-1}\),
and \(F_{-\varepsilon_{11}^{(0)}}\). The normalizations
\(\kappa_{1,0}^{(0)}=0\) and \(\kappa_{1,1}^{(0)}=1\) are known by assumption.
Moreover, \(F_{\mathbf X_{11}}\) is identified directly from the observed covariates in
cell \((1,1)\).

Next, under Assumptions~\ref{S4}(i)(b) and \ref{S5}, the treated potential outcome
\(Y_{11}(1)\) satisfies the same threshold-crossing structure as the observed
cells in the proof of Proposition~\ref{PROP3}. Repeating that identification
argument for the treated post-treatment cell yields identification of
\(\overline{\boldsymbol{\eta}}_{11}^{(1)}\),
\(\lambda_{11}^{(1)}\),
\(\{\kappa_{1,j}^{(1)}\}_{j=2}^{J-1}\),
and \(F_{-\varepsilon_{11}^{(1)}}\).

Hence, for each \(d\in\{0,1\}\), \(j\in\mathcal J\), and \(\mathbf x\in\mathcal X_{11}\),
\[
\Pr(Y_{11}(d)=j\mid \mathbf X_{11}=\mathbf x)
=
F_{-\varepsilon_{11}^{(d)}}\!\left(
\frac{\overline{\mathbf x}'\overline{\boldsymbol{\eta}}_{11}^{(d)}
-\kappa_{1,j-1}^{(d)}}
{\lambda_{11}^{(d)}}
\right)
-
F_{-\varepsilon_{11}^{(d)}}\!\left(
\frac{\overline{\mathbf x}'\overline{\boldsymbol{\eta}}_{11}^{(d)}
-\kappa_{1,j}^{(d)}}
{\lambda_{11}^{(d)}}
\right),
\]
where the conventions \(\kappa_{1,-1}^{(d)}=-\infty\) and \(\kappa_{1,J}^{(d)}=+\infty\)
cover the cases \(j=0\) and \(j=J\). Subtracting the expressions for \(d=0\) and
\(d=1\) yields \(\tau(j\mid \mathbf x)\). Integrating with respect to the identified
distribution \(F_{\mathbf X_{11}}\) yields \(\tau(j)\).
\end{proof}

\begin{proof}[Proof of Proposition~\ref{PROP7}]
We first focus on the estimators for the conditional DTTs  and verify the conditions of Theorems~2.5 and 3.3 in \citet{NM1994}.
Under the maintained assumptions, Theorem~2.5(ii), (iv) and
Theorem~3.3(i), (iv), (v) hold immediately.  Besides, Theorem~3.3(iii) in \citet{NM1994} (see also the discussion following Example~1.2 therein) requires the integrability conditions
\begin{align*}
&\sum_{j\in\mathcal J}\sum_{g\in\{0,1\}}\sum_{t\in\{0,1\}}
\pi_{gt}
\int
\sup_{\boldsymbol{\theta}\in\mathcal N}
\left\|
\nabla_{\boldsymbol{\theta}}
p_{gt,j}^{(g\cdot t)}(\mathbf x,\mathbf z;\boldsymbol{\theta})
\right\|
\,dF_{\mathbf X_{gt},\mathbf Z_{gt}}(\mathbf x,\mathbf z)
<\infty,
\\
&\sum_{j\in\mathcal J}\sum_{g\in\{0,1\}}\sum_{t\in\{0,1\}}
\pi_{gt}
\int
\sup_{\boldsymbol{\theta}\in\mathcal N}
\left\|
\nabla_{\boldsymbol{\theta}\boldsymbol{\theta}}^2
p_{gt,j}^{(g\cdot t)}(\mathbf x,\mathbf z;\boldsymbol{\theta})
\right\|
\,dF_{\mathbf X_{gt},\mathbf Z_{gt}}(\mathbf x,\mathbf z)
<\infty.
\end{align*}
These conditions are immediate from Assumption~\ref{AE3}(iii), since the index sets are finite and each expectation is finite.

Since the model is correctly specified by Assumption~\ref{AE2}, by Propositions~\ref{PROP3}--\ref{PROP4}, all parameters are identified except for $\rho_0$. Since $\rho\mapsto \overline{C}(u, v ; \rho)$ is strictly increasing by Assumption~\ref{AE3}(ii), $\rho_0$ is identified from the observed conditional choice probabilities as well, so Theorem~2.5(i) is satisfied.  

Since \(\overline{C}(u,v;\rho)\) is twice continuously differentiable in
\((u,v,\rho)\) by Assumption~\ref{AE3}(ii), and since the likelihood functions are
composed of \(\overline{C}\), \(\Phi\), and elementary functions with strictly positive
denominators on the admissible parameter space, it follows that
\(\ell_i(\boldsymbol{\theta})\) is continuous in \(\boldsymbol{\theta}\) on
\(\boldsymbol{\Theta}\), and that \(L_i(\boldsymbol{\theta})\) is twice continuously
differentiable in \(\boldsymbol{\theta}\) on \(\mathcal N\). Moreover, because
\(\overline{C}\) is strictly increasing in each coordinate by
Assumptions~\ref{S3}(ii), \ref{S5}(ii), and \ref{AE3}(ii), and because
\(\mathcal N\) is contained in the admissible parameter space, we have
\(L_i(\boldsymbol{\theta})>0\) for all \(\boldsymbol{\theta}\in\mathcal N\).
Hence, Theorem~2.5(iii) and Theorem~3.3(ii) are satisfied.

Therefore, all conditions of Theorems~2.5 and 3.3 in \citet{NM1994} are
satisfied, so consequently,
\[
\sqrt{N}\left(\widehat{\boldsymbol{\theta}}-\boldsymbol{\theta}_{0}\right)
\rightsquigarrow
N\left(0,\mathbf J_0^{-1}\right).
\]

For the rest of the proof, we focus on $\widehat{\tau}(j\mid\mathbf x)$ and $\widehat{\tau}(j)$. The arguments for \(\widehat{\tau}^{r}(j\mid\mathbf z)\) and  \(\widehat{\tau}^{r}(j)\) are identical.
 Since \(\tau(j\mid \mathbf x;\boldsymbol{\theta})\) is continuously
differentiable in \(\boldsymbol{\theta}\) on \(\mathcal N\), the delta method
yields
\[
\sqrt{N}\left(\widehat{\tau}(j\mid\mathbf x)-\tau(j\mid\mathbf x)\right)
\rightsquigarrow
N\left(
0,
\nabla_{\boldsymbol{\theta}}\tau(j\mid\mathbf x;\boldsymbol{\theta}_0)^\prime
\mathbf J_0^{-1}
\nabla_{\boldsymbol{\theta}}\tau(j\mid\mathbf x;\boldsymbol{\theta}_0)
\right).
\]

For the unconditional DTT, write
\[
\widehat{\tau}(j)
=
\frac{
\frac{1}{N}\sum_{i=1}^{N}\mathbb{I}\{G_i=1,T_i=1\}
\tau\left(j\mid \mathbf X_i;\widehat{\boldsymbol{\theta}}\right)
}{
\frac{1}{N}\sum_{i=1}^{N}\mathbb{I}\{G_i=1,T_i=1\}
}.
\]
Using $\frac{1}{N}\sum_{i=1}^{N}\mathbb{I}\{G_i=1,T_i=1\}
=\pi_{11}+o_{\Pr}(1)$ and a first-order expansion around $\boldsymbol{\theta}_0$,
\[
\sqrt{N}\left(\widehat{\tau}(j)-\tau(j)\right)
=
\overline{\nabla}_{\boldsymbol{\theta}}\tau_j^\prime
\sqrt{N}\left(\widehat{\boldsymbol{\theta}}-\boldsymbol{\theta}_0\right)
+
\frac{1}{\sqrt{N}}
\sum_{i=1}^{N}
\frac{\mathbb{I}\{G_i=1,T_i=1\}}{\pi_{11}}
\left[
\tau\left(j\mid \mathbf X_i;\boldsymbol{\theta}_0\right)-\tau(j)
\right]
+o_{\Pr}(1).
\]
Substituting the standard asymptotic linearization of the MLE
 \citep[see, e.g.,][Theorem 5.39]{VanderVaart1998} yields
\[
\sqrt{N}\left(\widehat{\tau}(j)-\tau(j)\right)
=
\frac{1}{\sqrt{N}}
\sum_{i=1}^{N}
\psi_{i,\tau_j}
+o_{\Pr}(1),
\]
where
$
\psi_{i,\tau_j}
:=
\overline{\nabla}_{\boldsymbol{\theta}}\tau_j^\prime
\mathbf J_0^{-1}
\nabla_{\boldsymbol{\theta}}\ell_i\left(\boldsymbol{\theta}_0\right)
+
\frac{\mathbb{I}\{G_i=1,T_i=1\}}{\pi_{11}}
\left(
\tau\left(j\mid \mathbf X_i;\boldsymbol{\theta}_0\right)-\tau(j)
\right).
$ Moreover, $\mathrm{E}\left[\psi_{i,\tau_j}\right]=0$, since
$
\mathrm{E}\left[
\nabla_{\boldsymbol{\theta}} \ell_i\left(\boldsymbol{\theta}_0\right)
\mid
\mathbf{X}_i, \mathbf{Z}_i, G_i, T_i
\right]=0
$
a.s. and
$
\mathrm{E}\left[
\frac{\mathbb{I}\left\{G_i=1,T_i=1\right\}}{\pi_{11}}
\left(
\tau\left(j\mid \mathbf X_i;\boldsymbol{\theta}_0\right)-\tau(j)
\right)
\right]
=0.
$ Hence, by the central limit theorem,
\[
\sqrt{N}\left(\widehat{\tau}(j)-\tau(j)\right)
\rightsquigarrow
N\left(0,\operatorname{Var}(\psi_{i,\tau_j})\right).
\]
Using the two mean-zero properties above, $\operatorname{Var}(\psi_{i,\tau_j})$ simplifies to the desired expression.
\end{proof}

\section{Monte Carlo Simulation: Bounds}\label{sec:MCBounds}

In this section, we assess the finite-sample performance of the estimation and inference procedure outlined in Section~\ref{sec:estimation} through a small Monte Carlo simulation for a structural misreporting model. We will only consider Mean Bias (MB), Root Mean Squared Error (RMSE), and empirical coverage rates of the uniform confidence bands for the counterfactual distribution. We will also leave a thorough power analysis to future work.

The structural model for observed consumption at levels $\{0,1,2\}$, which is akin to Section \ref{sec:parametricmodel}, models the observed consumption of interest for each { $(g,t)\neq(1,1)$} as:
\[
C_{gt}(0)=\min\{Y_{gt}(0),R_{gt}(0)\}.
\]
Here, $R_{gt}(0)$ is the potential ``reporting intention'' in the absence of treatment, while $Y_{gt}(0)$ denotes actual consumption in the absence of treatment. Specifically, for each { $(g,t)\neq(1,1)$}, the Data Generating Process (DGP) is given by a bivariate ordered probit model where:
\[
Y_{gt}(0) = 
\begin{cases}
0, & \text{if } Y^{\ast}_{gt}(0) \le 0,\\
1, & \text{if } 0 < Y^{\ast}_{gt}(0) \le 1,\\
2, & \text{if } Y^{\ast}_{gt}(0) > 1,
\end{cases}
\]
and
\[
R_{gt}(0) = 
\begin{cases}
0, & \text{if } R^{\ast}_{gt}(0) \le 0,\\
1, & \text{if } 0 < R^{\ast}_{gt}(0) \le 1,\\
2, & \text{if } R^{\ast}_{gt}(0) > 1.
\end{cases}
\]
We model $Y^{\ast}_{gt}(0)$ as $Y^{\ast}_{gt}(0)=\eta_{Y,gt}- V$ and $R^{\ast}_{gt}(0)$ as $R^{\ast}_{gt}(0)=\eta_{R,gt} + {\beta_{R,gt}} Z - E$, where:
\[
\begin{pmatrix}
V\\
E
\end{pmatrix}
\sim
N\left(
\begin{pmatrix}
0\\
0
\end{pmatrix},
\begin{pmatrix}
1 & \rho\\
\rho & 1
\end{pmatrix}
\right).
\]
Thus, $\rho$ captures the dependence between actual consumption and misreporting behavior. The variable $Z$, on the other hand, is the instrumental variable that only affects the reporting intention and takes on values $\{-\sqrt{3/2}, 0, \sqrt{3/2}\}$ with equal probability so that its mean is normalized to zero and its variance to one. For simplicity, we set { $\eta_{R,gt}=1$, $\beta_{R,00}=\beta_{R,01}=0.5$, and consider $\beta_{R,10}=0.5$ and $\beta_{R,10}=0$, with the latter producing a tie across the conditional CDFs of $C_{10}(0)$}. { In each design, we use the DGP-implied maximum cumulative misreporting probability as $\alpha$. Thus, the relevant bound maps are Hadamard differentiable in the no-tie design and only Hadamard directionally differentiable in the exact-tie design.} For the constants $\{\eta_{Y,00},\eta_{Y,01},\eta_{Y,10}\}$ we choose different parameter values, namely $\{-0.3,{-0.05},-0.2\}$. For the correlation parameter $\rho$, we also use different correlation patterns ranging from no correlation to relatively strong positive (negative) correlation. Specifically, we select $\rho\in\{-0.5,-0.25,0,0.25,0.5\}$. The resulting theoretical bounds {on $F_{Y_{11}(0)}$ in equation~(\ref{final2BOUND2})} are displayed in Table \ref{tab:theoretical_bounds} for the different $\rho$ parameter values.

\begin{table}[htbp!]
\centering
\caption{Theoretical Lower and Upper Bounds}
\label{tab:theoretical_bounds}
{\small
\setlength{\tabcolsep}{6pt}
\renewcommand{\arraystretch}{0.95}
\begin{tabular}{cc|ccc}
\hline\hline
$\rho$ & Bound & $y=0$ & $y=1$ & $y=2$ \\
\hline
\multicolumn{5}{l}{\textit{Panel A: No-tie design ($\beta_{R,10}=0.5$)}} \\
\hline
-0.50 & $L_{11,\alpha}^{(0)}$ & 0.0000 & 0.0000 & 1.0000 \\
 & $U_{11,\alpha}^{(0)}$ & 0.9541 & 1.0000 & 1.0000 \\[3pt]
-0.25 & $L_{11,\alpha}^{(0)}$ & 0.0000 & 0.0767 & 1.0000 \\
 & $U_{11,\alpha}^{(0)}$ & 0.9335 & 1.0000 & 1.0000 \\[3pt]
0 & $L_{11,\alpha}^{(0)}$ & 0.0000 & 0.3159 & 1.0000 \\
 & $U_{11,\alpha}^{(0)}$ & 0.9160 & 0.9160 & 1.0000 \\[3pt]
0.25 & $L_{11,\alpha}^{(0)}$ & 0.0000 & 0.4279 & 1.0000 \\
 & $U_{11,\alpha}^{(0)}$ & 0.9015 & 0.9015 & 1.0000 \\[3pt]
0.50 & $L_{11,\alpha}^{(0)}$ & 0.0000 & 0.5003 & 1.0000 \\
 & $U_{11,\alpha}^{(0)}$ & 0.5838 & 0.8908 & 1.0000 \\
\hline
\multicolumn{5}{l}{\textit{Panel B: Exact-tie design ($\beta_{R,10}=0$)}} \\
\hline
-0.50 & $L_{11,\alpha}^{(0)}$ & 0.0000 & 0.0000 & 1.0000 \\
 & $U_{11,\alpha}^{(0)}$ & 0.9802 & 1.0000 & 1.0000 \\[3pt]
-0.25 & $L_{11,\alpha}^{(0)}$ & 0.0000 & 0.0084 & 1.0000 \\
 & $U_{11,\alpha}^{(0)}$ & 0.9617 & 1.0000 & 1.0000 \\[3pt]
0 & $L_{11,\alpha}^{(0)}$ & 0.0000 & 0.2920 & 1.0000 \\
 & $U_{11,\alpha}^{(0)}$ & 0.9425 & 0.9425 & 1.0000 \\[3pt]
0.25 & $L_{11,\alpha}^{(0)}$ & 0.0000 & 0.4092 & 1.0000 \\
 & $U_{11,\alpha}^{(0)}$ & 0.9232 & 0.9232 & 1.0000 \\[3pt]
0.50 & $L_{11,\alpha}^{(0)}$ & 0.0000 & 0.4859 & 1.0000 \\
 & $U_{11,\alpha}^{(0)}$ & 0.6016 & 0.9047 & 1.0000 \\
\hline\hline
\end{tabular}
}
\end{table}

For the Monte Carlo analysis, we use $2,000$ replications for each of the three different sample sizes, { $n_{gt}\in\{500,1{,}000,10{,}000\}$}. For inference, we implement the bootstrap procedure from Section~\ref{sec:estimation} using the warp-speed procedure \citep{Giacomini2013}. As outlined in that section, we compute the numerical derivative for the bounds using the tuning parameter $\epsilon_{N}$. Here, we present a range of choices for $\epsilon_{N}$ scaled relative to { $n_{gt}^{-1/2}$}, which we label $\epsilon_{N}^{Naive}$ in analogy to \citet{MP2020}. We consider { seven} different scaling choices of $\epsilon_{N}^{Naive}$, namely $\epsilon_{N}=K\cdot\epsilon_{N}^{Naive}$ with { $K\in\{0.25,0.5,1,1.5,5,10,20\}$}. { Finally, we set $\gamma=0.1$, corresponding to a nominal coverage level of $90\%$ for $\mathrm{CS}_{\alpha}^{(0)}(y;1-\gamma)$.}

Turning to the results, we start with the MB and RMSE results for the counterfactual bound estimators in Tables \ref{tab:bias_results} and \ref{tab:rmse_results}, respectively. In particular, we observe that the bias is generally small for the chosen { designs}.\footnote{It is by construction zero for the lower bound at $y=0$ and for both bounds at $y=2$.} Moreover, together with the RMSE, the evidence suggests that the estimators in equation~(\ref{ESTLUBOUND}) are consistent for the lower and upper bound, respectively, and that the contribution of estimation error from $\underline{k}_{\alpha}(y)$ and $\overline{k}_{\alpha}(y)$ is not too large in finite to moderate samples. { Moving to the empirical coverage rates in Table \ref{tab:coverage}, we note that in the no-tie design, the coverage results are, as expected, largely insensitive to the choice of $K$. On the contrary,  in the exact-tie case there is more variation across the choice of $K$.} Moreover, we observe that the coverage results generally appear to be sensitive to the tuning parameter $\epsilon_{N}$ with too large choices of $\epsilon_{N}$ leading to over-coverage in small samples, while too small choices can sometimes lead to under-coverage. This is similar to the findings in \citet{MP2020}, but is difficult to address due to the non-differentiability of the bounds (see their paper for details). { Overall, $K=1$ performs reasonably well in most cases across both designs, particularly at larger sample sizes,  albeit being sometimes a bit too liberal in the exact-tie case.}

\begin{table}[htbp!]
\centering
\caption{Empirical MB Results for $\widehat{L}_{11,\alpha}^{(0)}$ and $\widehat{U}_{11,\alpha}^{(0)}$}
\label{tab:bias_results}
{\small
\setlength{\tabcolsep}{4pt}
\renewcommand{\arraystretch}{0.92}
\begin{tabular}{cc|ccc|ccc}
\hline\hline
 & & \multicolumn{3}{c}{$\widehat{L}_{11,\alpha}^{(0)}$} & \multicolumn{3}{c}{$\widehat{U}_{11,\alpha}^{(0)}$} \\
\cline{3-5} \cline{6-8}
$\rho$ & $n_{gt}$ & $y=0$ & $y=1$ & $y=2$ & $y=0$ & $y=1$ & $y=2$ \\
\hline
\multicolumn{8}{l}{\textit{Panel A: No-tie design ($\beta_{R,10}=0.5$)}} \\
\hline
-0.50 & 500 & 0.0000 & 0.0009 & 0.0000 & -0.0006 & -0.0080 & 0.0000 \\
 & 1,000 & 0.0000 & 0.0000 & 0.0000 & -0.0003 & -0.0044 & 0.0000 \\
 & 10,000 & 0.0000 & 0.0000 & 0.0000 & 0.0000 & 0.0000 & 0.0000 \\[3pt]
-0.25 & 500 & 0.0000 & 0.0057 & 0.0000 & -0.0012 & -0.0246 & 0.0000 \\
 & 1,000 & 0.0000 & 0.0023 & 0.0000 & -0.0007 & -0.0222 & 0.0000 \\
 & 10,000 & 0.0000 & 0.0005 & 0.0000 & 0.0000 & -0.0047 & 0.0000 \\[3pt]
0 & 500 & 0.0000 & 0.0045 & 0.0000 & -0.0031 & 0.0271 & 0.0000 \\
 & 1,000 & 0.0000 & -0.0003 & 0.0000 & -0.0010 & 0.0247 & 0.0000 \\
 & 10,000 & 0.0000 & 0.0003 & 0.0000 & 0.0000 & 0.0057 & 0.0000 \\[3pt]
0.25 & 500 & 0.0000 & 0.0107 & 0.0000 & -0.0510 & 0.0120 & 0.0000 \\
 & 1,000 & 0.0000 & 0.0004 & 0.0000 & -0.0179 & 0.0065 & 0.0000 \\
 & 10,000 & 0.0000 & 0.0000 & 0.0000 & -0.0001 & 0.0000 & 0.0000 \\[3pt]
0.50 & 500 & 0.0000 & 0.0176 & 0.0000 & 0.0673 & -0.0025 & 0.0000 \\
 & 1,000 & 0.0000 & 0.0003 & 0.0000 & 0.0589 & -0.0028 & 0.0000 \\
 & 10,000 & 0.0000 & 0.0000 & 0.0000 & 0.0149 & -0.0001 & 0.0000 \\
\hline
\multicolumn{8}{l}{\textit{Panel B: Exact-tie design ($\beta_{R,10}=0$)}} \\
\hline
-0.50 & 500 & 0.0000 & 0.0005 & 0.0000 & -0.0092 & -0.0030 & 0.0000 \\
 & 1,000 & 0.0000 & 0.0000 & 0.0000 & -0.0066 & -0.0012 & 0.0000 \\
 & 10,000 & 0.0000 & 0.0000 & 0.0000 & -0.0020 & 0.0000 & 0.0000 \\[3pt]
-0.25 & 500 & 0.0000 & 0.0240 & 0.0000 & -0.0128 & -0.0166 & 0.0000 \\
 & 1,000 & 0.0000 & 0.0153 & 0.0000 & -0.0092 & -0.0129 & 0.0000 \\
 & 10,000 & 0.0000 & 0.0028 & 0.0000 & -0.0028 & -0.0009 & 0.0000 \\[3pt]
0 & 500 & 0.0000 & 0.0042 & 0.0000 & -0.0157 & 0.0098 & 0.0000 \\
 & 1,000 & 0.0000 & 0.0004 & 0.0000 & -0.0112 & 0.0096 & 0.0000 \\
 & 10,000 & 0.0000 & 0.0004 & 0.0000 & -0.0035 & 0.0006 & 0.0000 \\[3pt]
0.25 & 500 & 0.0000 & 0.0112 & 0.0000 & -0.0642 & -0.0026 & 0.0000 \\
 & 1,000 & 0.0000 & 0.0004 & 0.0000 & -0.0280 & -0.0048 & 0.0000 \\
 & 10,000 & 0.0000 & 0.0000 & 0.0000 & -0.0040 & -0.0039 & 0.0000 \\[3pt]
0.50 & 500 & 0.0000 & 0.0182 & 0.0000 & 0.0505 & -0.0141 & 0.0000 \\
 & 1,000 & 0.0000 & 0.0003 & 0.0000 & 0.0447 & -0.0121 & 0.0000 \\
 & 10,000 & 0.0000 & 0.0000 & 0.0000 & 0.0079 & -0.0044 & 0.0000 \\
\hline\hline
\end{tabular}
}
\end{table}

\begin{table}[htbp!]
\centering
\caption{Empirical RMSE Results for $\widehat{L}_{11,\alpha}^{(0)}$ and $\widehat{U}_{11,\alpha}^{(0)}$}
\label{tab:rmse_results}
{\small
\setlength{\tabcolsep}{4pt}
\renewcommand{\arraystretch}{0.92}
\begin{tabular}{cc|ccc|ccc}
\hline\hline
 & & \multicolumn{3}{c}{$\widehat{L}_{11,\alpha}^{(0)}$} & \multicolumn{3}{c}{$\widehat{U}_{11,\alpha}^{(0)}$} \\
\cline{3-5} \cline{6-8}
$\rho$ & $n_{gt}$ & $y=0$ & $y=1$ & $y=2$ & $y=0$ & $y=1$ & $y=2$ \\
\hline
\multicolumn{8}{l}{\textit{Panel A: No-tie design ($\beta_{R,10}=0.5$)}} \\
\hline
-0.50 & 500 & 0.0000 & 0.0273 & 0.0000 & 0.0152 & 0.0203 & 0.0000 \\
 & 1,000 & 0.0000 & 0.0000 & 0.0000 & 0.0111 & 0.0147 & 0.0000 \\
 & 10,000 & 0.0000 & 0.0000 & 0.0000 & 0.0036 & 0.0000 & 0.0000 \\[3pt]
-0.25 & 500 & 0.0000 & 0.0670 & 0.0000 & 0.0180 & 0.0421 & 0.0000 \\
 & 1,000 & 0.0000 & 0.0426 & 0.0000 & 0.0132 & 0.0393 & 0.0000 \\
 & 10,000 & 0.0000 & 0.0135 & 0.0000 & 0.0043 & 0.0176 & 0.0000 \\[3pt]
0 & 500 & 0.0000 & 0.0625 & 0.0000 & 0.0252 & 0.0515 & 0.0000 \\
 & 1,000 & 0.0000 & 0.0295 & 0.0000 & 0.0144 & 0.0478 & 0.0000 \\
 & 10,000 & 0.0000 & 0.0096 & 0.0000 & 0.0048 & 0.0224 & 0.0000 \\[3pt]
0.25 & 500 & 0.0000 & 0.0778 & 0.0000 & 0.1243 & 0.0437 & 0.0000 \\
 & 1,000 & 0.0000 & 0.0299 & 0.0000 & 0.0725 & 0.0320 & 0.0000 \\
 & 10,000 & 0.0000 & 0.0074 & 0.0000 & 0.0052 & 0.0061 & 0.0000 \\[3pt]
0.50 & 500 & 0.0000 & 0.0868 & 0.0000 & 0.1566 & 0.0323 & 0.0000 \\
 & 1,000 & 0.0000 & 0.0263 & 0.0000 & 0.1427 & 0.0198 & 0.0000 \\
 & 10,000 & 0.0000 & 0.0061 & 0.0000 & 0.0690 & 0.0054 & 0.0000 \\
\hline
\multicolumn{8}{l}{\textit{Panel B: Exact-tie design ($\beta_{R,10}=0$)}} \\
\hline
-0.50 & 500 & 0.0000 & 0.0205 & 0.0000 & 0.0132 & 0.0097 & 0.0000 \\
 & 1,000 & 0.0000 & 0.0000 & 0.0000 & 0.0092 & 0.0055 & 0.0000 \\
 & 10,000 & 0.0000 & 0.0000 & 0.0000 & 0.0028 & 0.0000 & 0.0000 \\[3pt]
-0.25 & 500 & 0.0000 & 0.0575 & 0.0000 & 0.0179 & 0.0299 & 0.0000 \\
 & 1,000 & 0.0000 & 0.0338 & 0.0000 & 0.0128 & 0.0250 & 0.0000 \\
 & 10,000 & 0.0000 & 0.0116 & 0.0000 & 0.0038 & 0.0063 & 0.0000 \\[3pt]
0 & 500 & 0.0000 & 0.0641 & 0.0000 & 0.0263 & 0.0376 & 0.0000 \\
 & 1,000 & 0.0000 & 0.0302 & 0.0000 & 0.0153 & 0.0342 & 0.0000 \\
 & 10,000 & 0.0000 & 0.0097 & 0.0000 & 0.0046 & 0.0157 & 0.0000 \\[3pt]
0.25 & 500 & 0.0000 & 0.0799 & 0.0000 & 0.1292 & 0.0379 & 0.0000 \\
 & 1,000 & 0.0000 & 0.0303 & 0.0000 & 0.0745 & 0.0276 & 0.0000 \\
 & 10,000 & 0.0000 & 0.0074 & 0.0000 & 0.0053 & 0.0058 & 0.0000 \\[3pt]
0.50 & 500 & 0.0000 & 0.0889 & 0.0000 & 0.1499 & 0.0337 & 0.0000 \\
 & 1,000 & 0.0000 & 0.0265 & 0.0000 & 0.1374 & 0.0216 & 0.0000 \\
 & 10,000 & 0.0000 & 0.0062 & 0.0000 & 0.0673 & 0.0059 & 0.0000 \\
\hline\hline
\end{tabular}
}
\end{table}

\begin{table}[htbp!]
\centering
\caption{Empirical Coverage Rates for $\mathrm{CS}_{\alpha}^{(0)}(y;1-\gamma)$, $y\in\{0,1,2\}$ with $\gamma=0.1$}
\label{tab:coverage}
{\small
\setlength{\tabcolsep}{3.5pt}
\renewcommand{\arraystretch}{0.92}
\begin{tabular}{cc|rrrrrrr}
\hline\hline
 & & \multicolumn{7}{c}{$\epsilon_N/\epsilon_N^{\mathrm{Naive}}$} \\
\cline{3-9}
$\rho$ & $n_{gt}$ & 0.25 & 0.5 & 1 & 1.5 & 5 & 10 & 20 \\
\hline
\multicolumn{9}{l}{\textit{Panel A: No-tie design ($\beta_{R,10}=0.5$)}} \\
\hline
-0.50 & 500 & 0.762 & 0.762 & 0.765 & 0.767 & 0.901 & 0.997 & 0.999 \\
 & 1,000 & 0.803 & 0.804 & 0.805 & 0.808 & 0.856 & 0.959 & 0.999 \\
 & 10,000 & 0.905 & 0.905 & 0.905 & 0.905 & 0.905 & 0.928 & 0.986 \\[3pt]
-0.25 & 500 & 0.774 & 0.821 & 0.821 & 0.821 & 0.821 & 0.821 & 0.821 \\
 & 1,000 & 0.635 & 0.635 & 0.635 & 0.635 & 0.635 & 0.635 & 0.635 \\
 & 10,000 & 0.832 & 0.832 & 0.832 & 0.832 & 0.832 & 0.832 & 0.832 \\[3pt]
0 & 500 & 0.901 & 0.901 & 0.901 & 0.901 & 0.901 & 0.901 & 0.901 \\
 & 1,000 & 0.900 & 0.900 & 0.900 & 0.900 & 0.900 & 0.900 & 0.900 \\
 & 10,000 & 0.886 & 0.886 & 0.886 & 0.886 & 0.886 & 0.890 & 0.891 \\[3pt]
0.25 & 500 & 0.776 & 0.776 & 0.776 & 0.776 & 0.777 & 0.783 & 0.785 \\
 & 1,000 & 0.855 & 0.855 & 0.855 & 0.855 & 0.865 & 0.865 & 0.866 \\
 & 10,000 & 0.911 & 0.911 & 0.911 & 0.911 & 0.911 & 0.916 & 0.920 \\[3pt]
0.50 & 500 & 0.842 & 0.853 & 0.855 & 0.868 & 0.885 & 0.892 & 0.894 \\
 & 1,000 & 0.893 & 0.897 & 0.906 & 0.916 & 0.930 & 0.936 & 0.942 \\
 & 10,000 & 0.898 & 0.900 & 0.904 & 0.904 & 0.916 & 0.929 & 0.936 \\
\hline
\multicolumn{9}{l}{\textit{Panel B: Exact-tie design ($\beta_{R,10}=0$)}} \\
\hline
-0.50 & 500 & 0.763 & 0.776 & 0.791 & 0.799 & 0.936 & 1.000 & 1.000 \\
 & 1,000 & 0.758 & 0.768 & 0.793 & 0.799 & 0.861 & 0.997 & 1.000 \\
 & 10,000 & 0.760 & 0.779 & 0.803 & 0.823 & 0.871 & 0.887 & 0.946 \\[3pt]
-0.25 & 500 & 0.769 & 0.817 & 0.824 & 0.851 & 0.882 & 0.884 & 0.885 \\
 & 1,000 & 0.706 & 0.752 & 0.789 & 0.789 & 0.834 & 0.853 & 0.853 \\
 & 10,000 & 0.842 & 0.850 & 0.857 & 0.861 & 0.866 & 0.866 & 0.866 \\[3pt]
0 & 500 & 0.898 & 0.898 & 0.898 & 0.898 & 0.898 & 0.898 & 0.898 \\
 & 1,000 & 0.903 & 0.903 & 0.903 & 0.903 & 0.903 & 0.904 & 0.904 \\
 & 10,000 & 0.890 & 0.890 & 0.890 & 0.890 & 0.895 & 0.897 & 0.897 \\[3pt]
0.25 & 500 & 0.768 & 0.768 & 0.768 & 0.768 & 0.781 & 0.781 & 0.781 \\
 & 1,000 & 0.834 & 0.834 & 0.845 & 0.846 & 0.848 & 0.848 & 0.848 \\
 & 10,000 & 0.871 & 0.871 & 0.879 & 0.879 & 0.892 & 0.892 & 0.897 \\[3pt]
0.50 & 500 & 0.764 & 0.779 & 0.786 & 0.799 & 0.838 & 0.849 & 0.856 \\
 & 1,000 & 0.806 & 0.814 & 0.833 & 0.849 & 0.874 & 0.880 & 0.892 \\
 & 10,000 & 0.773 & 0.783 & 0.797 & 0.816 & 0.858 & 0.870 & 0.876 \\
\hline\hline
\end{tabular}
}
\end{table}
\newpage

\section{Additional Empirical Results}\label{sec:ADDEMP}

\begin{table}[htbp!]
\begin{center}
\caption{Descriptive Statistics}

\begin{tabular}{lrrrrrrr}
\hline
\hline
Variable & N & Mean & Std. Dev. & Min & Pctl. 25 & Pctl. 75 & Max \\ 
\hline
\texttt{cons\_30days} & 46472 &  &  &  &  &  &  \\ 
... 1-2 occasions & 1290 & 3\% &  &  &  &  &  \\ 
... 3 or more occasions & 1375 & 3\% &  &  &  &  &  \\ 
... no use & 43807 & 94\% &  &  &  &  &  \\ 
\texttt{male} & 46472 & 0.48 & 0.5 & 0 & 0 & 1 & 1 \\ 
\texttt{age\_month} & 46472 & 169 & 6.2 & 122 & 165 & 173 & 231 \\ 
\texttt{white} & 46472 & 0.47 & 0.5 & 0 & 0 & 1 & 1 \\ 
\texttt{parent\_edu\_clg} & 46472 & 0.58 & 0.49 & 0 & 0 & 1 & 1 \\ 
\texttt{state\_unemp} & 46472 & 5.3 & 0.78 & 2.9 & 4.8 & 5.9 & 6.8 \\ 
\texttt{state\_median\_income} & 46472 & 56552 & 7024 & 40037 & 51983 & 60925 & 75675 \\ 
\texttt{state\_price\_m} & 46472 & 240 & 35 & 159 & 220 & 265 & 325 \\ 
\texttt{trust} & 46472 & 0.96 & 0.072 & 0.28 & 0.96 & 1 & 1 \\ 
\texttt{treated} & 46472 & 0.19 & 0.39 & 0 & 0 & 0 & 1 \\ 
\texttt{after} & 46472 & 0.47 & 0.5 & 0 & 0 & 1 & 1\\ 
\hline
\hline
\end{tabular}
\end{center}
\footnotesize Note: Data for 8th graders from the ``Monitoring the Future'' survey. The variables \texttt{male}, \texttt{white}, \texttt{parent\_edu\_clg} are binary indicators for self-reported gender status (male), ethnicity (white), and parental education level (college) equal to one if the category was true, and zero otherwise. For the other variables, please refer to the text for the definition.\normalsize

\end{table}

\begin{figure}[htbp]
  \begin{center}
  \caption{Estimated QTT Bounds for  Subpopulations W/ and W/O Underreporting.}\label{FIG3} 

    \includegraphics[width=0.45\linewidth]{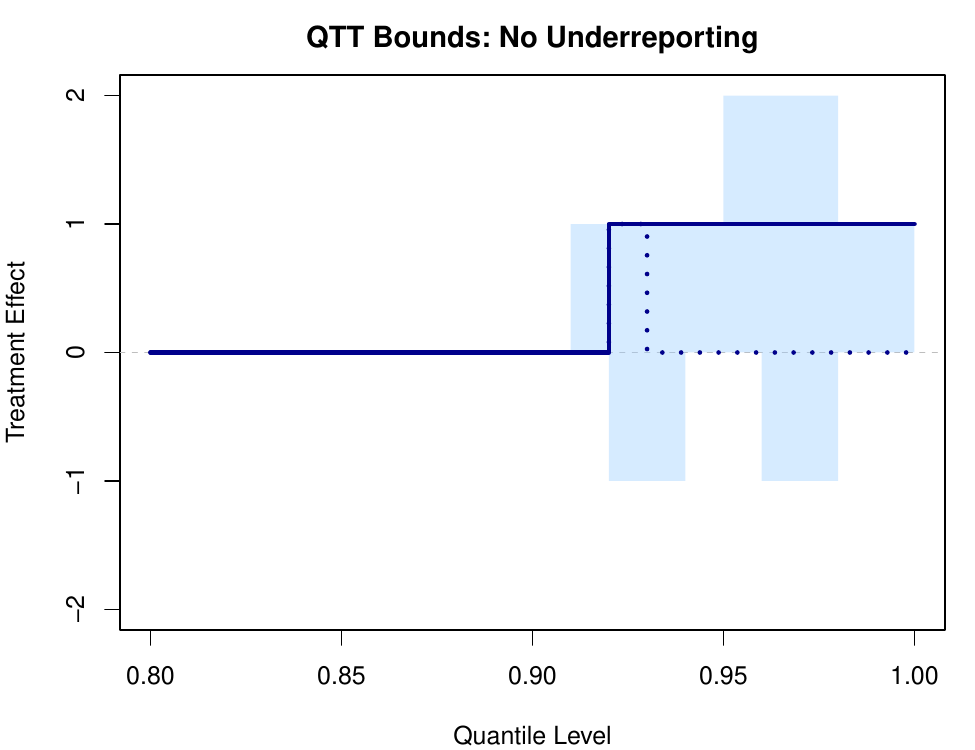}
    \includegraphics[width=0.45\linewidth]{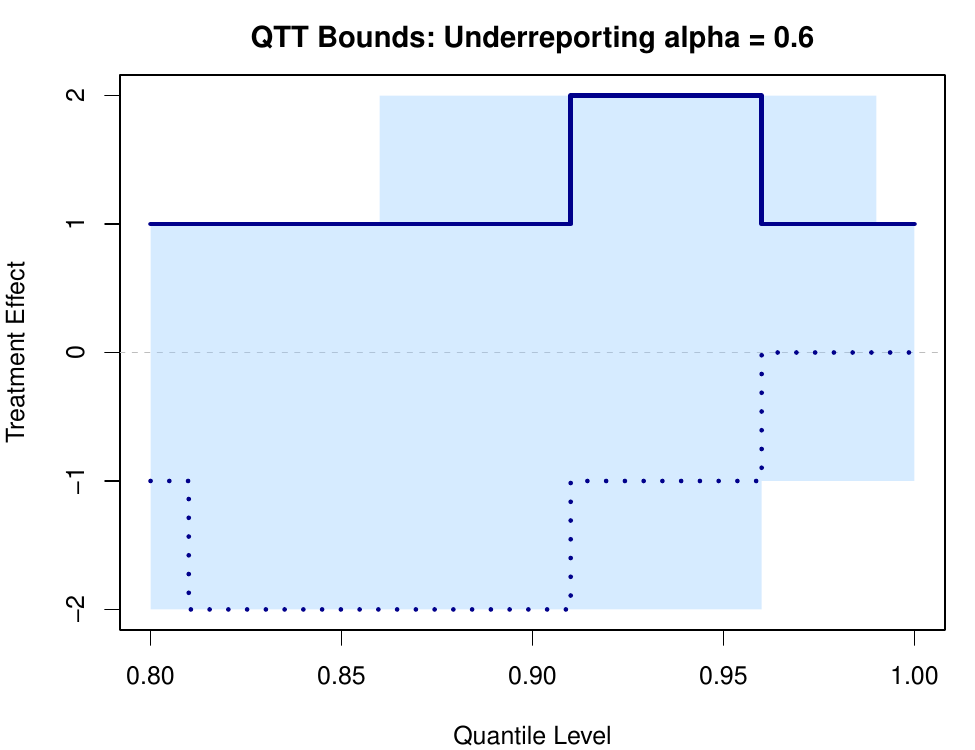}
   \\
    \includegraphics[width=0.45\linewidth]{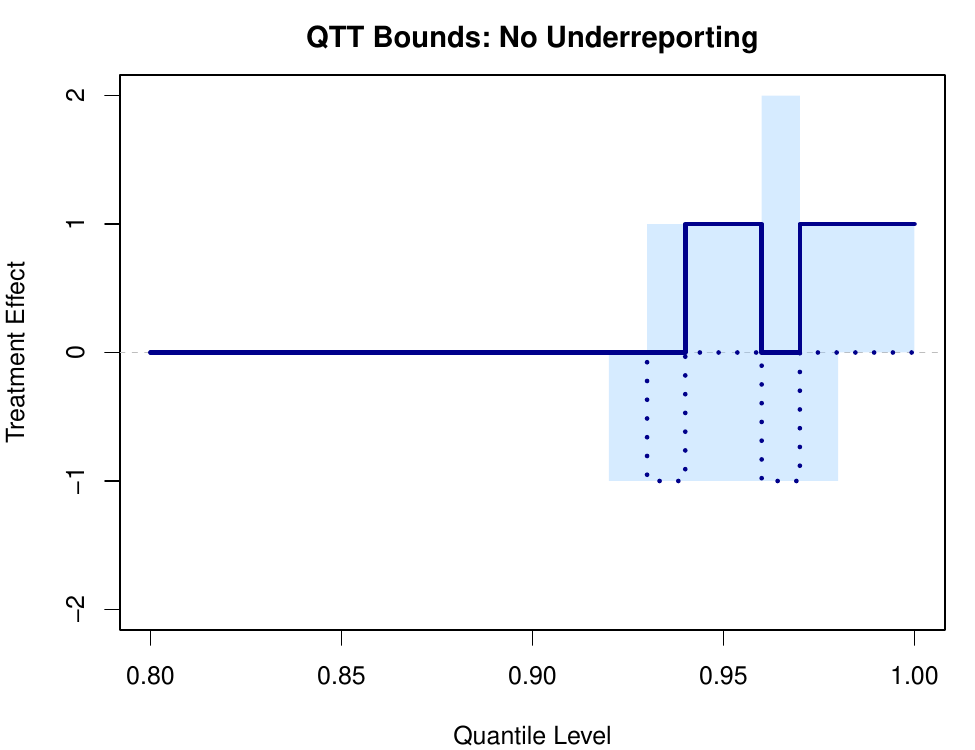}
    \includegraphics[width=0.45\linewidth]{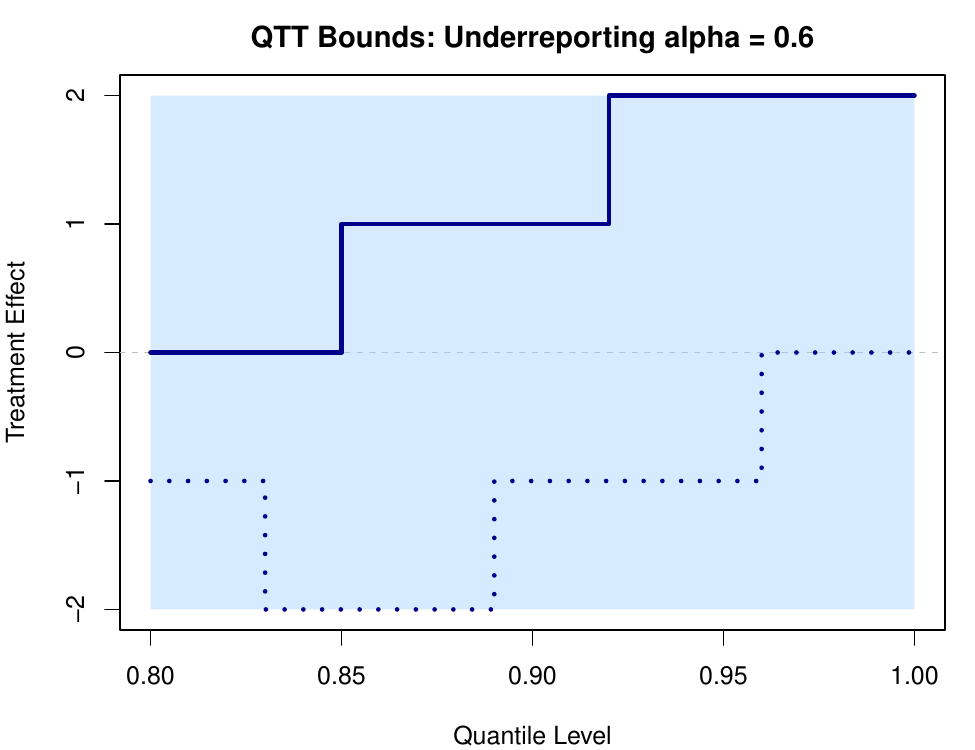}
  \end{center}

  {\footnotesize Note: The figure contains estimated  bounds for the QTTs $\Delta_{\mathrm{QTT}}(\tau)
$, $\tau\in(0.8,1)$, for two different subpopulations with and without underreporting. The upper panel displays the estimated QTT bounds for the subpopulation $\texttt{white}=0$ and $\texttt{male}=0$: the left figure exhibits the QTT bounds w/o underreporting, the right figure exhibits the QTT bounds w/ underreporting for $\alpha=0.6$. The lower panel  displays the estimated QTT bounds for the subpopulation $\texttt{white}=0$ and $\texttt{male}=1$: the left figure exhibits the QTT bounds w/o underreporting, the right figure exhibits the QTT bounds w/ underreporting for $\alpha=0.6$. Lower bound estimates appear in all figures as dotted lines, upper bound estimates as solid lines. Shaded areas illustrate uniform Confidence Sets $\mathrm{CS}_{\mathrm{QTT},\alpha}\left(\cdot;0.9\right)$. }
\end{figure}

\begin{figure}[htbp]
  \begin{center}
  \caption{Estimated QTT Bounds for Subpopulations W/ and W/O Underreporting.}\label{FIG4} 

    \includegraphics[width=0.45\linewidth]{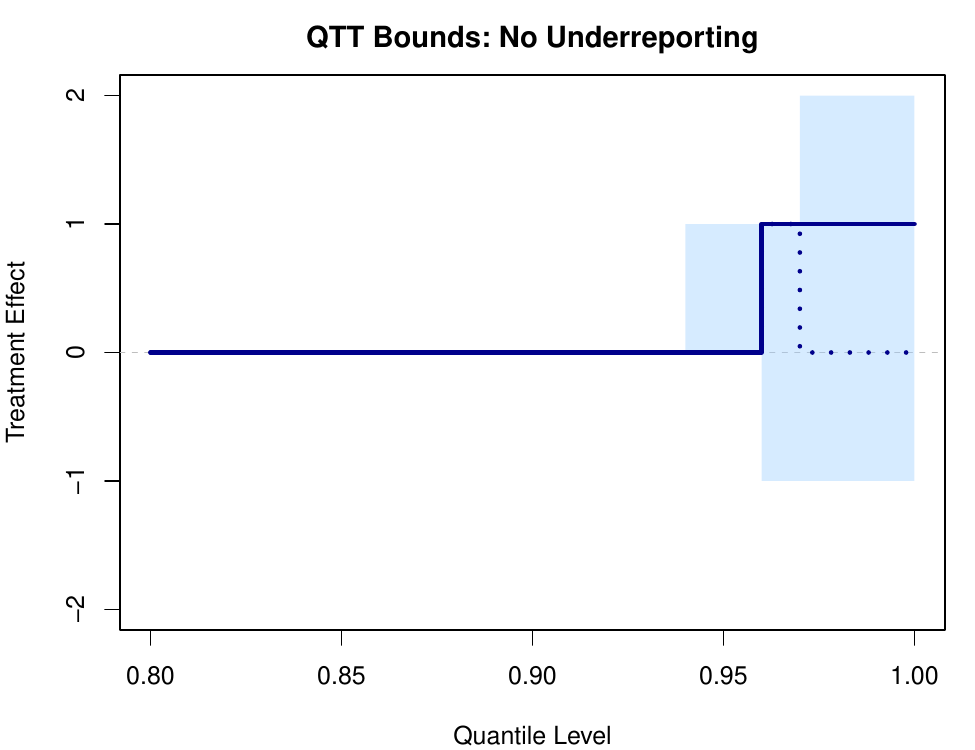}
    \includegraphics[width=0.45\linewidth]{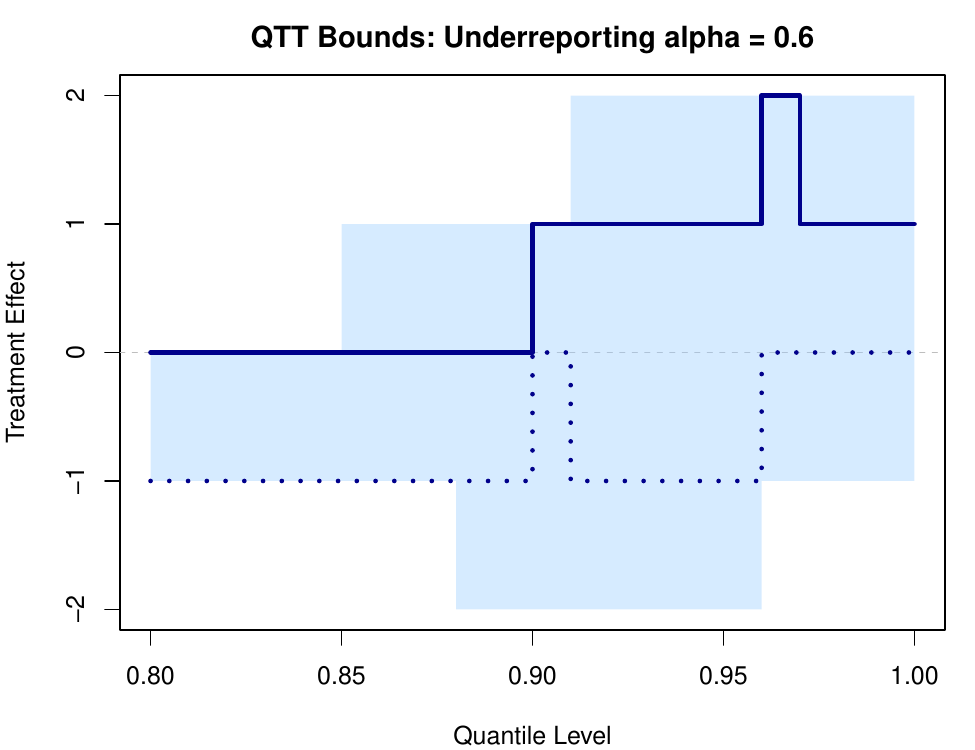}
   \\
    \includegraphics[width=0.45\linewidth]{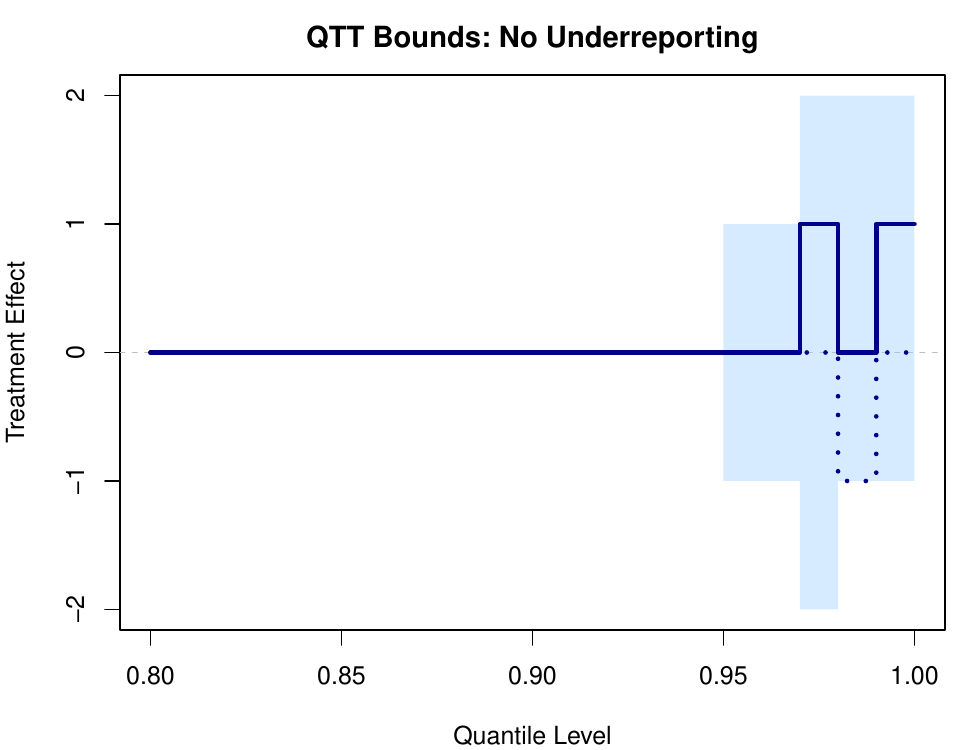}
    \includegraphics[width=0.45\linewidth]{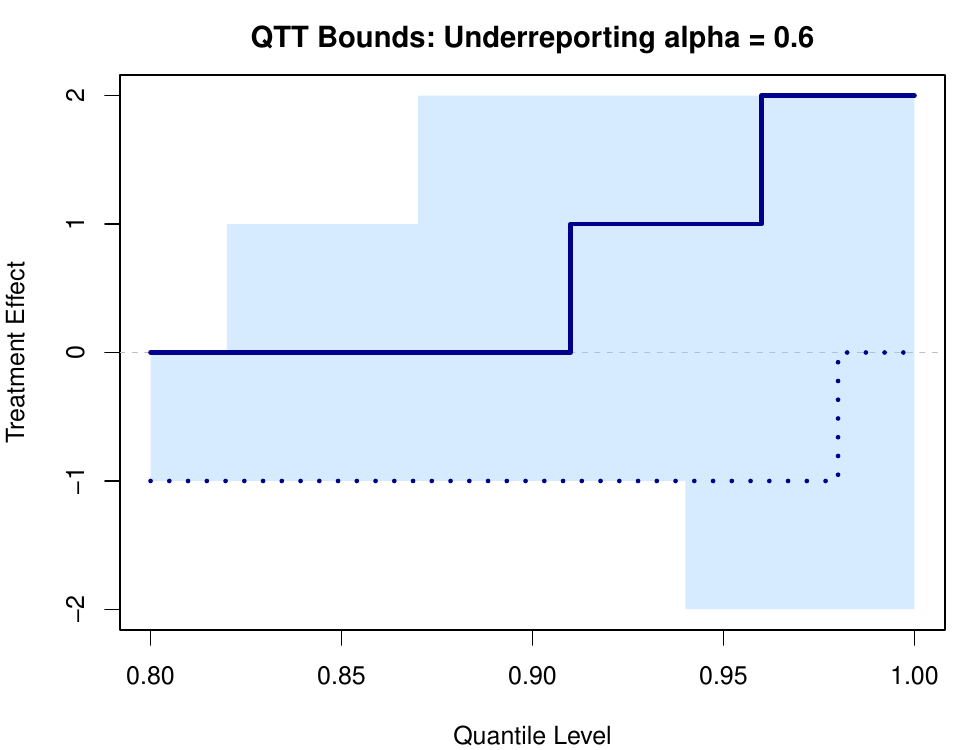}
  \end{center}

  {\footnotesize Note: The figure contains estimated  bounds for the QTTs $\Delta_{\mathrm{QTT}}(\tau)
$, $\tau\in(0.8,1)$, for two different subpopulations with and without underreporting. The upper panel displays the estimated QTT bounds for the subpopulation $\texttt{white}=1$ and $\texttt{male}=0$: the left figure exhibits the QTT bounds w/o underreporting, the right figure exhibits the QTT bounds w/ underreporting for $\alpha=0.6$. The lower panel  displays the estimated QTT bounds for the subpopulation $\texttt{white}=1$ and $\texttt{male}=1$: the left figure exhibits the QTT bounds w/o underreporting, the right figure exhibits the QTT bounds w/ underreporting for $\alpha=0.6$. Lower bound estimates appear in all figures as dotted lines, upper bound estimates as solid lines. Shaded areas illustrate uniform Confidence Sets $\mathrm{CS}_{\mathrm{QTT},\alpha}\left(\cdot;0.9\right)$. }
\end{figure}

\newpage
\spacingset{1}

\putbib[references/GL1]
\end{bibunit}

\end{document}